\documentclass[a4paper,9pt]{article}
\usepackage{mathtools}
\usepackage{mathrsfs} % Красивый матшрифт
\usepackage{dsfont}
\usepackage{xcolor}
\definecolor{burgundy}{rgb}{0.5, 0.0, 0.13}
\definecolor{olive}{rgb}{0.50, 0.50, 0.0}
\usepackage[linktocpage=true,colorlinks=true,linkcolor=burgundy,citecolor=black!20!blue,urlcolor=violet]{hyperref}
\usepackage{soul}

\usepackage[pdftex]{pict2e}

\usepackage[natbib=true, backend = bibtex, style=numeric-comp, sorting=none,maxbibnames=99]{biblatex}
\def\fD{\mathfrak{D}}

\def\MOY{\Gamma}

\RequirePackage{tikz-cd}
\RequirePackage{amssymb}
\usetikzlibrary{calc}
\usetikzlibrary{decorations.pathmorphing}
\usepackage{graphicx}
\usepackage{blindtext}

\tikzset{curve/.style={settings={#1},to path={(\tikztostart)
    .. controls ($(\tikztostart)!\pv{pos}!(\tikztotarget)!\pv{height}!270:(\tikztotarget)$)
    and ($(\tikztostart)!1-\pv{pos}!(\tikztotarget)!\pv{height}!270:(\tikztotarget)$)
    .. (\tikztotarget)\tikztonodes}},
    settings/.code={\tikzset{quiver/.cd,#1}
        \def\pv##1{\pgfkeysvalueof{/tikz/quiver/##1}}},
    quiver/.cd,pos/.initial=0.35,height/.initial=0}
\def\ba{\begin{equation}\begin{aligned}}
\def\ea{\end{aligned}\end{equation}}
\tikzset{tail reversed/.code={\pgfsetarrowsstart{tikzcd to}}}
\tikzset{2tail/.code={\pgfsetarrowsstart{Implies[reversed]}}}
\tikzset{2tail reversed/.code={\pgfsetarrowsstart{Implies}}}
\tikzset{no body/.style={/tikz/dash pattern=on 0 off 1mm}}
\usepackage{amsmath,tikz-cd}
\usepackage[english]{babel}
\usepackage[utf8x]{inputenc}
\usepackage[T1]{fontenc}
\usepackage[T2A]{fontenc}
\usepackage[a4paper,top=1.5 cm,bottom=2cm,left=1cm,right=1cm,marginparwidth=1.75cm]{geometry}

\def\myblue{white!40!blue}

\def\myred{black!40!red}
\usepackage{amsmath, amssymb}
\usepackage{lipsum}
\usepackage{ntheorem}

\newtheorem*{theorem-non}{Theorem}
\usepackage{authblk}

\numberwithin{equation}{section}
\usepackage{wrapfig}
\usepackage{color}
\usepackage{tikz}
\usetikzlibrary{arrows}
\usetikzlibrary{decorations.pathreplacing}
\usepackage[T2A]{fontenc}
\usepackage{circuitikz} 
\graphicspath{{pic/}}
\definecolor{water} {rgb} {0.667, 0.855, 1}
\usepackage{pgfplots}
\usepackage{pgfplotstable}
\usetikzlibrary{circuits}
\usetikzlibrary{circuits.ee}
\usetikzlibrary{circuits.ee.IEC}
\usetikzlibrary{circuits.logic.IEC}
\usetikzlibrary{intersections}
\usepackage{tikz-cd} 
\usepackage{floatrow,calc}
\usepackage{wrapfig}
\newcommand*\upd{\vcenter{\hbox{\includegraphics[width=1.2em]{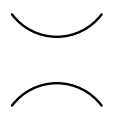}}}}
\newcommand*\ler{\vcenter{\hbox{\includegraphics[width=1.2em]{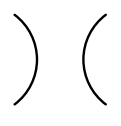}}}}
\date{}

\begin{document}

\title{\bf Khovanov--Rozansky \\ matrix factorization reduction for bipartite links}

%\title{\bf Khovanov-Rozansky cohomology - Matrix Factorizations}

\author[1,2,3]{{\bf E. Lanina}\thanks{\href{mailto:lanina.en@phystech.edu}{lanina.en@phystech.edu}}}
\author[1]{{\bf R. Stepanov}\thanks{\href{mailto:stepanov.rt@phystech.edu}{stepanov.rt@phystech.edu}}}
%\author[1,2,3]{{\bf A. Morozov}\thanks{\href{mailto:morozov@itep.ru}{ morozov@itep.ru}}}

\vspace{5cm}

\affil[1]{Moscow Institute of Physics and Technology, 141700, Dolgoprudny, Russia}
%\affil[2]{Institute for Theoretical and Experimental Physics, 117218, Moscow, Russia}
\affil[2]{Institute for Information Transmission Problems, 127051, Moscow, Russia}
\affil[3]{NRC "Kurchatov Institute", 123182, Moscow, Russia}
\affil[4]{Institute for Theoretical and Experimental Physics, 117218, Moscow, Russia}
\renewcommand\Affilfont{\itshape\small}

\maketitle

\vspace{-7.5cm}

\begin{center}
	\hfill MIPT/TH-19/25\\
	\hfill ITEP/TH-27/25\\
	\hfill IITP/TH-24/25
\end{center}

\vspace{4.7cm}

\begin{abstract}

The Khovanov--Rozansky (KR) link polynomial is a certain $t$-deformation of Wilson loops in 3-dimensional $SU(N)$ Chern--Simons topological field theory, believed to be an observable in the refined Chern–Simons theory, probably described in terms of 4d or 5d QFT and related by a certain procedure to the triply-graded link <<superpolynomial>>.
This link invariant was originally introduced by M. Khovanov and L. Rozansky through a sophisticated matrix factorization technique based on the bicomplex structure, which depends on entire link diagrams and rapidly increases in complexity with the growth of a link.  
However, for particular link diagrams a local reduction is possible, allowing to eliminate vertices in a regular way,
and thus, simplifying the KR polynomial and making it as simple as the Khovanov polynomial in the $N=2$ case.
In particular, for a distinguished family of bipartite links, matrix factorization defined on MOY diagrams
reduces just to planar cycles -- very similar to the original Kauffman--Khovanov construction at $N=2$ for the Jones polynomial and its $t$-deformation.
In the bipartite case, this can be done for any $N$.
We make a further step of simplification and reduce from cohomology factor-rings in even variables crucially depending on a MOY diagram to
vector spaces spanned by odd variables, so that the initial bicomplex of matrix factorizations becomes a monocomplex of just tensor products of $N$-dimensional vector spaces. 
We also find the explicit form of three universal morphisms
which were guessed in a recent paper on this subject. Universality means independence of the other edges of the diagram, and we explain why this works
in this particular case.
%To explain the universality???, we also prove a theorem on excluding variables in initial matrix factorization rings.
%a regular way??? abstract??? 

%The resulting cycle calculus for the Khovanov--Rozansky polynomial for bipartite links is in parallel with the Khovanov polynomial ($N=2$) construction, and we show that the $N=2$ matrix factorization approach reduces to the Khovanov planar technique. This similarity also allows for writing a computer program analogous to one in katlas.org.  

\end{abstract}

\tableofcontents
\section{Introduction}

In this paper, we follow~\cite{K} and show how to obtain the Khovanov--Rozansky cycle calculus for bipartite links from~\cite{2506.08721}. The Khovanov--Rozansky cycle calculus for an arbitrary $N$ for bipartite links is a lift from the Khovanov technique for $N=2$, and first, we explain how to obtain this technique from the Khovanov--Rozansky matrix factorization approach. 

Then, we move to the Khovanov--Rozansky case for bipartite links~\cite{BipKnots}. First, we explain that there are no MOY vertices in the Khovanov–Rozansky complex for bipartite links; instead, all the resolutions are cycles. This fact is actually a simple consequence of the corresponding categorified MOY relation~\eqref{MOYloc_V}, which has been proved in~\cite{KhR,K}. Second, we find the morphisms Sh, $\Delta$, $m$ guessed in~\cite{2506.08721} that appear in the planar Khovanov--Rozansky technique, and this is the main result of this text which is categorification of our planar calculus for the Euler characteristic~\cite{ALM,ALM2,ALM3}.  

These morphisms come from cycle resolutions of a link. The number of variables standing on edges can be arbitrary. Actually, these morphisms could depend on these variables. However, we prove the theorem on the exclusion of variables. It relates an edge with one variable to an edge with two variables via an operator. The morphisms between resolutions become conjugated by this operator, but due to the locality of the morphisms, the conjugation operators commute with the morphisms so that the last ones do not change. In this way, we prove the universality of the found morphisms Sh, $\Delta$, $m$. 

Now, let us comment on the objects under study. The {\it Khovanov--Rozansky polynomial}~\cite{KhR} is a categorification ($t$-deformation) of the {\it HOMFLY knot invariant}~\cite{HOMFLY,PT} being the Wilson loop in the $3d$ topological Chern--Simons theory~\cite{CS,Witten} with the $SU(N)$ gauge group:
\begin{equation}
     \label{WilsonLoopExpValue}
\begin{aligned}
             &H_{R}^{\mathcal{K}}(q, A=q^N) = \left\langle \text{tr}_{R} \ P \exp \left( \oint_{\mathcal{K}} {\cal A} \right) \right\rangle_{\text{CS}}, \\
             &S_{\text{CS}}[{\cal A}] = \frac{\kappa}{4 \pi} \int_{S^3} \text{tr} \left( {\cal A} \wedge d{\cal A} +  \frac{2}{3} {\cal A} \wedge {\cal A} \wedge {\cal A} \right).
\end{aligned}
\end{equation}
Here, the integration contour can be tied in an arbitrary knot $\cal K$, and the gauge fields are taken in an arbitrary $\mathfrak{sl}(N)$ representation $R$: ${\cal A} = {\cal A}_\mu^a T_a^R dx^\mu$ with $T_a$ being generators of $\mathfrak{sl}(N)$ Lie algebra. An interesting feature of this theory is that the Wilson loop is a polynomial in the variables $q = \exp\left(\frac{2\pi i}{\kappa +N}\right)$ and $A=q^N$. The HOMFLY polynomial in the fundamental representation is calculated in practice by the 2-hypercube of resolutions where each crossing is substituted with two types of resolutions, see Section~\ref{sec:HOMFLY}. A whole link resolution is called a {\it MOY graph}~\cite{MOY} or a {\it MOY diagram}.

For $N = 0,\, 2,\, 3$\footnote{The $N=1$ case is trivial because all $H_{R}^{\mathcal{L}}(q, q^N)=1$.} there exists a doubly-graded homology theory of links
whose Euler characteristic is $H_{\Box}^{\mathcal{L}}(q, q^N)$\footnote{We denote the fundamental representation as $\Box$.}. The $N=0$ case corresponds to the {\it Alexander polynomial}~\cite{alexander1928topological} $Al^{\mathcal{L}}(q) = H_{\Box}^{\mathcal{L}}(q, q^0)$, and the homology theory was constructed in~\cite{OS}, and independently, in~\cite{Ra}. The categorified polynomial is called the {\it Floer polynomial}. The celebrated {\it Jones polynomial} is the HOMFLY polynomial for $N=2$: $J^{\mathcal{L}}(q) = H_{\Box}^{\mathcal{L}}(q, q^2)$, its $t$-deformation is the {\it Khovanov polynomial}~\cite{Kh}. The $N=3$ homologies were also constructed by M. Khovanov in~\cite{Kh-sl3}. 

The generalized theory for an arbitrary $N$ was constructed by M. Khovanov and L. Rozansky~\cite{KhR}. It is based on the bicomplex structure:

\begin{equation}\label{KRbicompl}
		\begin{array}{c}
			\begin{tikzpicture}
				\node(A) at (0,0) {$M^0(\MOY_1)$};
				\node(B) at (4,0) {$M^0(\MOY_2)$};
				\node(C) at (8,0) {$M^0(\MOY_3)$};
				\node(E) at (-3,0) {$\ldots$};
				\node(F) at (11,0) {$\ldots$};
				\node(A1) at (0,-1.5) {$M^1(\MOY_1)$};
				\node(B1) at (4,-1.5) {$M^1(\MOY_2)$};
				\node(C1) at (8,-1.5) {$M^1(\MOY_3)$};
				\node(E1) at (-3,-1.5) {$\ldots$};
				\node(F1) at (11,-1.5) {$\ldots$};
				%\node(A0) at (0,1) {$\ldots$};
				%\node(B0) at (4,1) {$\ldots$};
				%\node(C0) at (8,1) {$\ldots$};
				%\node(A2) at (0,-2.5) {$\ldots$};
				%\node(B2) at (4,-2.5) {$\ldots$};
				%\node(C2) at (8,-2.5) {$\ldots$};
				\path (E) edge[->] (A) (A) edge[->] node[above] {$\scriptstyle \fD$} (B) (B) edge[->]  node[above] {$\scriptstyle \fD$}  (C) (C) edge[->] (F) (E1) edge[->] (A1) (A1) edge[->] node[above] {$\scriptstyle \fD$} (B1) (B1) edge[->]  node[above] {$\scriptstyle \fD$}  (C1) (C1) edge[->] (F1) (A) edge[->] node[right] {$\scriptstyle d_0(\Gamma_1)$} (A1) (B) edge[->] node[right] {$\scriptstyle d_0(\Gamma_2)$} (B1) (C) edge[->] node[right] {$\scriptstyle d_0(\Gamma_3)$} (C1) ([shift={(-0.2,0)}]C1.north) edge[->] node[left] {$\scriptstyle d_1(\Gamma_3)$} ([shift={(-0.2,0)}]C.south) (C) (B) (B) ([shift={(-0.2,0)}]B1.north) edge[->] node[left] {$\scriptstyle d_1(\Gamma_2)$} ([shift={(-0.2,0)}]B.south) (B) (A) (A) (A1) ([shift={(-0.2,0)}]A1.north) edge[->] node[left] {$\scriptstyle d_1(\Gamma_1)$} ([shift={(-0.2,0)}]A.south);
			\end{tikzpicture}
		\end{array}
	\end{equation}
where $\Gamma_i$ are MOY diagrams corresponding to a link, vertical complexes are called {\it matrix factorizations} $M(\Gamma_i) = \{ M^j(\Gamma_i),\, d^j(\Gamma_i) \}$, $j=0,1$. For the Khovanov--Rozansky construction, spaces $M^j(\Gamma_i)$ are rings in even variables. Matrix factorizations appeared in commutative algebra~\cite{E1,B,Kn,S,BEH} in the study of isolated hypersurface singularities, and much more recently in string theory, as boundary conditions for strings in Landau-Ginzburg models~\cite{KL1,KL2,KL3}.

The Khovanov--Rozansky cohomologies and polynomials have many applications in theoretical physics. In particular, the Khovanov--Rozansky polynomials can be reformulated in terms of integers which capture the spectrum of BPS states in the string Hilbert space~\cite{Gukov}. They also determine the oriented topological amplitudes of strings. Another example is the fact that the Khovanov--Rozansky homology theory can be derived from the theory of framed BPS states bound to domain walls separating two-dimensional Landau--Ginzburg models with (2,2) supersymmetry~\cite{Gukov2,Galakhov}. Moreover, there is an approach to Khovanov homology of knots and links based on counting the solutions of certain elliptic partial differential equations appearing in four and five dimensional gauge theories~\cite{Witten11,Witten45,GaiottoWitten}.

% ..................

% ???? 

\bigskip

\noindent The paper is organized as follows. In Section~\ref{sec:HOMFLY}, we introduce the notions of the HOMFLY polynomial, the MOY diagrams and MOY relations. Section~\ref{sec:KhR-hom} is devoted to a detailed description of the Khovanov--Rozansky cohomology. Namely, in Section~\ref{sec:mat-fact}, we introduce the definition of a matrix factorization and prove the theorem on the exclusion of variables in matrix factorizations. In Section~\ref{sec:KhRcomplex}, we build the Khovanov--Rozansky complex and describe the procedure for computing the Khovanov--Rozansky polynomial. Section~\ref{sec:excl-var-strand} is devoted to the application of the theorem from Section~\ref{sec:mat-fact} --- we show how to exclude variables on edges. We provide examples of calculation of the Khovanov--Rozansky polynomials in Section~\ref{sec:examples}. We demonstrate how to reduce the matrix factorization technique in the $N=2$ case to the Khovanov cycle calculus~\cite{Kh} in Section~\ref{sec:KhR-N=2}. In Section~\ref{sec:bip-red}, we explain how to reduce the matrix factorization technique for an arbitrary $N$ for a special family of bipartite links to the Khovanov--Rozansky cycle calculus introduced in our recent paper~\cite{2506.08721}. Section~\ref{sec:conclusion} is devoted to the concluding remarks.

\section{HOMFLY--PT polynomial}\label{sec:HOMFLY}

The HOMFLY--PT polynomial~\cite{freyd1985new,przytycki1987kobe,Guadagnini:1989kr,GUADAGNINI1990575,Alvarez_1993,Alvarez1994tt,Alvarez_1997,Labastida1997uw} is a quantum knot invariant associated with the Lie algebra $\mathfrak{su}_N$. Namely, the colored (by a representation $R$ of the $\mathfrak{su}_N$ algebra) HOMFLY--PT polynomials are observables in the 3-dimensional Chern--Simons topological field theory with the $SU(N)$ gauge group. In practice, it is usually computed via the Reshetikhin--Turaev approach~\cite{Reshetikhin,reshetikhin1990ribbon,reshetikhin1991invariants,turaev1990yang} --- using quantum $\cal R$-matrices and grading $\cal M$-operators from quantum universal enveloping algebra $U_q(\mathfrak{su}_N)$. 

In this paper, we deal with the HOMFLY--PT polynomial in the fundamental representation. Mentioning of the fundamental representation is usually omitted, and this polynomial is called just the HOMFLY--PT polynomial. In this case, the Reshetikhin--Turaev approach is analogous to replacing each crossing of a link by the two resolutions of type 0 and type 1 with some weights:
\begin{equation}\label{Hres}
\begin{picture}(300,50)(30,-15)

\linethickness{0.26mm}

    \setlength{\unitlength}{0.5pt}{
    \put(-60,-20){\vector(1,1){40}}
    \put(-19,-21){\line(-1,1){18}}
    \put(-43,3){\vector(-1,1){17}}
    }

    \put(0,-8){\mbox{\Large $:$}}

    \put(30,-8){\mbox{$q^N$}}

    \put(70,40){\mbox{type 0}}

    \put(140,-10){\mbox{\Large $,$}}

    \put(225,40){\mbox{type 1}}

    \put(170,-8){\mbox{$q^{N-1}$}}

    \put(290,-10){\mbox{\Large $,$}}
    
    \setlength{\unitlength}{1.3pt}{
    \put(-105,-8){
   \put(135,0){\vector(1,1){15}}
   \put(150,0){\vector(-1,1){15}}
   \put(142.5,7.5){\color{black}\circle{4.5}} 
   \put(142.5,7.5){\color{\myred}\circle*{4.1}}
   }
   }

    \setlength{\unitlength}{0.5pt}{
    \put(110,-2){
   \put(130,15){\vector(-1,1){5}}
    \put(155,15){\vector(1,1){5}}
    \qbezier(125,20)(145,0)(125,-20)
    \qbezier(160,20)(140,0)(160,-20)
    }
    }

    \setlength{\unitlength}{0.5pt}{
    \put(450,0){
    \put(-61,-21){\line(1,1){17}}
    \put(-20,-20){\vector(-1,1){40}}
    \put(-36,4){\vector(1,1){17}}
    }
    }

    \put(450,0){
    
    \put(0,-8){\mbox{\Large $:$}}

    \put(30,-8){\mbox{$q^{1-N}$}}

    \put(10,0){

    \put(70,40){\mbox{type 1}}

    \put(140,-10){\mbox{\Large $,$}}

    \put(210,40){\mbox{type 0}}

    \put(170,-8){\mbox{$q^{-N}$}}

    \put(280,-10){\mbox{\Large $.$}}
    
    \setlength{\unitlength}{1.3pt}{
    \put(-105,-8){
   \put(135,0){\vector(1,1){15}}
   \put(150,0){\vector(-1,1){15}}
   \put(142.5,7.5){\color{black}\circle{4.5}} 
   \put(142.5,7.5){\color{\myred}\circle*{4.1}}
   }
   }

    \setlength{\unitlength}{0.5pt}{
    \put(100,-2){
   \put(130,15){\vector(-1,1){5}}
    \put(155,15){\vector(1,1){5}}
    \qbezier(125,20)(145,0)(125,-20)
    \qbezier(160,20)(140,0)(160,-20)
    }
    }
    }
    }
    
\end{picture}
\end{equation}
The black vertex is called the MOY vertex, and graphs of a link resolution are called the MOY graphs. Thus, a link with $n$ crossings has $2^n$ resolutions that can be organized in a hypercube. Its vertices are some graphs of resolutions, and they have weights which are defined by multiplying the resolution weights from the (\ref{Hres}).

For MOY graphs, there are relations called MOY-relations~\cite{MOY} which follow from the Reidemeister invariance: %\textbf{(Вставить другой рисунок??)}

\begin{subequations}
	\begin{align}
		\label{MOYloc_II}&\begin{array}{c}
			\begin{tikzpicture}[scale=0.8]
				\draw[thick,postaction={decorate},decoration={markings, mark= at position 0.5 with {\arrow{stealth}}}] (0,0) circle (0.35);
%				\node[right] at (0.35,0) {$\scriptstyle x$};
			\end{tikzpicture}
		\end{array}= \ [N] \overset{\rm def}{=} \frac{q^N - q^{-N}}{q-q^{-1}} \,, \\
		\label{MOYloc_III}&\begin{array}{c}
			\begin{tikzpicture}[scale=0.65]
				\draw[thick,stealth-,postaction={decorate},decoration={markings, mark= at position 0.7 with {\arrow{stealth}}}] (0,1) -- (0,0.5) to[out=180,in=90] (-0.5,0) to[out=270,in=180] (0,-0.5) -- (0,-1);
				\draw[fill=\myred] (-0.1,0.5) to[out=90,in=180] (0,0.6) to[out=0,in=90] (0.6,0) to[out=270,in=0] (0,-0.6) to[out=180,in=270] (-0.1,-0.5) to[out=90,in=180] (0,-0.4) to[out=0,in=270] (0.4,0) to[out=90,in=0] (0,0.4) to[out=180,in=270] (-0.1,0.5);
%				\node[right] at (0,1) {$\scriptstyle x$};
%				\node[right] at (0,-1) {$\scriptstyle z$};
%				\node[left] at (-0.5,0) {$\scriptstyle y$};
			\end{tikzpicture}
		\end{array}=\begin{array}{c}
			\begin{tikzpicture}
				\draw[thick, -stealth] (0,-0.5) -- (0,0.5);
%				\node[right] at (0,0.5) {$\scriptstyle x$};
%				\node[right] at (0,-0.5) {$\scriptstyle z$};
			\end{tikzpicture}
		\end{array}\cdot \ [N-1]\,,\\
		\label{MOYloc_IV}&\begin{array}{c}
			\begin{tikzpicture}[scale=0.65]
				\begin{scope}[shift={(0,1)}]
					\draw[thick,-stealth] (-0.5,-0.5) to[out=90,in=270] (0.5,0.5);
					\draw[thick,-stealth] (0.5,-0.5) to[out=90,in=270] (-0.5,0.5);
					\draw[fill=\myred] (0,0) circle (0.1);
				\end{scope}
				\draw[thick,-stealth] (-0.5,-0.5) to[out=90,in=270] (0.5,0.5) -- (0.5,0.6);
				\draw[thick,-stealth] (0.5,-0.5) to[out=90,in=270] (-0.5,0.5) -- (-0.5,0.6);
				\draw[fill=\myred] (0,0) circle (0.1);
				%%%%%%%%%%%%%%%%%%%%%%%%%%%%%%%%%%%%%%%%%%%%%%%%%%%%%%%%%%%%5
				% \node[left] at (-0.5,-0.5) {$\scriptstyle i'$};
				% \node[right] at (0.5,-0.5) {$\scriptstyle j'$};
				% \node[left] at (-0.5,0.5) {$\scriptstyle k$};
				% \node[right] at (0.5,0.5) {$\scriptstyle l$};
				% \node[left] at (-0.5,1.5) {$\scriptstyle i$};
				% \node[right] at (0.5,1.5) {$\scriptstyle j$};
			\end{tikzpicture}
		\end{array} = \begin{array}{c}
			\begin{tikzpicture}[scale=0.7]
				\draw[thick,-stealth] (-0.5,-0.5) to[out=90,in=270] (0.5,0.5);
				\draw[thick,-stealth] (0.5,-0.5) to[out=90,in=270] (-0.5,0.5);
				\draw[fill=\myred] (0,0) circle (0.1);
				% \node[left] at (-0.5,-0.5) {$\scriptstyle i'$};
				% \node[right] at (0.5,-0.5) {$\scriptstyle j'$};
				% \node[left] at (-0.5,0.5) {$\scriptstyle i$};
				% \node[right] at (0.5,0.5) {$\scriptstyle j$};
			\end{tikzpicture}
		\end{array}\cdot \ [2]\,,\\
		&\label{MOYloc_V}\begin{array}{c}
			\begin{tikzpicture}[scale=0.65]
				\draw[thick,postaction={decorate},decoration={markings, mark= at position 0.7 with {\arrow{stealth}}}] (-1,-1) -- (-0.5,-0.5);
				\draw[thick,postaction={decorate},decoration={markings, mark= at position 0.7 with {\arrow{stealth}}}] (1,1) -- (0.5,0.5);
				\draw[thick,postaction={decorate},decoration={markings, mark= at position 0.7 with {\arrow{stealth}}}] (-0.5,0.5) -- (-1,1);
				\draw[thick,postaction={decorate},decoration={markings, mark= at position 0.7 with {\arrow{stealth}}}] (0.5,-0.5) -- (1,-1);
				\draw[thick,postaction={decorate},decoration={markings, mark= at position 0.7 with {\arrow{stealth}}}] (-0.5,0.5) -- (0.5,0.5);
				\draw[thick,postaction={decorate},decoration={markings, mark= at position 0.7 with {\arrow{stealth}}}] (0.5,-0.5) -- (-0.5,-0.5);
				\begin{scope}[shift={(-0.5,0)}]
					\draw[fill=\myred] (0.1,0.5) to[out=90,in=0] (0,0.6) to[out=180,in=90] (-0.1,0.5) -- (-0.1,-0.5) to[out=270,in=180] (0,-0.6) to[out=0,in=270] (0.1,-0.5) -- (0.1,0.5);
				\end{scope}
				\begin{scope}[shift={(0.5,0)}]
					\draw[fill=\myred] (0.1,0.5) to[out=90,in=0] (0,0.6) to[out=180,in=90] (-0.1,0.5) -- (-0.1,-0.5) to[out=270,in=180] (0,-0.6) to[out=0,in=270] (0.1,-0.5) -- (0.1,0.5);
				\end{scope}
%				\node[left] at (-1,-1) {$\scriptstyle x_1$};
%				\node[left] at (-1,1) {$\scriptstyle x_2$};
%				\node[right] at (1,1) {$\scriptstyle x_3$};
%				\node[right] at (1,-1) {$\scriptstyle x_4$};
%				\node[above] at (0,0.5) {$\scriptstyle x_5$};
%				\node[below] at (0,-0.5) {$\scriptstyle x_6$};
			\end{tikzpicture}
		\end{array}\;=\;\begin{array}{c}
			\begin{tikzpicture}
				\draw[thick, stealth-] (-0.5,0.5) to[out=315,in=225] (0.5,0.5);
				\draw[thick, stealth-] (0.5,-0.5) to[out=135,in=45] (-0.5,-0.5);
%				\node[left] at (-0.5,-0.5) {$\scriptstyle x_1$};
%				\node[left] at (-0.5,0.5) {$\scriptstyle x_2$};
%				\node[right] at (0.5,0.5) {$\scriptstyle x_3$};
%				\node[right] at (0.5,-0.5) {$\scriptstyle x_4$};
			\end{tikzpicture}
		\end{array} + \ \left(\begin{array}{c}
			\begin{tikzpicture}
				\draw[thick, stealth-] (-0.5,0.5) to[out=315,in=45] (-0.5,-0.5);
				\draw[thick, stealth-] (0.5,-0.5) to[out=135,in=225] (0.5,0.5);
%				\node[left] at (-0.5,-0.5) {$\scriptstyle x_1$};
%				\node[left] at (-0.5,0.5) {$\scriptstyle x_2$};
%				\node[right] at (0.5,0.5) {$\scriptstyle x_3$};
%				\node[right] at (0.5,-0.5) {$\scriptstyle x_4$};
			\end{tikzpicture}
		\end{array} \cdot \ [N-2]\right)\,,\\
		&\label{MOYloc_VI}\begin{array}{c}
			\begin{tikzpicture}[scale=0.55]
				\draw[thick,-stealth] (0,0) to[out=90,in=225] (1.5,1.5) to[out=45,in=270] (2,3);
				\draw[thick,-stealth] (2,0) to[out=90,in=315] (1.5,1.5) to[out=135,in=270] (0,3);
				\draw[thick,-stealth] (1,0) to[out=90,in=270] (0,1.5) to[out=90,in=270] (1,3);
%				\node[left] at (0,0) {$\scriptstyle x_1$};
%				\node[left] at (1,0) {$\scriptstyle x_2$};
%				\node[right] at (2,0) {$\scriptstyle x_3$};
%				\node[left] at (0,3) {$\scriptstyle x_4$};
%				\node[left] at (1,3) {$\scriptstyle x_5$};
%				\node[right] at (2,3) {$\scriptstyle x_6$};
				\draw[fill=\myred] (1.5,1.5) circle (0.13) (0.45,0.78) circle (0.13) (0.45,2.22) circle (0.13);
			\end{tikzpicture}
		\end{array}+ \ \begin{array}{c}
			\begin{tikzpicture}[scale=0.6]
				\draw[thick,-stealth] (0,0) -- (0,3);
				\draw[thick,-stealth] (1,0) to[out=90,in=270] (2,3);
				\draw[thick,-stealth] (2,0) to[out=90,in=270] (1,3);
%				\node[left] at (0,0) {$\scriptstyle x_1$};
%				\node[left] at (1,0) {$\scriptstyle x_2$};
%				\node[right] at (2,0) {$\scriptstyle x_3$};
%				\node[left] at (0,3) {$\scriptstyle x_4$};
%				\node[left] at (1,3) {$\scriptstyle x_5$};
%				\node[right] at (2,3) {$\scriptstyle x_6$};
				\draw[fill=\myred] (1.5,1.5) circle (0.13);
			\end{tikzpicture}
		\end{array}= \ \begin{array}{c}
			\begin{tikzpicture}[scale=0.6,xscale=-1]
				\draw[thick,-stealth] (0,0) to[out=90,in=225] (1.5,1.5) to[out=45,in=270] (2,3);
				\draw[thick,-stealth] (2,0) to[out=90,in=315] (1.5,1.5) to[out=135,in=270] (0,3);
				\draw[thick,-stealth] (1,0) to[out=90,in=270] (0,1.5) to[out=90,in=270] (1,3);
%				\node[right] at (0,0) {$\scriptstyle x_3$};
%				\node[left] at (1,0) {$\scriptstyle x_2$};
%				\node[left] at (2,0) {$\scriptstyle x_1$};
%				\node[right] at (0,3) {$\scriptstyle x_6$};
%				\node[left] at (1,3) {$\scriptstyle x_5$};
%				\node[left] at (2,3) {$\scriptstyle x_4$};
				\draw[fill=\myred] (1.5,1.5) circle (0.13) (0.45,0.78) circle (0.13) (0.45,2.22) circle (0.13);
			\end{tikzpicture}
		\end{array}+ \ \begin{array}{c}
			\begin{tikzpicture}[scale=0.6,xscale=-1]
				\draw[thick,-stealth] (0,0) -- (0,3);
				\draw[thick,-stealth] (1,0) to[out=90,in=270] (2,3);
				\draw[thick,-stealth] (2,0) to[out=90,in=270] (1,3);
%				\node[right] at (0,0) {$\scriptstyle x_3$};
%				\node[left] at (1,0) {$\scriptstyle x_2$};
%				\node[left] at (2,0) {$\scriptstyle x_1$};
%				\node[right] at (0,3) {$\scriptstyle x_6$};
%				\node[left] at (1,3) {$\scriptstyle x_5$};
%				\node[left] at (2,3) {$\scriptstyle x_4$};
				\draw[fill=\myred] (1.5,1.5) circle (0.13);
			\end{tikzpicture}
		\end{array}\,.
	\end{align}
    \label{MOY}
\end{subequations}
The HOMFLY-PT polynomial for a link $\cal L$ can be calculated as follows:
\begin{equation}\label{HOMFLY-gen}
    H^{\cal L}(q,q^N) = \sum_{r}^{2^n}(-)^{\alpha(r)}H(r)
\end{equation}
where $\alpha(r)$ is the number of MOY-vertices in a resolution $r$. The value of $H(r)$ is defined by (\ref{MOY}) and multiplied by its weight according to~\eqref{Hres}. %\textbf{Примеры??}

Let us consider the example of the hypercube and the calculation of the HOMFLY polynomial for the Hopf link~\eqref{Hopf-ex-HOMFLY}. Here we also give details on a hypercube construction. 

\begin{equation}\label{Hopf-ex-HOMFLY}
\begin{aligned}
    \begin{array}{c}
		\begin{tikzpicture}[scale=0.7]
			\begin{scope}[shift={(0,1)}]
				\draw[thick] (0.5,-0.5) to[out=90,in=270] (-0.5,0.5);
				\draw[white, line width = 1.5mm] (-0.5,-0.5) to[out=90,in=270] (0.5,0.5);
				\draw[thick] (-0.5,-0.5) to[out=90,in=270] (0.5,0.5);
			\end{scope}
			\draw[thick,-stealth] (0.5,-0.5) to[out=90,in=270] (-0.5,0.5) -- (-0.5,0.6);
			\draw[white, line width = 1.5mm] (-0.5,-0.5) to[out=90,in=270] (0.5,0.5);
			\draw[thick,-stealth] (-0.5,-0.5) to[out=90,in=270] (0.5,0.5) -- (0.5,0.6);
			\draw[thick] (-0.5,1.5) to[out=90,in=90] (-1.2,1.5) -- (-1.2,-0.5) to[out=270,in=270] (-0.5,-0.5);
			\begin{scope}[xscale=-1]
				\draw[thick] (-0.5,1.5) to[out=90,in=90] (-1.2,1.5) -- (-1.2,-0.5) to[out=270,in=270] (-0.5,-0.5);
			\end{scope}
			% \node[left] at (-0.5,-0.5) {$\scriptstyle x_1$};
			% \node[right] at (0.5,-0.5) {$\scriptstyle x_2$};
			% \node[left] at (-0.5,0.5) {$\scriptstyle x_3$};
			% \node[right] at (0.5,0.5) {$\scriptstyle x_4$};
			% \node[left] at (-0.5,1.5) {$\scriptstyle x_1$};
			% \node[right] at (0.5,1.5) {$\scriptstyle x_2$};
			\node[above] at (0,0) {$\scriptstyle 1$};
			\node[above] at (0,1) {$\scriptstyle 2$};
			% \node[left] at (-0.3,0) {$\scriptstyle \theta_1$};
			% \node[right] at (0.3,0) {$\scriptstyle \theta_2$};
			% \node[left] at (-0.3,1) {$\scriptstyle \theta_3$};
			% \node[right] at (0.3,1) {$\scriptstyle \theta_4$};
		\end{tikzpicture}
	\end{array}&=\left[\begin{array}{c}
		\begin{tikzpicture}
			\node(A) at (0,0) {
			$\begin{tikzpicture}[scale=0.5]
				\draw[thick] (0.5,-0.5) to[out=90,in=270] (0.3,0) to[out=90,in=270] (0.5,0.5) -- (0.5,0.6);
				\draw[thick] (-0.5,-0.5) to[out=90,in=270] (-0.3,0) to[out=90,in=270] (-0.5,0.5) -- (-0.5,0.6);
				\begin{scope}[shift={(0,1)}]
					\draw[thick] (0.5,-0.5) to[out=90,in=270] (0.3,0) to[out=90,in=270] (0.5,0.5);
					\draw[thick] (-0.5,-0.5) to[out=90,in=270] (-0.3,0) to[out=90,in=270] (-0.5,0.5);
				\end{scope}
				\draw[thick, postaction={decorate},decoration={markings, mark= at position 0.6 with {\arrow{stealth}}}] (-0.5,1.5) to[out=90,in=90] (-1.2,1.5) -- (-1.2,-0.5) to[out=270,in=270] (-0.5,-0.5);
				\begin{scope}[xscale=-1]
					\draw[thick, postaction={decorate},decoration={markings, mark= at position 0.6 with {\arrow{stealth}}}] (-0.5,1.5) to[out=90,in=90] (-1.2,1.5) -- (-1.2,-0.5) to[out=270,in=270] (-0.5,-0.5);
				\end{scope}
			\end{tikzpicture}_{\,00}$};
			\node(B) at (3,0.7) {$\begin{tikzpicture}[scale=0.5]
				\draw[thick] (0.5,-0.5) to[out=90,in=270] (-0.5,0.5);
				\draw[thick] (-0.5,-0.5) to[out=90,in=270] (0.5,0.5);
				\begin{scope}[shift={(0,1)}]
					\draw[thick] (0.5,-0.5) to[out=90,in=270] (0.3,0) to[out=90,in=270] (0.5,0.5);
					\draw[thick] (-0.5,-0.5) to[out=90,in=270] (-0.3,0) to[out=90,in=270] (-0.5,0.5);
				\end{scope}
				\draw[thick, postaction={decorate},decoration={markings, mark= at position 0.6 with {\arrow{stealth}}}] (-0.5,1.5) to[out=90,in=90] (-1.2,1.5) -- (-1.2,-0.5) to[out=270,in=270] (-0.5,-0.5);
				\begin{scope}[xscale=-1]
					\draw[thick, postaction={decorate},decoration={markings, mark= at position 0.6 with {\arrow{stealth}}}] (-0.5,1.5) to[out=90,in=90] (-1.2,1.5) -- (-1.2,-0.5) to[out=270,in=270] (-0.5,-0.5);
				\end{scope}
				\draw[fill=\myred] (0,0) circle (0.13);
			\end{tikzpicture}_{\,10}$};
			\node(C) at (3,-0.7) {$\begin{tikzpicture}[scale=0.5]
				\draw[thick] (0.5,-0.5) to[out=90,in=270] (0.3,0) to[out=90,in=270] (0.5,0.5);
				\draw[thick] (-0.5,-0.5) to[out=90,in=270] (-0.3,0) to[out=90,in=270] (-0.5,0.5);
				\begin{scope}[shift={(0,1)}]
					\draw[thick] (0.5,-0.5) to[out=90,in=270] (-0.5,0.5);
					\draw[thick] (-0.5,-0.5) to[out=90,in=270] (0.5,0.5);
				\end{scope}
				\draw[thick, postaction={decorate},decoration={markings, mark= at position 0.6 with {\arrow{stealth}}}] (-0.5,1.5) to[out=90,in=90] (-1.2,1.5) -- (-1.2,-0.5) to[out=270,in=270] (-0.5,-0.5);
				\begin{scope}[xscale=-1]
					\draw[thick, postaction={decorate},decoration={markings, mark= at position 0.6 with {\arrow{stealth}}}] (-0.5,1.5) to[out=90,in=90] (-1.2,1.5) -- (-1.2,-0.5) to[out=270,in=270] (-0.5,-0.5);
				\end{scope}
				\draw[fill=\myred] (0,1) circle (0.13);
			\end{tikzpicture}_{\,01}$};
			\node(D) at (6,0) {$\begin{tikzpicture}[scale=0.5]
				\draw[thick,-stealth] (0.5,-0.5) to[out=90,in=270] (-0.5,0.5) -- (-0.5,0.6);
				\draw[thick,-stealth] (-0.5,-0.5) to[out=90,in=270] (0.5,0.5) -- (0.5,0.6);
				\begin{scope}[shift={(0,1)}]
					\draw[thick] (0.5,-0.5) to[out=90,in=270] (-0.5,0.5);
					\draw[thick] (-0.5,-0.5) to[out=90,in=270] (0.5,0.5);
				\end{scope}
				\draw[thick, postaction={decorate},decoration={markings, mark= at position 0.6 with {\arrow{stealth}}}] (-0.5,1.5) to[out=90,in=90] (-1.2,1.5) -- (-1.2,-0.5) to[out=270,in=270] (-0.5,-0.5);
				\begin{scope}[xscale=-1]
					\draw[thick, postaction={decorate},decoration={markings, mark= at position 0.6 with {\arrow{stealth}}}] (-0.5,1.5) to[out=90,in=90] (-1.2,1.5) -- (-1.2,-0.5) to[out=270,in=270] (-0.5,-0.5);
				\end{scope}
				\draw[fill=\myred] (0,0) circle (0.13) (0,1) circle (0.13);
			\end{tikzpicture}_{\,11}$};
			\path (A) edge[-] (B) (A) edge[-] (C) (B) edge[-] (D) (C) edge[-] (D);
		\end{tikzpicture}
	\end{array}\right]\,, \\
    H^{\rm Hopf}(q,q^N) &= \ \ \ q^{2(N-1)}[N]^2 \ - 2 q^{2N-1}[N-1][N] \ + q^{2N} [2][N-1][N]\,.
\end{aligned}
\end{equation}
Enumerate crossings in a link. We label each resolution of a link by a sequence of 0 and 1 according to labels of the elementary resolutions in~\eqref{Hres}. So that one of the numbers in the resolution enumerator corresponds to a type of the elementary resolution in~\eqref{Hres} that is replaced instead by the corresponding crossing. For example, in the Hopf link, the first crossing is the bottom one and the second crossing is the upper one, and we have $2^2 = 4$ resolutions enumerated by $00$, $10$, $01$ and $11$. 

Organize resolutions in a hypercube so that each vertical is shared by resolutions having the same sum of numbers in their labels. This sum increases from left to right. We draw an edge between resolutions connected by an elementary flip, i.e. differing at exactly one elementary resolution. In the Hopf link in~\eqref{Hopf-ex-HOMFLY}, there stands the resolution $00$ (the sum of numbers is $0$) on the right; the middle vertical is shared by the resolutions $10$ and $01$ (the sum is $1$); the resolution $11$ stands on the left (the sum is $2$). When transferring from the resolution $00$ to the resolution $10$, the first vertex changes the type of elementary resolution, and thus, these resolutions are connected by an edge. Other edges are arranged according to the same rule.  

Then, the HOMFLY polynomial is calculated according to~\eqref{HOMFLY-gen} where the role of $\alpha(r)$ is played by the sum of numbers inside labels of a link resolution. We write the corresponding summands below the picture in~\eqref{Hopf-ex-HOMFLY}. The final correct answer for the HOMFLY invariant for the Hopf link is
\begin{equation}
	H^{\rm Hopf}(q,A=q^N) = \frac{A-A^{-1}}{(q-q^{-1})^2}(A^3-A q^2-Aq^{-2}+A)\,.
\end{equation}

% \begin{equation}
%     \begin{tikzpicture}[scale=0.5]
% 				\draw[thick,-stealth] (-0.5,-0.5) -- (0.5,0.5);
% 				\draw[thick,-stealth] (0.5,-0.5) -- (-0.5,0.5);
% 				\draw[fill=black] (0,0) circle (0.15);
% 			\end{tikzpicture}
%     \begin{tikzpicture}[scale=0.5]
% 				\draw[thick,-stealth] (-0.5,-0.5) -- (0.5,0.5);
% 				\draw[thick,-stealth] (-0.1,0.1) -- (-0.5,0.5);
% 				\draw[thick] (0.5,-0.5) -- (0.1,-0.1);
% 			\end{tikzpicture}
%             \begin{tikzpicture}[scale=0.5]
% 				\draw[thick,-stealth] (-0.5,-0.5) to[out=45,in=270] (-0.15,0) to[out=90,in=315] (-0.5,0.5);
% 				\draw[thick,-stealth] (0.5,-0.5) to[out=135,in=270] (0.15,0) to[out=90,in=225] (0.5,0.5);
% 			\end{tikzpicture}
%             \begin{tikzpicture}[scale=0.5, xscale=-1]
% 				\draw[thick,-stealth] (-0.5,-0.5) -- (0.5,0.5);
% 				\draw[thick,-stealth] (-0.1,0.1) -- (-0.5,0.5);
% 				\draw[thick] (0.5,-0.5) -- (0.1,-0.1);
% 			\end{tikzpicture}
% \end{equation}

\section{Khovanov--Rozansky cohomology}\label{sec:KhR-hom}

In this section, we introduce constituents of the Khovanov--Rozansky cohomologies and polynomials construction. Khovanov--Rozansky polynomial is calculated according to a bicomplex~\eqref{KRbicompl}. We describe the vertical morhisms structure in Section~\ref{sec:mat-fact}. The horizontal morphisms, the description of the whole Khovanov--Rozansky bicomplex and algorithm of Khovanov--Rozansky polynomial calculation are given in Section~\ref{sec:KhRcomplex}.

\subsection{Graded matrix factorizations}\label{sec:mat-fact}

Consider a $q$-graded polynomial ring $R = \mathbb{Q}[x_1,...,x_m]$ with ${\rm deg}_q(x_i)=2$. A matrix factorization $M$
with a potential $\omega \in R$ is called a set of free $R$-modules $M^i$ with operators $d_i$:
\begin{equation}
    \begin{split}
        M=[M^0 \xrightarrow{d_0} M^1 \xrightarrow{d_1} M^0]
    \end{split}
\end{equation}
such that
\begin{equation}
    d_0 d_1 =\omega\cdot {\rm id}_{M^0},\quad d_1 d_0 =\omega\cdot {\rm id}_{M^1}\,.
\end{equation}
Note that a matrix factorization, in addition to $q$-grading, also has $\mathbb{Z}_2$-grading $i=0,\,1\,.$

The tensor product of two matrix factorizations $M=\{M^i,d^{M}_i\}$ and $N=\{N^i,d^{N}_i\}$ is a factorization $M\otimes N=\{(M\otimes N)^i,d^{M\otimes N}_i\}$ defined by the following rules:
\begin{equation}
    \begin{split}
        (M\otimes N)^{k} = \bigoplus_{(i+j) \,{\rm mod}\, 2 \,=\,k}M^{i}\otimes N^{j},\qquad i,j,k\in \mathbb{Z}_2\,, \\
        d^{M\otimes N}_k(m\otimes n) = d^{M}_k(m)\otimes n + (-1)^{\deg(m)} m\otimes d^{N}_k(n),\quad m\in M,n\in N\,.
    \end{split}
\end{equation}
where $\deg(m)$ is the $\mathbb{Z}_2$-grading of $m\in M$.\\
In the matrix form, we have
\begin{equation}
    M\otimes N = \bigg[\begin{pmatrix}
        M^0\otimes N^0\\
        M^1\otimes N^1
    \end{pmatrix} \xrightarrow{\begin{pmatrix}
        {\rm id}\otimes d^{N}_0 & d^{M}_1 \otimes {\rm id} \\
        d^{M}_0 \otimes {\rm id} & -{\rm id} \otimes d^{N}_1
    \end{pmatrix}} \begin{pmatrix}
        M^0\otimes N^1\\
        M^1 \otimes N^0
    \end{pmatrix} \xrightarrow{\begin{pmatrix}
        {\rm id} \otimes d^{N}_1 &d^{M}_1 \otimes {\rm id} \\
        d^{M}_0 \otimes {\rm id} & - {\rm id} \otimes d^{N}_0
    \end{pmatrix}} \begin{pmatrix}
        M^0\otimes N^0\\
        M^1\otimes N^1
    \end{pmatrix}\bigg].
    \label{tensprod}
\end{equation}
It is easy to check that a potential of the tensor product of factorizations with potentials $\omega_M$ and $\omega_N$ is equal to $\omega_M+\omega_N$.
For simplicity, we will not write $\otimes$ in tensor products of factorizations further if there is no misunderstanding.

Define the operation of $q$-grading shift $\{\cdot\}$ as ${\rm deg}_q(x)\{k\} = k + {\rm deg}_q(x)$,
and $\mathbb{Z}_2$-grading shift $\langle\cdot\rangle$ by the following rules: 
\begin{equation}
    \begin{split}
    M^i \langle 1\rangle = M^{i+1}\,, \\
    d^{M\langle 1\rangle}_i = -d_{i+1}^{M}\,,\quad i\in \mathbb{Z}_2\,.
    \end{split}    
\end{equation}
If a factorization has the zero potential, it becomes a 2-complex being a periodic complex with
period two, and we can calculate its homology, which splits into a direct sum $\mathcal{H}=\mathcal{H}^0\oplus \mathcal{H}^1$.

Let us introduce the necessary definitions for maps between matrix factorizations. A {\it homomorphism} $f:M\rightarrow N$ of matrix factorizations is a pair of homomorphisms $f^0: M^0 \rightarrow N^0$ and $f^1: M^1\rightarrow N^1$ that make the diagram below commutative:
\begin{equation}
    % https://tikzcd.yichuanshen.de/#N4Igdg9gJgpgziAXAbVABwnAlgFyxMJZABgBpiBdUkANwEMAbAVxiRDpAF9T1Nd9CKAIzkqtRizYAjLjxAZseAkQBMo6vWatEIAMazeigUTJCxmyTqgH5fJYOQizGidpCtuh-spRrn4rTYAMy4xGCgAc3giUCCAJwgAWyQyEBwIJCFPEHikzOp0pBVs3OTEAGYCjMQAFhKEspqqpABWerzENTTqtrlSlOaK9rKRbqQ6ik4gA
\begin{tikzcd}
M^0 \arrow{r}{d^M_0} \arrow{d}{f^0} & M^1 \arrow{r}{d^M_1} \arrow{d}{f^1} & M^0 \arrow{d}{f^0} \\
N^0 \arrow{r}{d^N_0}           & N^1 \arrow{r}{d^N_1}          & N^0          
\end{tikzcd}
\label{homom}
\end{equation}
A {\it homotopy} $h$ between homomorphisms $f,\,g:M\rightarrow N$ of matrix factorizations is a pair of maps $h^0:M^0 \rightarrow N^1$ and $h^1: M^1 \rightarrow N^0$ such that
\begin{equation}
	f^i - g^i = h^{i+1} \cdot d_{i}^M + d_{i+1}^N \cdot h^{i}\,, \quad i\in \mathbb{Z}_2\,.
\end{equation}
In this case, the maps $f$ and $g$ are called {\it homotopic}.

A homomorphism $f:M\xrightarrow{} N$ is сalled a {\it quasi-isomorphism} of matrix factorizations if there is a homomorphism of factorizations $g:N\xrightarrow{}M$ such that $f\cdot g$ and $g \cdot f$ are homotopic to the identity maps.

In what follows, we will often be interested in matrix factorizations with the zero potential, as they appear in the Khovanov--Rozansky complex for a link. In this case, the factorizations become 2-complexes, and a quasi-isomorphism of matrix factorizations becomes a quasi-isomorphism of complexes. A homomorphism $f:M\rightarrow N$ of 2-complexes (i.e., matrix factorizations with the zero potential) is called a {\it quasi-isomorphism} if it induces an isomorphism of cohomologies $f^*:\mathcal{H}(M)\rightarrow\mathcal{H}(N)$.

\begin{theorem-non}[Excluding variables,~\cite{KhR}]
    Introduce the notations for matrix factorizations:
\begin{equation}\label{a,b-defs}
    \begin{split}
        \{a,b\}_R := [ R\xrightarrow{a}R\xrightarrow{b}R ]\,, \\
        \{\textbf{a,b}\}_R :=\bigotimes_{i=1}^{n} [R\xrightarrow{a_i}R\xrightarrow[]{b_i}R]\,, \\
        \{\textbf{a,b}\}^j_R:=\{\textbf{a,b}\}_R\backslash \{a_j,b_j\} = \bigotimes_{i \neq j} [R\xrightarrow{a_i}R\xrightarrow[]{b_i}R]\,.
    \end{split}
\end{equation}
Suppose that a factorization $\{\textbf{a,b}\}_R$ has the zero potential $\omega=\sum_i a_ib_i=0\,$, and there exists $j$: $b_j=x_1-c$ and $c\in R'$, where $R=\mathbb{Q}[x_1,x_2,...,x_m]$, $R'=\mathbb{Q}[x_1,x_2,...,x_m]/\langle x_1-c\rangle=\mathbb{Q}[x_2,...,x_m]$.

Define a map $\Pi:R\xrightarrow{}R'$:
\begin{equation}\label{Pi(x)}
    \Pi(x_i)=x_i\,,\quad \Pi(x_1)=c\,,\quad i \in 2,...,\,m\,.
\end{equation}
Then, we have following quasi-isomorphism of factorizations (or quasi-isomorphism of complexes)\footnote{By writing $\Pi(\boldsymbol{a})$ we mean that we substitute $c$ instead of $x_1$ in all $a_i$ (according to rule~\eqref{Pi(x)}).}:
\begin{equation}
    \{\textbf{a,b}\}_R\cong \{\Pi(\textbf{a}),\Pi(\textbf{b})\}^j_{R'}\,.
\end{equation}

\end{theorem-non}

\paragraph{Proof.} 
Let us describe in more detail the matrix factorizations featured in the theorem:
\begin{equation}
    \begin{split}
    \{\boldsymbol{a},\boldsymbol{b}\}_R= V^{(0)}\xrightarrow{D_0}V^{(1)}\xrightarrow{D_1}V^{(0)}\,, \\
    \{\boldsymbol{a},\boldsymbol{b}\}^j_R=U^{(0)}\xrightarrow{P_0}U^{(1)}\xrightarrow{P_1}U^{(0)}\,, \\
    \{\Pi(\boldsymbol{a}),\Pi(\boldsymbol{b})\}^j_{R'}=W^{(0)}\xrightarrow{\Pi(P_0)}W^{(1)}\xrightarrow{\Pi(P_1)}W^{(0)}\,.
    \end{split}
\end{equation}
Here $V^{(k)}$, $U^{(k)}$ are free $R$-modules and $W^{(k)}$ is a free $R'$-module where in the superscripts we indicate $\mathbb{Z}_2$-gradings; the operators $D_{0,1}$, $P_{0,1}$, $\Pi(P_{0,1})$ are differentials obtained by expanding the tensor products in~\eqref{a,b-defs}. Note that the factorization $ \{\boldsymbol{a},\boldsymbol{b}\}^j_R$ has the potential $\omega^j = \omega-a_jb_j = -a_jb_j$ and the factorization $\{\Pi(\boldsymbol{a}),\Pi(\boldsymbol{b})\}^j_{R'}$ has the potential $\Omega = 0$, which means it becomes a complex.

In a proper basis, we can write the first factorization as:
\begin{equation}
    \{\boldsymbol{a},\boldsymbol{b}\}_R= V^{(0)} \xrightarrow[]{
  D_0=\left(\begin{array}{@{}c|c@{}}
    P_0 & b_j \cdot\mathbf{1}\\ \hline
    a_j \cdot\mathbf{1} & -P_1 
  \end{array}\right)
}V^{(1)} \xrightarrow[]{D_1=\left(\begin{array}{@{}c|c@{}}
    P_1 & b_j \cdot\mathbf{1}\\ \hline
    a_j \cdot\mathbf{1} & -P_0 
  \end{array}\right)} V^{(0)}\,.
\end{equation}
Accordingly, using these block matrices, we decompose the $R$-module $V$ into the direct sum $V=V_1\oplus V_2$. \\
\begin{equation}
	\{\boldsymbol{a},\boldsymbol{b}\}_R= \begin{pmatrix}
		V^{(0)}_1 \\ \hline
		V^{(0)}_2
	\end{pmatrix} \xrightarrow[]{
		\left(\begin{array}{@{}c|c@{}}
			P_0 & b_j \cdot\mathbf{1}\\ \hline
			a_j \cdot\mathbf{1} & -P_1 
		\end{array}\right)
	}\begin{pmatrix}
	V^{(1)}_1 \\ \hline
	V^{(1)}_2
	\end{pmatrix} \xrightarrow[]{\left(\begin{array}{@{}c|c@{}}
			P_1 & b_j \cdot\mathbf{1}\\ \hline
			a_j \cdot\mathbf{1} & -P_0 
		\end{array}\right)} \begin{pmatrix}
		V^{(0)}_1 \\ \hline
		V^{(0)}_2
	\end{pmatrix}.
\end{equation}
We would like to define a map $f:\{\boldsymbol{a},\boldsymbol{b}\}_R\rightarrow \{\Pi(\boldsymbol{a}),\Pi(\boldsymbol{b})\}^j_{R'}$ and $g:\{\Pi(\boldsymbol{a}),\Pi(\boldsymbol{b})\}^j_{R'}\rightarrow\{\boldsymbol{a},\boldsymbol{b}\}_R$ between two 2-complexes such that the following diagrams are commutative:
\begin{equation}
    \begin{tikzcd}
V^{(0)} \arrow{r}{D_0} \arrow{d}{f} & V^{(1)} \arrow{r}{D_1} \arrow{d}{f} & V^{(0)} \arrow{d}{f} \\
W^{(0)} \arrow{r}{\Pi(P_0)}           & W^{(1)} \arrow{r}{\Pi(P_1)}          & W^{(0)}          
\end{tikzcd}\qquad \begin{tikzcd}
W^{(0)} \arrow{r}{\Pi(P_0)} \arrow{d}{g} & W^{(1)} \arrow{r}{\Pi(P_1)} \arrow{d}{g} & W^{(0)} \arrow{d}{g} \\
V^{(0)} \arrow{r}{D_0}           & V^{(1)} \arrow{r}{D_0}          & V^{(0)}          
\end{tikzcd}
\label{excldiagr}
\end{equation}
In the notations $V^{(k)}\ni\boldsymbol{\upsilon}^{(k)}=\begin{pmatrix}
    \boldsymbol{\upsilon}^{(k)}_1\\
    \boldsymbol{\upsilon}^{(k)}_2
\end{pmatrix}$, $\boldsymbol{\upsilon}^{(k)}_j\in V_{j}^{(k)}$, and $W^{(k)}\ni \boldsymbol{w}^{(k)}$, we can write these maps as
\begin{equation}
    f: \boldsymbol{\upsilon}^{(k)}= \begin{pmatrix}
    	\boldsymbol{\upsilon}^{(k)}_1\\
    	\boldsymbol{\upsilon}^{(k)}_2
    \end{pmatrix} \mapsto \Pi (\boldsymbol{\upsilon}^{(k)}_1) \in W^{(k)}\,,\quad k\in\mathbb{Z}_2\,.
\end{equation}
\begin{equation}\label{inv-map-lin-red}
    g: \boldsymbol{w}^{(k)} \mapsto \begin{pmatrix}
    	\boldsymbol{w}^{(k)}\\
    	-\frac{P_k - \Pi (P_k)}{b_j} \boldsymbol{w}^{(k)}
    \end{pmatrix}\in V^{(k)}\,, \quad k\in\mathbb{Z}_2\,.
\end{equation}
To show that the map $f:\{\boldsymbol{a},\boldsymbol{b}\}_R\rightarrow \{\Pi(\boldsymbol{a}),\Pi(\boldsymbol{b})\}^j_{R'}$ induces an isomorphism on the corresponding cohomologies, it suffices to verify that for the maps $f$ and $g$ the diagrams in (\ref{excldiagr}) are commutative, and that the maps induced by $f$ and $g$ on cohomologies are inverses of each other. For definiteness, we will only consider the case $k=0\in \mathbb{Z}_2$ (the case $k=1$ is treated completely analogously).

Let us prove the commutativity of the diagram for $f$. We must show that $D_0\cdot f=f\cdot \Pi(P_0)$. Indeed, take an arbitrary vector $\boldsymbol{\upsilon} \in V^{(0)}$:
\begin{align}
	\boldsymbol{\upsilon}=\begin{pmatrix}
		\boldsymbol{\upsilon}_1\\
		\boldsymbol{\upsilon}_2
	\end{pmatrix} \xrightarrow{D_0} \begin{pmatrix}
	P_0 \boldsymbol{\upsilon}_1 + b_j \boldsymbol{\upsilon}_2\\
	a_j \boldsymbol{\upsilon}_1 - P_1 \boldsymbol{\upsilon}_2
	\end{pmatrix} \xrightarrow{f} \Pi(P_0 \boldsymbol{\upsilon}_1 + b_j \boldsymbol{\upsilon}_2)= \Pi(P_0) \Pi( \boldsymbol{\upsilon})\,, \\
	 \boldsymbol{\upsilon}=\begin{pmatrix}
	 	\boldsymbol{\upsilon}_1\\
	 	\boldsymbol{\upsilon}_2
	 \end{pmatrix} \xrightarrow{f} \boldsymbol{\upsilon}_1 \xrightarrow{\Pi(P_0)} \Pi (P_0) \Pi (	\boldsymbol{\upsilon}_1)\,. 
\end{align}
Let us now check the commutativity of the diagram for $g$: $g\cdot\Pi (P_0) = D_0 \cdot g$. Take an arbitrary vector $w\in W$:
\begin{align}
	\boldsymbol{w} \xrightarrow{\Pi(P_0)} \Pi(P_0) \boldsymbol{w} \xrightarrow{g} \begin{pmatrix}
		\Pi (P_0)\boldsymbol{w}\\
		\Pi (P_0)\cdot -\frac{P_1-\Pi(P_1)}{b_j} \Pi(P_0) \boldsymbol{w} 
	\end{pmatrix} = \begin{pmatrix}
	\Pi (P_0)\boldsymbol{w}\\
	-\Pi (P_0)\cdot \frac{P_1}{b_j}\boldsymbol{w} 
	\end{pmatrix}, \\
	\boldsymbol{w} \xrightarrow{g} \begin{pmatrix}
		\boldsymbol{w} \\
		-\frac{P_0-\Pi(P_0)}{b_j} \boldsymbol{w} 
	\end{pmatrix}\xrightarrow{D_0} \begin{pmatrix}
	P_0 \boldsymbol{w} + b_j \cdot -\frac{P_0-\Pi(P_0)}{b_j} \boldsymbol{w} \\
	a_j \boldsymbol{w} + P_1 \cdot \frac{P_0-\Pi(P_0)}{b_j} \boldsymbol{w}
	\end{pmatrix} = \begin{pmatrix}
	\Pi (P_0)\boldsymbol{w}\\
	-\Pi (P_0)\cdot \frac{P_1}{b_j}\boldsymbol{w} 
	\end{pmatrix}.
\end{align}
Here, we have used the equalities for the potentials of the corresponding factorizations, $\Pi(P_0)\Pi(P_1)=0$ and $P_1\cdot P_0 = - a_j b_j \cdot {\rm id}$.

It remains to check that the maps $f$ and $g$ induced on cohomologies are inverses of each other. The equality $f \cdot g|_{\mathcal{H}} = {\rm id}$ is obvious. Let us prove that $g \cdot f|_{\mathcal{H}} = {\rm id}$. To prove this, we need to show that if a vector $\boldsymbol{\upsilon}$ belongs to the cohomology group $\mathcal{H}(\{\boldsymbol{a},\boldsymbol{b}\}_R)$  then the vector $fg(\boldsymbol{\upsilon})$ is equal to $\boldsymbol{\upsilon}$ modulo an image of the differential, i.e., $\boldsymbol{\upsilon}= gf(\boldsymbol{\upsilon}) + \boldsymbol{u}$, $u\in {\rm Im}(D_1)$.
\begin{equation}
	\mathcal{H}(\{\boldsymbol{a},\boldsymbol{b}\}_R) \ni \boldsymbol{\upsilon} = \begin{pmatrix}
		\boldsymbol{\upsilon}_1\\
		\boldsymbol{\upsilon}_2
	\end{pmatrix} \xrightarrow{f} \Pi (\boldsymbol{\upsilon}_1) \xrightarrow{g} \begin{pmatrix}
	\Pi(\boldsymbol{\upsilon}_1)\\
	-\frac{P_0-\Pi(P_0)}{b_j}\Pi(\boldsymbol{\upsilon}_1)
	\end{pmatrix} = f g (\boldsymbol{\upsilon})\,,
\end{equation} 
\begin{equation}
	{\rm Im}(D_1) \ni \boldsymbol{u} = \begin{pmatrix}
		P_1 & b_j\\
		a_j & -P_0
	\end{pmatrix} \begin{pmatrix}
	0\\
	\frac{\boldsymbol{\upsilon}_1 - \Pi(\boldsymbol{\upsilon}_1)}{b_j}
	\end{pmatrix} = \begin{pmatrix}
	\boldsymbol{\upsilon}_1 - \Pi(\boldsymbol{\upsilon}_1)\\
	-P_0 \frac{\boldsymbol{\upsilon}_1 - \Pi(\boldsymbol{\upsilon}_1)}{b_j}
	\end{pmatrix}. 
\end{equation}
To show that $\boldsymbol{\upsilon}= gf(\boldsymbol{\upsilon}) + \boldsymbol{u}$, we use the fact that $D_0 \boldsymbol{\upsilon} = 0$, which implies $P_0 \boldsymbol{\upsilon}_1 + b_j \boldsymbol{\upsilon}_2 = 0$.

Note that here we choose $x_1$ in $b_j = x_1 - c$ without loss of generality because we can number edges by $x_i$ in an arbitrary way. $\qquad \qquad \qquad \qquad \qquad \qquad \qquad \qquad \qquad \qquad \qquad \qquad \qquad \qquad \qquad \qquad \qquad \qquad \qquad \qquad \qquad \qquad \qquad \Box$ 

\medskip

\noindent We can make a similar statement for $a_j=x_1-c$, but we must shift the $\mathbb{Z}_2$-grading of factorizations.

Despite the fact that this theorem was proved in~\cite{KhR}, we provide here the proof in another form and in more detail and also find the explicit form of the inverse map~\eqref{inv-map-lin-red} that we use in Sections~\ref{sec:excl-var-strand},~\ref{sec:Hopf},~\ref{sec:KhR-N=2},~\ref{sec:bip-red}.

\subsection{Khovanov--Rozansky complex}\label{sec:KhRcomplex}

The Khovanov--Rozansky polynomial is the categorification of the HOMFLY polynomial (in the fundamental representation). The corresponding cohomologies are defined as the cohomologies of a certain bicomplex~\eqref{KRbicompl}, which we describe below.

First, we mark variables $x_i$ to each edge of a link and fix the polynomial ring $R=\mathbb{Q}[x_1,...,x_m]$ with all marked variables (see the example in the picture~\eqref{Hopf-compl}). We then resolve the crossings of a link as follows:
%(before a crossing) the following complexes

\begin{equation}\label{res-cross}
\begin{array}{c}
     \begin{tikzpicture}[scale=0.5]
				\draw[thick,-stealth] (-0.5,-0.5) -- (0.5,0.5);
				\draw[thick,-stealth] (-0.1,0.1) -- (-0.5,0.5);
				\draw[thick] (0.5,-0.5) -- (0.1,-0.1);
	\end{tikzpicture}
\end{array} = \left[ \ 0 \ \xrightarrow{} \
\begin{array}{c}
     \begin{tikzpicture}[scale=0.5]
				\draw[thick,-stealth] (-0.5,-0.5) -- (0.5,0.5);
				\draw[thick,-stealth] (0.5,-0.5) -- (-0.5,0.5);
				\draw[fill=\myred] (0,0) circle (0.15);
	\end{tikzpicture}
\end{array}
\{N\}[-1] \ \xrightarrow{\chi_0} \ 
\begin{array}{c}
    \begin{tikzpicture}[scale=0.5]
				\draw[thick,-stealth] (-0.5,-0.5) to[out=45,in=270] (-0.15,0) to[out=90,in=315] (-0.5,0.5);
				\draw[thick,-stealth] (0.5,-0.5) to[out=135,in=270] (0.15,0) to[out=90,in=225] (0.5,0.5);
	\end{tikzpicture}
\end{array}
\{N-1\}[0] \ \xrightarrow{} \ 0 \ \right], \;
\begin{array}{c}
     \begin{tikzpicture}[scale=0.5, xscale=-1]
				\draw[thick,-stealth] (-0.5,-0.5) -- (0.5,0.5);
				\draw[thick,-stealth] (-0.1,0.1) -- (-0.5,0.5);
				\draw[thick] (0.5,-0.5) -- (0.1,-0.1);
	\end{tikzpicture}
\end{array}
 = \left[ \ 0 \ \xrightarrow{} \ 
 \begin{array}{c}
       \begin{tikzpicture}[scale=0.5]
				\draw[thick,-stealth] (-0.5,-0.5) to[out=45,in=270] (-0.15,0) to[out=90,in=315] (-0.5,0.5);
				\draw[thick,-stealth] (0.5,-0.5) to[out=135,in=270] (0.15,0) to[out=90,in=225] (0.5,0.5);
	   \end{tikzpicture}
 \end{array}
\{1-N\}[0] \ \xrightarrow{\chi_1} \ 
\begin{array}{c}
    \begin{tikzpicture}[scale=0.5]
				\draw[thick,-stealth] (-0.5,-0.5) -- (0.5,0.5);
				\draw[thick,-stealth] (0.5,-0.5) -- (-0.5,0.5);
				\draw[fill=\myred] (0,0) circle (0.15);
	\end{tikzpicture}
\end{array}
\{-N\}[1] \ \xrightarrow{} \ 0 \ \right]
\end{equation}
where $[k]$ denotes the cohomological grading of the complex, i.e. $t$-powers in the Khovanov--Rozansky polynomials. In what follows we omit them in complexes but restore in the final answers for the Khovanov--Rozansky polynomials.

To each resolution, we assign the matrix factorization with the potential $\omega = x_1^{N+1}+x_2^{N+1}-x_3^{N+1}-x_4^{N+1}$:

\begin{equation}\label{mfres0}
\setlength{\unitlength}{0.45pt}
\begin{picture}(300,30)(340,-5)

\linethickness{0.26mm}

    \put(115,30){\mbox{$\scriptstyle x_1$}}
    \put(155,30){\mbox{$\scriptstyle x_2$}}
    \put(115,-34){\mbox{$\scriptstyle x_4$}}
    \put(155,-34){\mbox{$\scriptstyle x_3$}}

    \put(130,15){\vector(-1,1){5}}
    \put(155,15){\vector(1,1){5}}
    \qbezier(125,20)(145,0)(125,-20)
    \qbezier(160,20)(140,0)(160,-20)

    \put(185,-10){\mbox{$= [R \xrightarrow{\pi_{14}} R\{1-N\} \xrightarrow{x_1 - x_4} R\,] \otimes [R \xrightarrow{\pi_{23}} R\{1-N\} \xrightarrow{x_2 - x_3} R\,]$}}\,,
\end{picture} 
\end{equation}

% \begin{equation}
% \begin{array}{c}
%      \begin{tikzpicture}[scale=0.5]
% 				\draw[thick,-stealth] (-0.5,-0.5) to[out=45,in=270] (-0.15,0) to[out=90,in=315] (-0.5,0.5);
% 				\draw[thick,-stealth] (0.5,-0.5) to[out=135,in=270] (0.15,0) to[out=90,in=225] (0.5,0.5);
% 	\end{tikzpicture} 
% \end{array} = [R \xrightarrow{\pi_{13}} R \xrightarrow{x_1 - x_3} R] \otimes [R \xrightarrow{\pi_{24}} R \xrightarrow{x_2 - x_4} R]
% \end{equation}

\begin{equation}\label{mfres1}
\setlength{\unitlength}{1.2pt}
\begin{picture}(300,30)(150,-5)

    \linethickness{0.26mm}

    \put(155,19){\mbox{$\scriptstyle x_1$}}
    \put(175,19){\mbox{$\scriptstyle x_2$}}
    \put(155,-6){\mbox{$\scriptstyle x_4$}}
    \put(175,-6){\mbox{$\scriptstyle x_3$}}

    \put(25,0){
   \put(135,0){\vector(1,1){15}}
   \put(150,0){\vector(-1,1){15}}
   \put(142.5,7.5){\color{black}\circle{4.5}} 
   \put(142.5,7.5){\color{\myred}\circle*{4.1}}
   }

   \put(185,4){\mbox{$= [R \xrightarrow{u_1} R\{1-N\} \xrightarrow{x_1+x_2-x_3-x_4} R\,] \otimes [R \xrightarrow{u_2} R\{3-N\} \xrightarrow{x_1 x_2 - x_3 x_4} R\,,]\{-1\}$}}\,.
\end{picture}
\end{equation}
 % \begin{equation}
 % \begin{array}{c}
 %     \begin{tikzpicture}[scale=0.5]
 % 				\draw[thick,-stealth] (-0.5,-0.5) -- (0.5,0.5);
 % 				\draw[thick,-stealth] (0.5,-0.5) -- (-0.5,0.5);
 % 				\draw[fill=black] (0,0) circle (0.15);
 % 	\end{tikzpicture}
 % \end{array} = [R \xrightarrow{u_1} R\{1-N\} \xrightarrow{x_1+x_2-x_3-x_4} R] \otimes [R \xrightarrow{u_2} R\{3-N\} \xrightarrow{x_1 x_2 - x_3 x_4} R]\{-1\}
 % \end{equation}
%where the superpotential 
%\begin{equation}\label{superpot}
%    \omega = x_1^{N+1}+x_2^{N+1}-x_3^{N+1}-x_4^{N+1}\,.
%\end{equation}
To have such a potential, the undefined polynomials above must satisfy the following relations:
\begin{equation}
\begin{aligned}
    \pi_{14}(x_1,x_4)(x_1 - x_4) + \pi_{23}(x_2,x_3)(x_2 - x_3) = x_1^{N+1}+x_2^{N+1}-x_3^{N+1}-x_4^{N+1}\,, \\  
    u_1(x_1,x_2,x_3,x_4)(x_1 + x_2 - x_3 - x_4) + u_2(x_1,x_2,x_3,x_4)(x_1 x_2 - x_3 x_4) = x_1^{N+1}+x_2^{N+1}-x_3^{N+1}-x_4^{N+1}\,.
\end{aligned}
\end{equation}
The solutions of Khovanov and Rozansky~\cite{KhR} are
%Here we need the following functions:
\begin{equation}
    \pi_{ij}(x_i,x_j) = \frac{x_{i}^{N+1} - x_{j}^{N+1}}{x_{i} - x_{j}}\,,
\end{equation}
\begin{equation}
    g(x,y) = \left(\frac{x - \sqrt{x^2 - 4y}}{2}\right)^{N+1} + \left(\frac{x + \sqrt{x^2 - 4y}}{2}\right)^{N+1}\,,
\end{equation}
\begin{equation}\label{u}
    \begin{split}
        u_{1}(x_1,x_2,x_3,x_4) = \frac{x_{1}^{N+1} + x_{2}^{N+1} - g(x_3 + x_4, x_1x_2)}{x_1 +x_2 - x_3 - x_4}\,, \\
        u_{2}(x_1,x_2,x_3,x_4) = \frac{g(x_3 + x_4, x_1x_2) - x_{3}^{N+1} - x_{4}^{N+1}}{x_1x_2-x_3x_4}\,.
    \end{split}
\end{equation}
After resolving all the crossings according to~\eqref{res-cross}, we obtain a hypercube as in Section~\ref{sec:HOMFLY}. At each vertex of this hypercube, there lies a closed MOY graph being the whole link resolution. To such a graph, we associate the tensor product of the matrix factorizations from (\ref{res0}) and (\ref{res1}). Note that a resulting factorization for a graph has the zero potential, which implies that it is a complex. We can calculate its cohomologies, which split in a direct sum $\mathcal{H}=\mathcal{H}^0\oplus\mathcal{H}^1$.
%, constructed from a number of resolutions

For convenience, we introduce the numbering of vertices of this hypercube. Consider a link with $n$ crossings, which we enumerate from 1 to $n$ in some fixed order. As in the HOMFLY polynomial case (see Section~\ref{sec:HOMFLY}), we define the 0- and 1-resolutions by the following scheme:

\begin{equation}\label{KhRres}
\begin{picture}(300,40)(30,-15)

\linethickness{0.26mm}

    \setlength{\unitlength}{0.5pt}{
    \put(-60,-20){\vector(1,1){40}}
    \put(-19,-21){\line(-1,1){18}}
    \put(-43,3){\vector(-1,1){17}}
    }

    \put(0,-8){\mbox{\Large $:$}}

    \put(90,-8){\mbox{$\{N\}$}}

    \put(30,40){\mbox{type 0}}

    \put(135,-10){\mbox{\Large $,$}}

    \put(160,40){\mbox{type 1}}

    \put(210,-8){\mbox{$\{N-1\}$}}

    \put(290,-10){\mbox{\Large $,$}}
    
    \setlength{\unitlength}{1.3pt}{
    \put(-120,-8){
   \put(135,0){\vector(1,1){15}}
   \put(150,0){\vector(-1,1){15}}
   \put(142.5,7.5){\color{black}\circle{4.5}} 
   \put(142.5,7.5){\color{\myred}\circle*{4.1}}
   }
   }

    \setlength{\unitlength}{0.5pt}{
    \put(40,-2){
   \put(130,15){\vector(-1,1){5}}
    \put(155,15){\vector(1,1){5}}
    \qbezier(125,20)(145,0)(125,-20)
    \qbezier(160,20)(140,0)(160,-20)
    }
    }

    \setlength{\unitlength}{0.5pt}{
    \put(450,0){
    \put(-61,-21){\line(1,1){17}}
    \put(-20,-20){\vector(-1,1){40}}
    \put(-36,4){\vector(1,1){17}}
    }
    }

    \put(450,0){
    
    \put(0,-8){\mbox{\Large $:$}}

    \put(75,-8){\mbox{$\{1-N\}$}}

    \put(10,0){

    \put(10,40){\mbox{type 1}}

    \put(145,-10){\mbox{\Large $,$}}

    \put(165,40){\mbox{type 0}}

    \put(220,-8){\mbox{$\{-N\}$}}

    \put(280,-10){\mbox{\Large $.$}}
    
    \setlength{\unitlength}{1.3pt}{
    \put(-127.5,-8){
   \put(135,0){\vector(1,1){15}}
   \put(150,0){\vector(-1,1){15}}
   \put(142.5,7.5){\color{black}\circle{4.5}} 
   \put(142.5,7.5){\color{\myred}\circle*{4.1}}
   }
   }

    \setlength{\unitlength}{0.5pt}{
    \put(50,-2){
   \put(130,15){\vector(-1,1){5}}
    \put(155,15){\vector(1,1){5}}
    \qbezier(125,20)(145,0)(125,-20)
    \qbezier(160,20)(140,0)(160,-20)
    }
    }
    }
    }
    
\end{picture}
\end{equation}
Then, to each MOY diagram, one can associate a vector $\boldsymbol{\alpha}=(\alpha_1,...,\alpha_n)$, where $\alpha_i=0,1$ indicates how the $i$-th crossing is resolved in a diagram and define a {\it height} of a vertex $|\boldsymbol{\alpha}|=\sum\alpha_i$. Thus, all vertices of a hypercube become indexed by such vectors.

Let us expand tensor products in factorizations for resolutions~\eqref{mfres0},~\eqref{mfres1}:
%write factorizations for resolutions in more detail

\begin{equation}\label{res0}
\setlength{\unitlength}{0.5pt}
\begin{picture}(300,50)(180,20)

\linethickness{0.26mm}

	\put(-10,10){\mbox{$\Gamma_0=$}}

    \put(40,45){\mbox{$\scriptstyle x_1$}}
    \put(80,45){\mbox{$\scriptstyle x_2$}}
    \put(40,-20){\mbox{$\scriptstyle x_4$}}
    \put(80,-20){\mbox{$\scriptstyle x_3$}}
    
    \put(-75,15){
    \put(130,15){\vector(-1,1){5}}
    \put(155,15){\vector(1,1){5}}
    \qbezier(125,20)(145,0)(125,-20)
    \qbezier(160,20)(140,0)(160,-20)
    }

    \put(100,10){\mbox{$= \bigg[\begin{pmatrix}
        R \\
        R\{2-2N\}
    \end{pmatrix} \xrightarrow{P_0} \begin{pmatrix}
        R\{1-N\}\\
        R\{1-N\}
    \end{pmatrix} \xrightarrow{P_1} \begin{pmatrix}
        R \\
        R\{2-2N\}
    \end{pmatrix}\bigg]$}}
    
\end{picture}
\end{equation}

% \begin{equation}
% \begin{array}{c}
%      \begin{tikzpicture}[scale=0.5]
% 				\draw[thick,-stealth] (-0.5,-0.5) to[out=45,in=270] (-0.15,0) to[out=90,in=315] (-0.5,0.5);
% 				\draw[thick,-stealth] (0.5,-0.5) to[out=135,in=270] (0.15,0) to[out=90,in=225] (0.5,0.5);
% 	\end{tikzpicture}
% \end{array} = \begin{pmatrix}
%         R \\
%         R
%     \end{pmatrix} \xrightarrow{\begin{pmatrix}
%         \pi_{14} & x_2-x_3\\
%         \pi_{23} & x_4 - x_1
%     \end{pmatrix}} \begin{pmatrix}
%         R\\
%         R
%     \end{pmatrix} \xrightarrow{\begin{pmatrix}
%         x_1-x_4 & x_2-x_3 \\
%         \pi_{23} & -\pi_{14}
%     \end{pmatrix}} \begin{pmatrix}
%         R \\
%         R
%     \end{pmatrix}
% \end{equation}

\begin{equation}\label{res1}
\setlength{\unitlength}{1.2pt}
\begin{picture}(200,50)(98,-10)
    \linethickness{0.26mm}
    
    \put(55,10){\mbox{$\Gamma_1=$}}

    \put(75,25){\mbox{$\scriptstyle x_1$}}
    \put(95,25){\mbox{$\scriptstyle x_2$}}
    \put(75,-2.5){\mbox{$\scriptstyle x_4$}}
    \put(95,-2.5){\mbox{$\scriptstyle x_3$}}

    \put(-55,5){
   \put(135,0){\vector(1,1){15}}
   \put(150,0){\vector(-1,1){15}}
   \put(142.5,7.5){\color{black}\circle{4.5}} 
   \put(142.5,7.5){\color{\myred}\circle*{4.1}}
   }

    \put(100,10){\mbox{$= \bigg[\begin{pmatrix}
         R\{-1\} \\
         R\{3-2N\}
     \end{pmatrix} \xrightarrow{Q_0} \begin{pmatrix}
         R\{-N\} \\
         R\{2-N\}
     \end{pmatrix} \xrightarrow{Q_1} \begin{pmatrix}
         R\{-1\} \\
         R\{3-2N\}
     \end{pmatrix}\bigg]$}}
    
\end{picture}
\end{equation}
where matrices $P_0,P_1$ and $Q_0,Q_1$ are equal to:
\begin{equation}\label{Ps}
    \begin{split}
        P_0=\begin{pmatrix}
        \pi_{14} & x_2-x_3\\
        \pi_{23} & x_4 - x_1
    \end{pmatrix},\quad P_1=\begin{pmatrix}
        x_1-x_4 & x_2-x_3 \\
        \pi_{23} & -\pi_{14}
    \end{pmatrix},
    \end{split}
\end{equation}
\begin{equation}\label{Qs}
     Q_0=\begin{pmatrix}
         u_1 & x_1x_2-x_3x_4\\
         u_2 & x_3 +x_4-x_1-x_2
     \end{pmatrix},\quad Q_1=\begin{pmatrix}
         x_1 +x_2 - x_3 -x_4 & x_1x_2 - x_3x_4 \\
         u_2 & 
         -u_1
     \end{pmatrix}.
\end{equation}
For tensor products of these matrix factorizations, there exist quasi-isomorphisms that we call categorified MOY-relations which also follow from the Reidemeister invariance:

\begin{subequations}
	\begin{align}
		&\label{MOYloc_II}\begin{array}{c}
			\begin{tikzpicture}
				\draw[thick,postaction={decorate},decoration={markings, mark= at position 0.5 with {\arrow{stealth}}}] (0,0) circle (0.35);
				\node[right] at (0.35,0) {$\scriptstyle x$};
			\end{tikzpicture}
		\end{array}\cong V_{N},\quad{\rm dim}_q\,V_N=[N]\,, \\
		&\label{MOYloc_III}\begin{array}{c}
			\begin{tikzpicture}[scale=0.7]
				\draw[thick,stealth-,postaction={decorate},decoration={markings, mark= at position 0.7 with {\arrow{stealth}}}] (0,1) -- (0,0.5) to[out=180,in=90] (-0.5,0) to[out=270,in=180] (0,-0.5) -- (0,-1);
				\draw[fill=\myred] (-0.1,0.5) to[out=90,in=180] (0,0.6) to[out=0,in=90] (0.6,0) to[out=270,in=0] (0,-0.6) to[out=180,in=270] (-0.1,-0.5) to[out=90,in=180] (0,-0.4) to[out=0,in=270] (0.4,0) to[out=90,in=0] (0,0.4) to[out=180,in=270] (-0.1,0.5);
				\node[right] at (0,1) {$\scriptstyle x$};
				\node[right] at (0,-1) {$\scriptstyle z$};
				\node[left] at (-0.5,0) {$\scriptstyle y$};
			\end{tikzpicture}
		\end{array}\cong\begin{array}{c}
			\begin{tikzpicture}[scale=1.0]
				\draw[thick, -stealth] (0,-0.5) -- (0,0.5);
				\node[right] at (0,0.5) {$\scriptstyle x$};
				\node[right] at (0,-0.5) {$\scriptstyle z$};
			\end{tikzpicture}
		\end{array}\otimes V_{N-1}, \quad {\rm dim}_q\,V_{N-1}=[N-1]\,,\\
		&\label{MOYloc_IV}\begin{array}{c}
			\begin{tikzpicture}[scale=0.7]
				\begin{scope}[shift={(0,1)}]
					\draw[thick,-stealth] (-0.5,-0.5) to[out=90,in=270] (0.5,0.5);
					\draw[thick,-stealth] (0.5,-0.5) to[out=90,in=270] (-0.5,0.5);
					\draw[fill=\myred] (0,0) circle (0.1);
				\end{scope}
				\draw[thick,-stealth] (-0.5,-0.5) to[out=90,in=270] (0.5,0.5) -- (0.5,0.6);
				\draw[thick,-stealth] (0.5,-0.5) to[out=90,in=270] (-0.5,0.5) -- (-0.5,0.6);
				\draw[fill=\myred] (0,0) circle (0.1);
				%%%%%%%%%%%%%%%%%%%%%%%%%%%%%%%%%%%%%%%%%%%%%%%%%%%%%%%%%%%%5
				\node[left] at (-0.5,-0.5) {$\scriptstyle x_4$};
				\node[right] at (0.5,-0.5) {$\scriptstyle x_3$};
				\node[left] at (-0.5,0.5) {$\scriptstyle y$};
				\node[right] at (0.5,0.5) {$\scriptstyle z$};
				\node[left] at (-0.5,1.5) {$\scriptstyle x_1$};
				\node[right] at (0.5,1.5) {$\scriptstyle x_2$};
			\end{tikzpicture}
		\end{array}\cong \begin{array}{c}
			\begin{tikzpicture}[scale=0.7]
				\draw[thick,-stealth] (-0.5,-0.5) to[out=90,in=270] (0.5,0.5);
				\draw[thick,-stealth] (0.5,-0.5) to[out=90,in=270] (-0.5,0.5);
				\draw[fill=\myred] (0,0) circle (0.1);
				\node[left] at (-0.5,-0.5) {$\scriptstyle x_4$};
				\node[right] at (0.5,-0.5) {$\scriptstyle x_3$};
				\node[left] at (-0.5,0.5) {$\scriptstyle x_1$};
				\node[right] at (0.5,0.5) {$\scriptstyle x_2$};
			\end{tikzpicture}
		\end{array}\otimes V_2,\quad {\rm dim}_q\,V_2=[2]\,,\\
		&\label{MOYloc_V}\begin{array}{c}
			\begin{tikzpicture}[scale=0.7]
				\draw[thick,postaction={decorate},decoration={markings, mark= at position 0.7 with {\arrow{stealth}}}] (-1,-1) -- (-0.5,-0.5);
				\draw[thick,postaction={decorate},decoration={markings, mark= at position 0.7 with {\arrow{stealth}}}] (1,1) -- (0.5,0.5);
				\draw[thick,postaction={decorate},decoration={markings, mark= at position 0.7 with {\arrow{stealth}}}] (-0.5,0.5) -- (-1,1);
				\draw[thick,postaction={decorate},decoration={markings, mark= at position 0.7 with {\arrow{stealth}}}] (0.5,-0.5) -- (1,-1);
				\draw[thick,postaction={decorate},decoration={markings, mark= at position 0.7 with {\arrow{stealth}}}] (-0.5,0.5) -- (0.5,0.5);
				\draw[thick,postaction={decorate},decoration={markings, mark= at position 0.7 with {\arrow{stealth}}}] (0.5,-0.5) -- (-0.5,-0.5);
				\begin{scope}[shift={(-0.5,0)}]
					\draw[fill=\myred] (0.1,0.5) to[out=90,in=0] (0,0.6) to[out=180,in=90] (-0.1,0.5) -- (-0.1,-0.5) to[out=270,in=180] (0,-0.6) to[out=0,in=270] (0.1,-0.5) -- (0.1,0.5);
				\end{scope}
				\begin{scope}[shift={(0.5,0)}]
					\draw[fill=\myred] (0.1,0.5) to[out=90,in=0] (0,0.6) to[out=180,in=90] (-0.1,0.5) -- (-0.1,-0.5) to[out=270,in=180] (0,-0.6) to[out=0,in=270] (0.1,-0.5) -- (0.1,0.5);
				\end{scope}
				\node[left] at (-1,-1) {$\scriptstyle x_1$};
				\node[left] at (-1,1) {$\scriptstyle x_2$};
				\node[right] at (1,1) {$\scriptstyle x_3$};
				\node[right] at (1,-1) {$\scriptstyle x_4$};
				\node[above] at (0,0.5) {$\scriptstyle x_5$};
				\node[below] at (0,-0.5) {$\scriptstyle x_6$};
			\end{tikzpicture}
		\end{array}\;\cong\;\begin{array}{c}
			\begin{tikzpicture}
				\draw[thick, stealth-] (-0.5,0.5) to[out=315,in=225] (0.5,0.5);
				\draw[thick, stealth-] (0.5,-0.5) to[out=135,in=45] (-0.5,-0.5);
				\node[left] at (-0.5,-0.5) {$\scriptstyle x_1$};
				\node[left] at (-0.5,0.5) {$\scriptstyle x_2$};
				\node[right] at (0.5,0.5) {$\scriptstyle x_3$};
				\node[right] at (0.5,-0.5) {$\scriptstyle x_4$};
			\end{tikzpicture}
		\end{array} \oplus \left(\begin{array}{c}
			\begin{tikzpicture}
				\draw[thick, stealth-] (-0.5,0.5) to[out=315,in=45] (-0.5,-0.5);
				\draw[thick, stealth-] (0.5,-0.5) to[out=135,in=225] (0.5,0.5);
				\node[left] at (-0.5,-0.5) {$\scriptstyle x_1$};
				\node[left] at (-0.5,0.5) {$\scriptstyle x_2$};
				\node[right] at (0.5,0.5) {$\scriptstyle x_3$};
				\node[right] at (0.5,-0.5) {$\scriptstyle x_4$};
			\end{tikzpicture}
		\end{array} \otimes V_{N-2}\right),\quad {\rm dim}_q\,V_{N-2}=[N-2]\,,\\
		&\label{MOYloc_VI}\begin{array}{c}
			\begin{tikzpicture}[scale=0.6]
				\draw[thick,-stealth] (0,0) to[out=90,in=225] (1.5,1.5) to[out=45,in=270] (2,3);
				\draw[thick,-stealth] (2,0) to[out=90,in=315] (1.5,1.5) to[out=135,in=270] (0,3);
				\draw[thick,-stealth] (1,0) to[out=90,in=270] (0,1.5) to[out=90,in=270] (1,3);
				\node[left] at (0,0) {$\scriptstyle x_1$};
				\node[left] at (1,0) {$\scriptstyle x_2$};
				\node[right] at (2,0) {$\scriptstyle x_3$};
				\node[left] at (0,3) {$\scriptstyle x_4$};
				\node[left] at (1,3) {$\scriptstyle x_5$};
				\node[right] at (2,3) {$\scriptstyle x_6$};
				\draw[fill=\myred] (1.5,1.5) circle (0.13) (0.45,0.78) circle (0.13) (0.45,2.22) circle (0.13);
			\end{tikzpicture}
		\end{array}\oplus\begin{array}{c}
			\begin{tikzpicture}[scale=0.6]
				\draw[thick,-stealth] (0,0) -- (0,3);
				\draw[thick,-stealth] (1,0) to[out=90,in=270] (2,3);
				\draw[thick,-stealth] (2,0) to[out=90,in=270] (1,3);
				\node[left] at (0,0) {$\scriptstyle x_1$};
				\node[left] at (1,0) {$\scriptstyle x_2$};
				\node[right] at (2,0) {$\scriptstyle x_3$};
				\node[left] at (0,3) {$\scriptstyle x_4$};
				\node[left] at (1,3) {$\scriptstyle x_5$};
				\node[right] at (2,3) {$\scriptstyle x_6$};
				\draw[fill=\myred] (1.5,1.5) circle (0.13);
			\end{tikzpicture}
		\end{array}\cong \begin{array}{c}
			\begin{tikzpicture}[scale=0.6,xscale=-1]
				\draw[thick,-stealth] (0,0) to[out=90,in=225] (1.5,1.5) to[out=45,in=270] (2,3);
				\draw[thick,-stealth] (2,0) to[out=90,in=315] (1.5,1.5) to[out=135,in=270] (0,3);
				\draw[thick,-stealth] (1,0) to[out=90,in=270] (0,1.5) to[out=90,in=270] (1,3);
				\node[right] at (0,0) {$\scriptstyle x_3$};
				\node[left] at (1,0) {$\scriptstyle x_2$};
				\node[left] at (2,0) {$\scriptstyle x_1$};
				\node[right] at (0,3) {$\scriptstyle x_6$};
				\node[left] at (1,3) {$\scriptstyle x_5$};
				\node[left] at (2,3) {$\scriptstyle x_4$};
				\draw[fill=\myred] (1.5,1.5) circle (0.13) (0.45,0.78) circle (0.13) (0.45,2.22) circle (0.13);
			\end{tikzpicture}
		\end{array}\oplus\begin{array}{c}
			\begin{tikzpicture}[scale=0.6,xscale=-1]
				\draw[thick,-stealth] (0,0) -- (0,3);
				\draw[thick,-stealth] (1,0) to[out=90,in=270] (2,3);
				\draw[thick,-stealth] (2,0) to[out=90,in=270] (1,3);
				\node[right] at (0,0) {$\scriptstyle x_3$};
				\node[left] at (1,0) {$\scriptstyle x_2$};
				\node[left] at (2,0) {$\scriptstyle x_1$};
				\node[right] at (0,3) {$\scriptstyle x_6$};
				\node[left] at (1,3) {$\scriptstyle x_5$};
				\node[left] at (2,3) {$\scriptstyle x_4$};
				\draw[fill=\myred] (1.5,1.5) circle (0.13);
			\end{tikzpicture}
		\end{array}\,.
	\end{align}
\end{subequations}
% \begin{equation}
% \begin{array}{c}
%      \begin{tikzpicture}[scale=0.5]
% 				\draw[thick,-stealth] (-0.5,-0.5) -- (0.5,0.5);
% 				\draw[thick,-stealth] (0.5,-0.5) -- (-0.5,0.5);
% 				\draw[fill=black] (0,0) circle (0.15);
% 	\end{tikzpicture}
% \end{array} = \begin{pmatrix}
%          R \\
%          R\{4-2N\}
%      \end{pmatrix} \xrightarrow{\begin{pmatrix}
%          u_1 & x_1x_2-x_3x_4\\
%          u_2 & x_3 +x_4-x_1-x_2
%      \end{pmatrix}} \begin{pmatrix}
%          R\{1-N\} \\
%          R\{3-N\}
%      \end{pmatrix} \xrightarrow{\begin{pmatrix}
%          x_1 +x_2 - x_3 -x_4 & x_1x_2 - x_3x_4 \\
%          u_2 & 
%          -u_1
%      \end{pmatrix}} \begin{pmatrix}
%          R \\
%          R\{4-2N\}
%      \end{pmatrix}\{-1\}
% \end{equation}
Define the morphisms between resolutions $\chi_0: 
{\setlength{\unitlength}{0.4pt}
\begin{picture}(50,20)(40,5)

\linethickness{0.26mm}

    \put(-75,15){
    \put(130,15){\vector(-1,1){5}}
    \put(155,15){\vector(1,1){5}}
    \qbezier(125,20)(145,0)(125,-20)
    \qbezier(160,20)(140,0)(160,-20)
    }
\end{picture}}
% \begin{array}{c}
%      \begin{tikzpicture}[scale=0.5]
% 				\draw[thick,-stealth] (-0.5,-0.5) to[out=45,in=270] (-0.15,0) to[out=90,in=315] (-0.5,0.5);
% 				\draw[thick,-stealth] (0.5,-0.5) to[out=135,in=270] (0.15,0) to[out=90,in=225] (0.5,0.5);
% 	\end{tikzpicture}
% \end{array}
 \xrightarrow{} 
{\setlength{\unitlength}{1.0pt}
\begin{picture}(20,20)(57,3)
    \linethickness{0.26mm}

    \put(-75,0){
   \put(135,0){\vector(1,1){15}}
   \put(150,0){\vector(-1,1){15}}
   \put(142.5,7.5){\color{black}\circle{4.5}} 
   \put(142.5,7.5){\color{\myred}\circle*{4.1}}
   }
\end{picture}}
 % \begin{array}{c}
 %       \begin{tikzpicture}[scale=0.5]
	% 			\draw[thick,-stealth] (-0.5,-0.5) -- (0.5,0.5);
	% 			\draw[thick,-stealth] (0.5,-0.5) -- (-0.5,0.5);
	% 			\draw[fill=black] (0,0) circle (0.15);
	%    \end{tikzpicture}
 % \end{array}
 \, $ and $\chi_1: 
 {\setlength{\unitlength}{1.0pt}
\begin{picture}(20,20)(57,3)
    \linethickness{0.26mm}

    \put(-75,0){
   \put(135,0){\vector(1,1){15}}
   \put(150,0){\vector(-1,1){15}}
   \put(142.5,7.5){\color{black}\circle{4.5}} 
   \put(142.5,7.5){\color{\myred}\circle*{4.1}}
   }
\end{picture}}
 % \begin{array}{c}
 %       \begin{tikzpicture}[scale=0.5]
	% 			\draw[thick,-stealth] (-0.5,-0.5) -- (0.5,0.5);
	% 			\draw[thick,-stealth] (0.5,-0.5) -- (-0.5,0.5);
	% 			\draw[fill=black] (0,0) circle (0.15);
	% \end{tikzpicture}
 % \end{array} 
 \xrightarrow{} 
 {\setlength{\unitlength}{0.4pt}
\begin{picture}(50,20)(40,5)

\linethickness{0.26mm}

    \put(-75,15){
    \put(130,15){\vector(-1,1){5}}
    \put(155,15){\vector(1,1){5}}
    \qbezier(125,20)(145,0)(125,-20)
    \qbezier(160,20)(140,0)(160,-20)
    }
\end{picture}}
 % \begin{array}{c}
 %       \begin{tikzpicture}[scale=0.5]
	% 			\draw[thick,-stealth] (-0.5,-0.5) to[out=45,in=270] (-0.15,0) to[out=90,in=315] (-0.5,0.5);
	% 			\draw[thick,-stealth] (0.5,-0.5) to[out=135,in=270] (0.15,0) to[out=90,in=225] (0.5,0.5);
	% 	\end{tikzpicture}
 % \end{array}
\ $ 
by the following matrices:
\begin{equation}
    \chi_0(x_1,x_2,x_3,x_4):\quad U_0 = \begin{pmatrix}
        x_4-x_2+\mu(x_1+x_2-x_3-x_4) & 0 \\
        \frac{x_1 + x_1u_2 - \pi_{23}}{x_1-x_4} & 1
    \end{pmatrix},\quad U_1 = \begin{pmatrix}
        x_4+\mu(x_1-x_4) & \mu(x_2-x_3)-x_2\\
        -1 & 1
    \end{pmatrix},
\label{chi0}
\end{equation}
\begin{equation}
    \chi_1(x_1,x_2,x_3,x_4):\quad V_{0} = \begin{pmatrix}
        1&0\\
        \lambda u_2 - \frac{x_1 + x_1u_2 - \pi_{23}}{x_4-x_1} & \lambda(x_3+x_4-x_1-x_2) + x_1 -x_3
    \end{pmatrix},\quad V_{1} = \begin{pmatrix}
        1 &x_3 \lambda(x_2-x_3)\\
        1 & x_1 + \lambda(x_4-x_1)
    \end{pmatrix}.
\label{chi1}
\end{equation}
where $\mu,\lambda\in \mathbb{Z}$. Different choices of $\mu$ and $\lambda$ give homotopic maps.

The following diagrams are commutative:

\begin{equation}
	\begin{array}{c}
		\begin{tikzpicture}
			\node(A) at (0,0) {$\left(\begin{array}{c}
					R \\
					R\{2-2N\}\\
				\end{array}\right)$};
			\node(B) at (5,0) {$\left(\begin{array}{c}
					R\{1-N\}\\
					R\{1-N\}\\
				\end{array}\right)$};
			\node(C) at (0,-2) {$\left(\begin{array}{c}
					R\{-1\}\\
					R\{3-2N\}\\
				\end{array}\right)$};
			\node(D) at (5,-2) {$\left(\begin{array}{c}
					R\{-N\}\\
					R\{2-N\}\\
				\end{array}\right)$};
			\draw[-stealth] ([shift={(0,0.1)}]A.east) -- ([shift={(0,0.1)}]B.west) node[pos=0.5,above] {$\scriptstyle P_0$};
			\draw[stealth-] ([shift={(0,-0.1)}]A.east) -- ([shift={(0,-0.1)}]B.west) node[pos=0.5,below] {$\scriptstyle P_1$};
			\draw[-stealth] ([shift={(0,0.1)}]C.east) -- ([shift={(0,0.1)}]D.west) node[pos=0.5,above] {$\scriptstyle Q_0$};
			\draw[stealth-] ([shift={(0,-0.1)}]C.east) -- ([shift={(0,-0.1)}]D.west) node[pos=0.5,below] {$\scriptstyle Q_1$};
			\draw[-stealth] ([shift={(-0.1,0)}]A.south) -- ([shift={(-0.1,0)}]C.north) node[pos=0.5,left] {$\scriptstyle U_0$};
			\draw[stealth-] ([shift={(0.1,0)}]A.south) -- ([shift={(0.1,0)}]C.north) node[pos=0.5,right] {$\scriptstyle V_0$};
			\draw[-stealth] ([shift={(-0.1,0)}]B.south) -- ([shift={(-0.1,0)}]D.north) node[pos=0.5,left] {$\scriptstyle U_1$};
			\draw[stealth-] ([shift={(0.1,0)}]B.south) -- ([shift={(0.1,0)}]D.north) node[pos=0.5,right] {$\scriptstyle V_1$};
			\node[left] at (A.west) {$\Gamma_0=$};
			\node[left] at (C.west) {$\Gamma_1=$};
		\end{tikzpicture}
	\end{array}
\end{equation}
%\begin{equation}
%    % https://tikzcd.yichuanshen.de/#N4Igdg9gJgpgziAXAbVABwnAlgFyxMJZABgBpiBdUkANwEMAbAVxiRDpAF9T1Nd9CKMgCYqtRizYAjLjxAZseAkWHkx9Zq0QgAxrN6KBK0qOobJ2qFzEwoAc3hFQAMwBOEALZIyIHBCTC3C7uXog+fkgAjEEgbp5R1BGIAMwxcaGqvv4pnBScQA
%\begin{tikzcd}
%\begin{pmatrix}
%    R\\
%    R
%\end{pmatrix} \arrow{rr}{P_0} \arrow{dd}{U_0} &  & \begin{pmatrix}
%    R\\
%    R
%\end{pmatrix} \arrow{dd}{U_1} \\
%                        &  &              \\
%\begin{pmatrix}
%    R\{-1\}\\
%    R\{3-2N\}
%\end{pmatrix} \arrow{rr}{Q_0}            &  & \begin{pmatrix}
%    R\{-N\}\\
%    R\{2-N\}
%\end{pmatrix}           
%\end{tikzcd} \quad
%\begin{tikzcd}
%\begin{pmatrix}
%    R\\
%    R
%\end{pmatrix} \arrow{rr}{P_0} \arrow{dd}{V_0} &  & \begin{pmatrix}
%    P\\
%    P
%\end{pmatrix} \arrow{dd}{V_1} \\
%                        &  &              \\
%\begin{pmatrix}
%    R\{-1\}\\
%    R\{3-2N\}
%\end{pmatrix} \arrow{rr}{Q_0}            &  & \begin{pmatrix}
%    R\{-N\}\\
%    R\{2-N\}
%\end{pmatrix}           
%\end{tikzcd}
%\end{equation}
where matrices $P_i,\,Q_i$ are defined in (\ref{Ps}),~\eqref{Qs}. %Actually, the commutativity of these diagrams and explicit formulas for $P_i,\,Q_i$~(\ref{Ps}),~\eqref{Qs} completely fixes  

\medskip

\noindent We now describe the procedure for computing the Khovanov-Rozansky cohomologies.
\begin{enumerate}
    \item We resolve the crossings of the link $\mathcal{L}$, thus, obtaining a set of MOY graphs organized into a hypercube of resolutions. Its vertices are indexed by vectors $\boldsymbol{\alpha}$ and are equipped with a height $|\boldsymbol{\alpha}|=\sum \alpha_i$. See the example in~\eqref{Hopf-compl}.
    %$\Gamma_{\boldsymbol{\alpha}}$ with 
    
    \item To each MOY graph $\Gamma_{\boldsymbol{\alpha}}$, we assign the tensor product of matrix factorizations from (\ref{mfres0}) and (\ref{mfres1}), which itself is a 2-complex (see for example,~\eqref{G0-unknot},~\eqref{G1-unknot}). We denote its cohomology by $\mathcal{H}(\Gamma_{\boldsymbol{\alpha}})$ (this cohomology also splits into a direct sum $\mathcal{H}(\Gamma_{\boldsymbol{\alpha}})=\mathcal{H}^0(\Gamma_{\boldsymbol{\alpha}})\oplus\mathcal{H}^1(\Gamma_{\boldsymbol{\alpha}})$).

    \item We then obtain a hypercube whose vertices are the homologies $\mathcal{H}(\Gamma_{\boldsymbol{\alpha}})$ of the corresponding factorizations, indexed by the vectors $\boldsymbol{\alpha}$. The edges of this hypercube are the maps $\chi_j$ from (\ref{chi0}) and (\ref{chi1}), restricted to the cohomologies. This hypercube is the Khovanov--Rozansky complex. The example is given by~\eqref{Hom-Hopf-compl}.

    \item To write this complex in a more convenient form, we denote the direct sum of the cohomologies at vertices of height $j$ as $\mathcal{H}_j= \bigoplus_{|\boldsymbol{\alpha}|=j}\mathcal{H}_{\boldsymbol{\alpha}}$, and we denote the map between such direct sums as ${\cal D}_j: \mathcal{H}_j \rightarrow \mathcal{H}_{j+1}$. That is, our complex can be rewritten as:

\begin{equation}
    ...\xrightarrow{{\cal D}_{j-1}} \mathcal{H}_j \xrightarrow{{\cal D}_j} \mathcal{H}_{j+1} \xrightarrow{{\cal D}_{j+1}}...
\end{equation}

    \item To preserve the cohomological grading from \eqref{res-cross}, we must shift the numbering $j$ in $\mathcal{H}_j$ by $[-n_-]$, where $n_-$ is the number of negative crossings in a link. Cohomologies of this complex $\mathcal{H}^* = \mathcal{H}(\mathcal{H}_j,D_j)[-n_{-}]$ are Khovanov--Rozansky cohomologies. We also focus on the corresponding Poincare polynomial (it is called Khovanov-Rozansky polynomial):
    \begin{equation}
        {\rm KhR}^{\mathcal{L}}(t,q,q^N) = t^{-n_{-}}\sum_{j} t^j \dim_q \mathcal{H}^{*}_j\,.
        %=P^{\mathcal{L}}[t,q,q^N]
    \end{equation}
\end{enumerate}
Note that a $q$-dimension is equal to a sum of $q$-gradings of the corresponding basis elements.
%Note that the q-dimension, is defined in accordance with the q-grading.

The HOMFLY polynomial is the Euler characteristic of this complex:
\begin{equation}
    H^{\mathcal{L}}(q,q^N) = \sum_{j} (-1)^{j} \dim_q \mathcal{H}^{*}_j\,.
\end{equation}
We remark that Khovanov--Rozansky cohomology is triply-graded: it has the cohomologial $t$-grading, the quantum $q$-grading, and the matrix factorizations $\mathbb{Z}_2$-grading.

\section{Khovanov--Rozansky cohomology calculation}

In this section, we show how to calculate Khovanov--Rozansky cohomologies. In Section~\ref{sec:excl-var-strand}, we apply general method from Section~\ref{sec:mat-fact} to show that variables on edges can be excluded. Then, we provide some examples of practical calculations of Khovanov--Rozansky cohomologies and polynomials in Section~\ref{sec:examples}.

\subsection{Excluding variables on edges}\label{sec:excl-var-strand}

To compute the cohomology of complicated MOY graphs, we will remove some variables, thereby simplifying the graph. To achieve this, we demonstrate that if an arbitrary MOY graph has two variables on a single strand, they can be identified. Let us reiterate that our statements are formulated for closed MOY graphs, and hence, for matrix factorizations with the zero potential -- that is, for complexes.

Suppose we have a closed MOY graph with marked variables $\{x,y,z_1,...,z_n\}$ where the variables $x$ and $y$ lie on the same strand, and $z_1,...,z_n$ belong to some remaining graph $\Gamma$. Then, there exists the following quasi-isomorphism between the matrix factorizations corresponding to these graphs:
% where $R''=\mathbb{Q}[z_1,\dots,z_n]$  = [R \xrightarrow{D_0^\Gamma} R \xrightarrow{D_1^\Gamma} R\, ]
\begin{equation}
	\begin{array}{c}
		\begin{tikzpicture}[scale=0.7]
			\draw[thick, postaction={decorate},decoration={markings, mark = at position 0.25 with {\arrowreversed{stealth}}, mark = at position 0.75 with {\arrowreversed{stealth}}}] (0,0) to[out=45,in=180] (0.8,0.5) to[out=0,in=90] (1.2,0) to[out=270,in=0] (0.8,-0.5) to[out=180,in=315] (0,0);
			\draw[fill=\myblue] (0,0) circle (0.5);
			\node[white] at (0,0) {$\scriptstyle \MOY$};
			\node[right] at (1,0.5) {$\scriptstyle x$};
			\node[right] at (1,-0.5) {$\scriptstyle y$};
			%\draw[fill=\mygreen] (1.2,0) circle (0.1);
		\end{tikzpicture}
	\end{array}\cong\begin{array}{c}
		\begin{tikzpicture}[scale=0.7]
			\draw[thick, postaction={decorate},decoration={markings, mark = at position 0.4 with {\arrowreversed{stealth}}}] (0,0) to[out=45,in=180] (0.8,0.5) to[out=0,in=90] (1.2,0) to[out=270,in=0] (0.8,-0.5) to[out=180,in=315] (0,0);
			\draw[fill=\myblue] (0,0) circle (0.5);
			\node[white] at (0,0) {$\scriptstyle \MOY$};
			\node[right] at (1.2,0) {$\scriptstyle x$};
		\end{tikzpicture}
	\end{array}\,.
\end{equation}
Let us introduce the rings $R=\mathbb{Q}[x,y,z_1,...,z_n]$ and $R'=\mathbb{Q}[x,z_1,...,z_n]$ and write down the factorizations for the graphs explicitly:
\begin{equation}
    \begin{array}{c}
    	\begin{tikzpicture}[scale=0.7]
    		\draw[thick, postaction={decorate},decoration={markings, mark = at position 0.25 with {\arrowreversed{stealth}}, mark = at position 0.75 with {\arrowreversed{stealth}}}] (0,0) to[out=45,in=180] (0.8,0.5) to[out=0,in=90] (1.2,0) to[out=270,in=0] (0.8,-0.5) to[out=180,in=315] (0,0);
    		\draw[fill=\myblue] (0,0) circle (0.5);
    		\node[white] at (0,0) {$\scriptstyle \MOY$};
    		\node[right] at (1,0.5) {$\scriptstyle x$};
    		\node[right] at (1,-0.5) {$\scriptstyle y$};
    		%\draw[fill=\mygreen] (1.2,0) circle (0.1);
    	\end{tikzpicture}
    \end{array}= \Gamma\otimes[\mathbb{Q}[x,y]\xrightarrow{\pi_{xy}}\mathbb{Q}[x,y]\xrightarrow{x-y}\mathbb{Q}[x,y]\,]=\bigg[\begin{pmatrix}
        \boldsymbol{\upsilon}^{0}\\
        \boldsymbol{u}^{0}
    \end{pmatrix} \xrightarrow{\begin{pmatrix}
        D^{\Gamma}_0& x-y\\
        \pi_{xy} &-D^{\Gamma}_1
    \end{pmatrix}} \begin{pmatrix}
        \boldsymbol{\upsilon}^{1}\\
        \boldsymbol{u}^{1}
    \end{pmatrix} \xrightarrow{\begin{pmatrix}
        D^{\Gamma}_1& x-y\\
        \pi_{xy} &-D^{\Gamma}_0 
    \end{pmatrix}} \begin{pmatrix}
        \boldsymbol{\upsilon}^{0}\\
        \boldsymbol{u}^{0}
    \end{pmatrix}\bigg],
\end{equation}
\begin{equation}
    \begin{array}{c}
    	\begin{tikzpicture}[scale=0.7]
    		\draw[thick, postaction={decorate},decoration={markings, mark = at position 0.4 with {\arrowreversed{stealth}}}] (0,0) to[out=45,in=180] (0.8,0.5) to[out=0,in=90] (1.2,0) to[out=270,in=0] (0.8,-0.5) to[out=180,in=315] (0,0);
    		\draw[fill=\myblue] (0,0) circle (0.5);
    		\node[white] at (0,0) {$\scriptstyle \MOY$};
    		\node[right] at (1.2,0) {$\scriptstyle x$};
    	\end{tikzpicture}
    \end{array}=\Pi(\Gamma)_{R'} = \big[\begin{pmatrix}
        \boldsymbol{w}^0
    \end{pmatrix} \xrightarrow{\Pi(D_0^{\Gamma})} (\boldsymbol{w}^1)\xrightarrow{\Pi(D_1^{\Gamma})}(\boldsymbol{w}^0)\,\big],
\end{equation}
\begin{equation}
    \Pi(x)=x,\quad\Pi(y)=x, \quad\Pi(z_j)=z_j\,.
\end{equation}
Here, we denote both graphs and the corresponding matrix factorizations by the same letters. These factorizations are written in the form of block matrices in a suitable basis with some vectors $\boldsymbol{\upsilon}^{0,1},\,\boldsymbol{u}^{0,1}$ and $\boldsymbol{w}^{0,1}$ from $R$-modules and $R'$-module respectively. The block matrices $D^\Gamma_0$ and $D^\Gamma_1$ are taken from the following matrix factorization over the ring $\mathbb{Q}[z_1,\dots,z_n]$:
\begin{equation}
    \Gamma=\big[(V^0)\xrightarrow{D^\Gamma_0}(V^1)\xrightarrow{D^\Gamma_1}(V^0)\,\big]\,.
\end{equation}
Quasi-isomorphism $\Gamma\otimes[R\xrightarrow{\pi_{xy}}R\xrightarrow{x-y}R\,] \cong \Pi(\Gamma)_{R'}$ is a direct consequence of the theorem from Section~\ref{sec:mat-fact}, applied to $b=x-y$ and the rings $R$ and $R'$. The quasi-isomorphism $f$ and its inverse $g$ are defined as follows:
\begin{equation}
    f: \Gamma\otimes[R\xrightarrow{\pi_{xy}}R\xrightarrow{x-y}R] \rightarrow \Pi(\Gamma)_{R'}\,,\qquad  \begin{pmatrix}
        \boldsymbol{\upsilon}^k\\
        \boldsymbol{u}^k
    \end{pmatrix}\mapsto (\Pi(\boldsymbol{\upsilon}^k))\,,\quad k\in \mathbb{Z}_2\,,
\end{equation}

\begin{equation}\label{inv-map}
    g:\Pi(\Gamma)_{R'}\rightarrow \Gamma\otimes[R\xrightarrow{\pi_{xy}}R\xrightarrow{x-y}R]\,,\qquad (\boldsymbol{w}^k)\mapsto \begin{pmatrix}
        \boldsymbol{w}^k\\
        -\frac{D^{\Gamma}_k - \Pi(D^{\Gamma}_k)}{x-y} \boldsymbol{w^k}
    \end{pmatrix}\,,\quad k\in\mathbb{Z}_2\,.
\end{equation}
Note that $D^{\Gamma}_k - \Pi(D^{\Gamma}_k)$ is divisible by the polynomial $x-y$.

\paragraph{Example.} 
Let us demonstrate this quasi-isomorphism with a simple example. Consider a cycle with two marks:
\begin{equation}
\begin{picture}(300,20)(10,0)
    \put(77,7){\mbox{$\Gamma=$}}
    \put(125,10){\circle{20}}
    \put(112,10){\line(1,0){6}}
    \put(132,10){\line(1,0){6}}
    \put(100,10){\mbox{\small $x_1$}}
    \put(140,10){\mbox{\small $x_2$}}
\end{picture}
\end{equation}
Its matrix factorization (over $R = \mathbb
Q[x_1,x_2]$):
\begin{equation}
    \Gamma= [R\xrightarrow[]{\pi_{12}}R\xrightarrow[]{x_1-x_2}R\,]\otimes[R\xrightarrow[]{\pi_{21}}R\xrightarrow[]{x_2-x_1}R\,]\,.
\end{equation}
We can exclude the $x_2$ variable. So, after the map $\Pi$ we have the factorization over $R'=\mathbb{Q}[x_1]$:
\begin{equation}
    \Gamma' = [R'\xrightarrow[]{\pi_{11}}R'\xrightarrow{0}R'\,]\,.
\end{equation}
And its homologies $\mathcal{H}_{\Gamma'} = {\mathcal{H}^{1}_{\Gamma'}} = \mathbb{Q}[x_1]/x_1^N$. After the inverse map $g:\Gamma'\rightarrow\Gamma$ we have:
\begin{equation}
    \mathcal{H}_{\Gamma'} \quad \longrightarrow \quad \mathcal{H}_{\Gamma} = \mathcal{H}^1_{\Gamma} = \begin{pmatrix}
    	\mathbb{Q}[x_1]/x_1^N\\
    	-\frac{x_1 - x_2}{x_2 - x_1}\mathbb{Q}[x_1]/x_1^N
    \end{pmatrix} = \begin{pmatrix}
        \mathbb{Q}[x_1]/x_1^N\\
        \mathbb{Q}[x_1]/x_1^N
    \end{pmatrix}.
\end{equation}
Thus, we can identify variables on the same strand and compute the cohomology of simpler matrix factorizations. 

\medskip

\noindent Now we will write how the local maps between MOY graphs change when we exclude variables in both.

\paragraph{Transformation of horizontal morphism.} 
We are interested in the local map between matrix factorizations corresponding to MOY graphs, $\chi: \Gamma_0 \rightarrow \Gamma_1$. This map is defined by the components $\chi^0: M^0 \rightarrow N^0$ and $\chi^1: M^1 \rightarrow N^1$. Suppose the graphs $\Gamma_0$ and $\Gamma_1$ are embedded identically into a (already closed) MOY graph with marked variables $x_1, \ldots, x_n$. We now have two closed graphs with matrix factorizations $\Sigma_0, \Sigma_1$ over the ring $R = \mathbb{Q}[x_1, \ldots, x_n]$ constructed as tensor products of other factorizations. Assume we would like to eliminate the variable $x_1$, which corresponds to the factorization $\{a, b\}$, i.e., $b = x_1 - c$, where $c \in R' = \mathbb{Q}[x_2, \ldots, x_n]$. Let us write the factorizations $\Sigma_0, \Sigma_1$:\\
\begin{equation}
	\begin{split}
		\Sigma_0 = \Gamma_0 \otimes W \otimes \{a,b\}\,, \\
		\Sigma_1 = \Gamma_1 \otimes W \otimes \{a,b\}\,, \\
	\end{split}
\end{equation}
\begin{eqnarray}
	\Gamma_0 = M^0 \xrightarrow{d^{M}_0} M^1 \xrightarrow{d^{M}_1} M^0\,, \quad \Gamma_1 = N^0 \xrightarrow{d^{N}_0} N^1 \xrightarrow{d^{N}_1} N^0\,, \\
	W = W^0 \xrightarrow{P_0} W^1 \xrightarrow{P_1} W^0\,, \quad \{a,b\} = R^0 \xrightarrow{a} R^1 \xrightarrow{b} R^0\,.
\end{eqnarray}
Let us expand tensor products for $\Sigma_0$ (the structure of $\Sigma_1$ is analogous):
\begin{equation}
	\Gamma_0 \otimes W = \begin{pmatrix}
		M^0 W^0\\
		M^1 W^1
	\end{pmatrix} \xrightarrow{\begin{pmatrix}
			P_0 & d^{M}_1\\
			d^{M}_0 & - P_1
	\end{pmatrix}} \begin{pmatrix}
		M^0 W^1\\
		M^1 W^0
	\end{pmatrix} \xrightarrow{\begin{pmatrix}
			P_1 & d^{M}_1\\
			d^{M}_0 & -P_0
	\end{pmatrix}} \begin{pmatrix}
		M^0 W^0\\
		M^1 W^1
	\end{pmatrix},
\end{equation}
\begin{equation}
	\Gamma_0 \otimes W \otimes \{a,b\} = \begin{pmatrix}
		M^0 W^0 R^0\\
		M^1 W^1 R^0\\
		M^0 W^1 R^1\\
		M^1 W^0 R^1
	\end{pmatrix} \xrightarrow{\begin{pmatrix}
			a & 0 & P_1 & d^{M}_1\\ 
			0 & a & d^{M}_0 & -P_0\\
			P_0 & d^{M}_1 & -b& 0\\
			d^{M}_0 & -P_1 & 0 & -b
	\end{pmatrix}} \begin{pmatrix}
		M^0 W^0 R^1\\
		M^1 W^1 R^1\\
		M^0 W^1 R^0\\
		M^1 W^0 R^0
	\end{pmatrix} \xrightarrow{\begin{pmatrix}
			b & 0 & P_1 & d^{M}_1\\ 
			0 & b & d^{M}_0 & -P_0\\
			P_0 & d^{M}_1 & -a& 0\\
			d^{M}_0 & -P_1 & 0 & -a
	\end{pmatrix}} \begin{pmatrix}
		M^0 W^0 R^0\\
		M^1 W^1 R^0\\
		M^0 W^1 R^1\\
		M^1 W^0 R^1
	\end{pmatrix}.
\end{equation}
Describe how the local map $\chi$ acts on this tensor product:

\begin{equation}
	\begin{array}{c}
		\begin{tikzpicture}
			\node(A) at (0,0) {$\left(\begin{array}{c}
					M^0 W^0 R^0\\
					M^1 W^1 R^0\\
					M^0 W^1 R^1\\
					M^1 W^0 R^1
				\end{array}\right)$};
			\node(B) at (7,0) {$\left(\begin{array}{c}
					M^0 W^0 R^1\\
					M^1 W^1 R^1\\
					M^0 W^1 R^0\\
					M^1 W^0 R^0
				\end{array}\right)$};
			\node(C) at (0,-4) {$\left(\begin{array}{c}
					N^0 W^0 R^0\\
					N^1 W^1 R^0\\
					N^0 W^1 R^1\\
					N^1 W^0 R^1
				\end{array}\right)$};
			\node(D) at (7,-4) {$\left(\begin{array}{c}
					N^0 W^0 R^1\\
					N^1 W^1 R^1\\
					N^0 W^1 R^0\\
					N^1 W^0 R^0
				\end{array}\right)$};
			\draw[-stealth] ([shift={(0,0.1)}]A.east) -- ([shift={(0,0.1)}]B.west) node[pos=0.5,above] {$\scriptstyle D_0^{\Sigma_0}$};
			\draw[stealth-] ([shift={(0,-0.1)}]A.east) -- ([shift={(0,-0.1)}]B.west) node[pos=0.5,below] {$\scriptstyle D_1^{\Sigma_0}$};
			\draw[-stealth] ([shift={(0,0.1)}]C.east) -- ([shift={(0,0.1)}]D.west) node[pos=0.5,above] {$\scriptstyle D_0^{\Sigma_1}$};
			\draw[stealth-] ([shift={(0,-0.1)}]C.east) -- ([shift={(0,-0.1)}]D.west) node[pos=0.5,below] {$\scriptstyle D_1^{\Sigma_1}$};
			\draw[-stealth] ([shift={(-0.1,0)}]A.south) -- ([shift={(-0.1,0)}]C.north) node[pos=0.5,right] {$\begin{pmatrix}
					\chi_0 & & & \\
					& \chi_1 & & \\
					& & \chi_0 & \\
					& & & \chi_1
				\end{pmatrix}$};
			\draw[-stealth] ([shift={(-0.1,0)}]B.south) -- ([shift={(-0.1,0)}]D.north) node[pos=0.5,right] {$\begin{pmatrix}
					\chi_0 & & & \\
					& \chi_1 & & \\
					& & \chi_0 & \\
					& & & \chi_1
				\end{pmatrix}$};
			\node[left] at (A.west) {$\Sigma_0=$};
			\node[left] at (C.west) {$\Sigma_1=$};
		\end{tikzpicture}
	\end{array}
\end{equation}
As is known from Theorem from Section~\ref{sec:mat-fact}, there is an isomorphism $\Gamma_0 \otimes W \otimes \{a,b\} \cong \Pi(\Gamma_0 \otimes W)_{R'}$.  Let us write the last factorization explicitly:
\begin{equation}
	\Pi(\Gamma_0 \otimes W)_{R'} =  \begin{pmatrix}
		M^0 W^0\\
		M^1 W^1
	\end{pmatrix}_{R'} \xrightarrow{\begin{pmatrix}
			\Pi(P_0) & \Pi(d^{M}_1)\\
			\Pi(d^{M}_0) & - \Pi(P_1)
	\end{pmatrix}} \begin{pmatrix}
		M^0 W^1\\
		M^1 W^0
	\end{pmatrix}_{R'} \xrightarrow{\begin{pmatrix}
			\Pi(P_1) & \Pi(d^{M}_1)\\
			\Pi(d^{M}_0) & -\Pi(P_0)
	\end{pmatrix}} \begin{pmatrix}
		M^0 W^0\\
		M^1 W^1
	\end{pmatrix}_{R'}.
\end{equation}
An analogous isomorphism $\Gamma_1 \otimes W \otimes \{a,b\} \cong \Pi(\Gamma_1 \otimes W)_{R'}$.

We now consider the following sequence of maps:
\begin{equation}
	 \Pi(\Gamma_0 \otimes W)_{R'} \xrightarrow{g} \Gamma_0 \otimes W \otimes \{a,b\} \xrightarrow{\chi} \Gamma_1 \otimes W \otimes \{a,b\} \xrightarrow{f} \Pi(\Gamma_1 \otimes W)_{R'}\,.
	 \label{seq}
\end{equation}
Let us now write out (\ref{seq}) in more detail. For definiteness, consider a vector $\boldsymbol{\upsilon}\in\mathcal{H}^0(\Pi(\Gamma_0 \otimes W)_{R'})$:
\begin{equation}
	\boldsymbol{\upsilon} = \begin{pmatrix}
		\boldsymbol{\upsilon}_1\\
		\boldsymbol{\upsilon}_2
	\end{pmatrix} \xrightarrow{g} \begin{pmatrix}
		\boldsymbol{\upsilon}_1\\
		\boldsymbol{\upsilon}_2\\
		-\frac{P_0 - \Pi(P_0)}{b}\boldsymbol{\upsilon}_1-\frac{d_1^M - \Pi(d_1^M)}{b}\boldsymbol{\upsilon}_2\\
		-\frac{d_0^M - \Pi(d_0^M)}{b}\boldsymbol{\upsilon}_1+\frac{P_1 - \Pi(P_1)}{b}\boldsymbol{\upsilon}_2
	\end{pmatrix} \xrightarrow{\chi} \begin{pmatrix}
	\chi^0\cdot \boldsymbol{\upsilon}_1\\
	\chi^1\cdot\boldsymbol{\upsilon}_2\\
	\chi^0\cdot-\frac{P_0 - \Pi(P_0)}{b}\boldsymbol{\upsilon}_1-\frac{d_1^M - \Pi(d_1^M)}{b}\boldsymbol{\upsilon}_2\\
	\chi^1\cdot-\frac{d_0^M - \Pi(d_0^M)}{b}\boldsymbol{\upsilon}_1+\frac{P_1 - \Pi(P_1)}{b}\boldsymbol{\upsilon}_2
	\end{pmatrix} \xrightarrow{f} \begin{pmatrix}
		\Pi(\chi^0\cdot \boldsymbol{\upsilon}_1)\\
		\Pi(\chi^1\cdot\boldsymbol{\upsilon}_2)
	\end{pmatrix} = \Pi(\chi \boldsymbol{\upsilon})\,.
\end{equation}
The last equality implies that the map $\chi:\Gamma_0\otimes W \rightarrow \Gamma_1\otimes W$ acts on the tensor product via the block matrix $\chi = \begin{pmatrix}
	\chi^0 & 0\\
	0 & \chi^1
\end{pmatrix}$. Similarly, we can write for the vector $$\boldsymbol{\upsilon}\in\mathcal{H}^1(\Pi(\Gamma_0 \otimes W)_{R'})$$.\\
Thus, we have shown that under this quasi-isomorphism, the map $\chi' = \Pi (\chi)$ .

This procedure remains unchanged regardless of the number of excluded variables.

\subsection{Examples of calculation}\label{sec:examples}

In this subsection, we consider examples of calculation of Khovanov--Rozansky cohomologies and polynomial in order to clarify the general algorithm from Section~\ref{sec:KhRcomplex}. In Section~\ref{sec:unknot}, we consider the simplest example -- the 1-unknot. In this case, the Khovanov--Rozansky cohomologies are easily computed without any categorified MOY relations or exclusion of variables while the case of the Hopf link in Section~\ref{sec:Hopf} demonstates the excluding variables technique from Section~\ref{sec:excl-var-strand} and then raising to the full cohomologies via an inverse map~\eqref{inv-map}.

\subsubsection{Unknot}\label{sec:unknot}

We demonstrate how to compute the cohomologies of the Khovanov-Rozansky bicomplex using the following simple example.  Consider the complex for the unknot:
\begin{equation}
    \begin{array}{c}
    \begin{tikzpicture}[scale=0.7]
			\draw[thick,-stealth] (0.5,-0.5) to[out=90,in=270] (-0.5,0.5);
			\draw[thick,white,line width = 1.5mm] (-0.5,-0.5) to[out=90,in=270] (0.5,0.5);
			\draw[thick,-stealth] (-0.5,-0.5) to[out=90,in=270] (0.5,0.5);
			\draw[thick] (-0.5,0.5) to[out=90,in=90] (-0.8,0.5) -- (-0.8,-0.5) to[out=270,in=270] (-0.5,-0.5);
			\begin{scope}[xscale=-1]
				\draw[thick] (-0.5,0.5) to[out=90,in=90] (-0.8,0.5) -- (-0.8,-0.5) to[out=270,in=270] (-0.5,-0.5);
			\end{scope}
		\end{tikzpicture}
        \end{array} = 0 \xrightarrow{}\Gamma_0\{1-N\} \xrightarrow{\chi} \Gamma_1\{-N\}\xrightarrow{}0
\end{equation}
where
\begin{equation}\Gamma_0=
\begin{array}{c}
     \begin{tikzpicture}[scale=0.75]
				\draw[thick,-stealth] (0.5,-0.5) to[out=90,in=270] (0.3,0) to[out=90,in=270] (0.5,0.5);
				\draw[thick,-stealth] (-0.5,-0.5) to[out=90,in=270] (-0.3,0) to[out=90,in=270] (-0.5,0.5);
				\draw[thick] (-0.5,0.5) to[out=90,in=90] (-0.8,0.5) -- (-0.8,-0.5) to[out=270,in=270] (-0.5,-0.5);
				\begin{scope}[xscale=-1]
					\draw[thick] (-0.5,0.5) to[out=90,in=90] (-0.8,0.5) -- (-0.8,-0.5) to[out=270,in=270] (-0.5,-0.5);
				\end{scope}
				\node[right] at (-0.63,-0.65) {$\scriptstyle x_1$};
				\node[right] at (-0.63,0.65) {$\scriptstyle x_1$};
				\node[left] at (0.65,-0.65) {$\scriptstyle x_2$};
				\node[left] at (0.65,0.65) {$\scriptstyle x_2$};
		\end{tikzpicture} 
\end{array},\qquad \Gamma_1 = \begin{array}{c}
			\begin{tikzpicture}[scale=0.75]
				\draw[thick,-stealth] (0.5,-0.5) to[out=90,in=270] (-0.5,0.5);
				\draw[thick,-stealth] (-0.5,-0.5) to[out=90,in=270] (0.5,0.5);
				\draw[thick] (-0.5,0.5) to[out=90,in=90] (-0.8,0.5) -- (-0.8,-0.5) to[out=270,in=270] (-0.5,-0.5);
				\begin{scope}[xscale=-1]
					\draw[thick] (-0.5,0.5) to[out=90,in=90] (-0.8,0.5) -- (-0.8,-0.5) to[out=270,in=270] (-0.5,-0.5);
				\end{scope}
				\node[right] at (-0.63,-0.65) {$\scriptstyle x_1$};
				\node[right] at (-0.63,0.65) {$\scriptstyle x_1$};
				\node[left] at (0.65,-0.65) {$\scriptstyle x_2$};
				\node[left] at (0.65,0.65) {$\scriptstyle x_2$};
				\draw[fill=\myred] (0,0) circle (0.1);
			\end{tikzpicture}
		\end{array}.
\end{equation}
We have the following matrix factorizations:
\begin{equation}\label{G0-unknot}
    \Gamma_0=\bigg[\begin{pmatrix}
        R\\
        R\{2-2N\}
    \end{pmatrix} \xrightarrow[]{\begin{pmatrix}
        x_1^N &0\\
        x_2^N &0
    \end{pmatrix}} \begin{pmatrix}
        R\{1-N\} \\
        R\{1-N\}
    \end{pmatrix}\xrightarrow[]{\begin{pmatrix}
        0 & 0\\
        x_2^N & -x_1^N
    \end{pmatrix}}\begin{pmatrix}
        R\\
        R\{2-2N\}
    \end{pmatrix}\bigg],
\end{equation}
\begin{equation}\label{G1-unknot}
    \Gamma_1=\bigg[\begin{pmatrix}
        R\{-1\} \\
        R\{3-2N\}
    \end{pmatrix}\xrightarrow[]{\begin{pmatrix}
        u_1 & 0\\
        u_2 & 0
    \end{pmatrix}}\begin{pmatrix}
        R\{-N\} \\
        R\{2-N\}
    \end{pmatrix}\xrightarrow[]{\begin{pmatrix}
        0 & 0\\
        u_2 & -u_1
    \end{pmatrix}}\begin{pmatrix}
        R\{-1\} \\
        R\{3-2N\}
    \end{pmatrix}\bigg].
\end{equation}
In this subsection, we omit the explicit dependence on $(x_1,x_2,x_2,x_1)$ of $u_{1,2}(x_1,x_2,x_2,x_1)$. The corresponding homologies of the matrix factorizations are 
\begin{equation}
    \mathcal{H}_{\Gamma_0}=\mathcal{H}_{\Gamma_0}^0=\begin{pmatrix}
            0\\
            (\mathbb{Q}[x_1,x_2]/\langle x_1^N,x_2^N\rangle)\{2-2N\}
        \end{pmatrix},
\end{equation}
\begin{equation}
    \mathcal{H}_{\Gamma_1}=\mathcal{H}_{\Gamma_1}^0=\begin{pmatrix}
            0\\
            \mathbb{Q}[x_1,x_2]/\langle u_1,u_2\rangle\{3-2N\}
        \end{pmatrix}.
\end{equation}
Now we need to apply the map $\chi_0:\Gamma_0\xrightarrow[]{}\Gamma_1\,$:
\begin{equation}
    % https://tikzcd.yichuanshen.de/#N4Igdg9gJgpgziAXAbVABwnAlgFyxMJZABgBpiBdUkANwEMAbAVxiRDpAF9T1Nd9CKMgEYqtRizYAjLjxAZseAkQBM5MfWatEIAMazeigatKjqmyTqhcxMKAHN4RUADMAThAC2SMiBwQkFW5XD29EX38kYWCQdy9A6kjEAGYYuLDhRICUzgpOIA
\begin{tikzcd}
 \begin{pmatrix}
            0 \\
            (\mathbb{Q}[x_1,x_2]/\langle x_1^N,x_2^N\rangle)\{2-2N\}
        \end{pmatrix} \arrow[rr] \arrow{d}{U_0} &  & \begin{pmatrix}
            0\\
            0
        \end{pmatrix} \arrow{d}{U_1} \\
\begin{pmatrix}
            0\\
            (\mathbb{Q}[x_1,x_2]/\langle u_1,u_2\rangle)\{3-2N\}
        \end{pmatrix} \arrow[rr]           &  & \begin{pmatrix}
            0\\
            0
        \end{pmatrix}           
\end{tikzcd}
\end{equation}
Inducing the map $\chi_0$ on the cohomologies component, we obtain the final complex:
\begin{equation}
    0\xrightarrow{} \mathbb{Q}[x_1,x_2]/\langle x_1^N,x_2^N\rangle\{3-3N\} \xrightarrow[]{\mathds{1}} \mathbb{Q}[x_1,x_2]/\langle u_1,u_2\rangle\{3-3N\} \xrightarrow[]{}0
\end{equation}
where $\mathds{1}$ is just the multiplication by the unity, and not the identity map.

The corresponding Poincare polynomial is
\begin{equation}
    {\rm KhR}(t,q,A=q^N)= q^{3(1-N)}\cdot q^{2(N-1)} \cdot q^{N-1} \cdot [N] = [N]\,.
\end{equation}
what is the right answer.

\subsubsection{Hopf link}\label{sec:Hopf}
We demonstrate how the excluding variables theorem can simplify the computation of matrix factorizations for difficult MOY diagrams. Consider the complex for the Hopf link:
\begin{equation}\label{Hopf-compl}
	\begin{array}{c}
		\begin{tikzpicture}[scale=0.75]
			\begin{scope}[shift={(0,1)}]
				\draw[thick] (0.5,-0.5) to[out=90,in=270] (-0.5,0.5);
				\draw[white, line width = 1.5mm] (-0.5,-0.5) to[out=90,in=270] (0.5,0.5);
				\draw[thick] (-0.5,-0.5) to[out=90,in=270] (0.5,0.5);
			\end{scope}
			\draw[thick,-stealth] (0.5,-0.5) to[out=90,in=270] (-0.5,0.5) -- (-0.5,0.6);
			\draw[white, line width = 1.5mm] (-0.5,-0.5) to[out=90,in=270] (0.5,0.5);
			\draw[thick,-stealth] (-0.5,-0.5) to[out=90,in=270] (0.5,0.5) -- (0.5,0.6);
			\draw[thick] (-0.5,1.5) to[out=90,in=90] (-1.2,1.5) -- (-1.2,-0.5) to[out=270,in=270] (-0.5,-0.5);
			\begin{scope}[xscale=-1]
				\draw[thick] (-0.5,1.5) to[out=90,in=90] (-1.2,1.5) -- (-1.2,-0.5) to[out=270,in=270] (-0.5,-0.5);
			\end{scope}
			\node[left] at (-0.5,-0.5) {$\scriptstyle x_1$};
			\node[right] at (0.5,-0.5) {$\scriptstyle x_2$};
			\node[left] at (-0.5,0.5) {$\scriptstyle x_3$};
			\node[right] at (0.5,0.5) {$\scriptstyle x_4$};
			\node[left] at (-0.5,1.5) {$\scriptstyle x_1$};
			\node[right] at (0.5,1.5) {$\scriptstyle x_2$};
			\node[above] at (0,0) {$\scriptstyle 2$};
			\node[above] at (0,1) {$\scriptstyle 1$};
		\end{tikzpicture}
	\end{array}=\left[\begin{array}{c}
		\begin{tikzpicture}
			\node(A) at (0,0) {$\MOY_{00}\{2-2N\}$};
			\node(B) at (3,0.7) {$\MOY_{01}\{1-2N\}$};
			\node(C) at (3,-0.7) {$\MOY_{10}\{1-2N\}$};
			\node(D) at (6,0) {$\MOY_{11}\{-2N\}$};
			\path (A) edge[-stealth] node[above left] {$\scriptstyle\chi_0 \otimes {\rm id}$} (B) (A) edge[-stealth] node[below left] {$\scriptstyle {\rm id} \otimes \chi_0$} (C) (B) edge[-stealth] node[above right] {$\scriptstyle {\rm id} \otimes \chi_0$} (D) (C) edge[-stealth] node[below right] {$\scriptstyle -\chi_0\otimes {\rm id}$} (D);
		\end{tikzpicture}
	\end{array}\right]
\end{equation}
where the following MOY diagrams are used:
\begin{equation}
	\begin{aligned}
		& \MOY_{00}=\begin{array}{c}
			\begin{tikzpicture}[scale=0.5]
				\draw[thick] (0.5,-0.5) to[out=90,in=270] (0.3,0) to[out=90,in=270] (0.5,0.5) -- (0.5,0.6);
				\draw[thick] (-0.5,-0.5) to[out=90,in=270] (-0.3,0) to[out=90,in=270] (-0.5,0.5) -- (-0.5,0.6);
				\begin{scope}[shift={(0,1)}]
					\draw[thick] (0.5,-0.5) to[out=90,in=270] (0.3,0) to[out=90,in=270] (0.5,0.5);
					\draw[thick] (-0.5,-0.5) to[out=90,in=270] (-0.3,0) to[out=90,in=270] (-0.5,0.5);
				\end{scope}
				\draw[thick, postaction={decorate},decoration={markings, mark= at position 0.6 with {\arrow{stealth}}}] (-0.5,1.5) to[out=90,in=90] (-1.2,1.5) -- (-1.2,-0.5) to[out=270,in=270] (-0.5,-0.5);
				\begin{scope}[xscale=-1]
					\draw[thick, postaction={decorate},decoration={markings, mark= at position 0.6 with {\arrow{stealth}}}] (-0.5,1.5) to[out=90,in=90] (-1.2,1.5) -- (-1.2,-0.5) to[out=270,in=270] (-0.5,-0.5);
				\end{scope}
			\end{tikzpicture}
		\end{array},\quad \MOY_{10}=\begin{array}{c}
			\begin{tikzpicture}[scale=0.5]
				\draw[thick] (0.5,-0.5) to[out=90,in=270] (-0.5,0.5);
				\draw[thick] (-0.5,-0.5) to[out=90,in=270] (0.5,0.5);
				\begin{scope}[shift={(0,1)}]
					\draw[thick] (0.5,-0.5) to[out=90,in=270] (0.3,0) to[out=90,in=270] (0.5,0.5);
					\draw[thick] (-0.5,-0.5) to[out=90,in=270] (-0.3,0) to[out=90,in=270] (-0.5,0.5);
				\end{scope}
				\draw[thick, postaction={decorate},decoration={markings, mark= at position 0.6 with {\arrow{stealth}}}] (-0.5,1.5) to[out=90,in=90] (-1.2,1.5) -- (-1.2,-0.5) to[out=270,in=270] (-0.5,-0.5);
				\begin{scope}[xscale=-1]
					\draw[thick, postaction={decorate},decoration={markings, mark= at position 0.6 with {\arrow{stealth}}}] (-0.5,1.5) to[out=90,in=90] (-1.2,1.5) -- (-1.2,-0.5) to[out=270,in=270] (-0.5,-0.5);
				\end{scope}
				\draw[fill=\myred] (0,0) circle (0.13);
			\end{tikzpicture}
		\end{array},\quad\MOY_{01}=\begin{array}{c}
			\begin{tikzpicture}[scale=0.5]
				\draw[thick] (0.5,-0.5) to[out=90,in=270] (0.3,0) to[out=90,in=270] (0.5,0.5);
				\draw[thick] (-0.5,-0.5) to[out=90,in=270] (-0.3,0) to[out=90,in=270] (-0.5,0.5);
				\begin{scope}[shift={(0,1)}]
					\draw[thick] (0.5,-0.5) to[out=90,in=270] (-0.5,0.5);
					\draw[thick] (-0.5,-0.5) to[out=90,in=270] (0.5,0.5);
				\end{scope}
				\draw[thick, postaction={decorate},decoration={markings, mark= at position 0.6 with {\arrow{stealth}}}] (-0.5,1.5) to[out=90,in=90] (-1.2,1.5) -- (-1.2,-0.5) to[out=270,in=270] (-0.5,-0.5);
				\begin{scope}[xscale=-1]
					\draw[thick, postaction={decorate},decoration={markings, mark= at position 0.6 with {\arrow{stealth}}}] (-0.5,1.5) to[out=90,in=90] (-1.2,1.5) -- (-1.2,-0.5) to[out=270,in=270] (-0.5,-0.5);
				\end{scope}
				\draw[fill=\myred] (0,1) circle (0.13);
			\end{tikzpicture}
		\end{array},\quad\MOY_{11}=\begin{array}{c}
			\begin{tikzpicture}[scale=0.5]
				\draw[thick,-stealth] (0.5,-0.5) to[out=90,in=270] (-0.5,0.5) -- (-0.5,0.6);
				\draw[thick,-stealth] (-0.5,-0.5) to[out=90,in=270] (0.5,0.5) -- (0.5,0.6);
				\begin{scope}[shift={(0,1)}]
					\draw[thick] (0.5,-0.5) to[out=90,in=270] (-0.5,0.5);
					\draw[thick] (-0.5,-0.5) to[out=90,in=270] (0.5,0.5);
				\end{scope}
				\draw[thick, postaction={decorate},decoration={markings, mark= at position 0.6 with {\arrow{stealth}}}] (-0.5,1.5) to[out=90,in=90] (-1.2,1.5) -- (-1.2,-0.5) to[out=270,in=270] (-0.5,-0.5);
				\begin{scope}[xscale=-1]
					\draw[thick, postaction={decorate},decoration={markings, mark= at position 0.6 with {\arrow{stealth}}}] (-0.5,1.5) to[out=90,in=90] (-1.2,1.5) -- (-1.2,-0.5) to[out=270,in=270] (-0.5,-0.5);
				\end{scope}
				\draw[fill=\myred] (0,0) circle (0.13) (0,1) circle (0.13);
			\end{tikzpicture}
		\end{array}\,.
	\end{aligned}
\end{equation}
Introduce the ring $R=\mathbb{Q}[x_1,x_2,x_3,x_4]$. Let us compute the cohomologies of the matrix factorizations corresponding to these MOY graphs.

Here, we again identify graphs and their matrix factorizations denoting them by the same letter. The factorization $\Gamma_{00}$ is given by the following tensor product:
\begin{equation}\label{gamma00}
	\begin{split}
		\Gamma_{00} = [R \xrightarrow{\pi_{13}} R(a) \xrightarrow{x_1 - x_3} R\,]\otimes [R \xrightarrow{\pi_{31}} R(b) \xrightarrow{x_3 - x_1} R\,]\otimes [R \xrightarrow{\pi_{24}} R(c) \xrightarrow{x_2 - x_4} R\,]\otimes[R \xrightarrow{\pi_{42}} R(d) \xrightarrow{x_4 - x_2} R\,]\,.
	\end{split}
\end{equation}
Here and hereafter, we write ring letters in round brackets to distinguish bases in their tensor products and imply that $R(a) \otimes R(b) := R(ab)$. We also omit the ring gradings, which we restore at the end using \eqref{mfres0} and \eqref{mfres1}.

Let us expand the tensor product of factorizations:
\begin{eqnarray}
	\Gamma_{00}=\begin{pmatrix}
		R\\
		R(ac)\\
		R(ab)\\
		R(bc)\\
		R(ad)\\
		R(cd)\\
		R(bd)\\
		R(abcd)
	\end{pmatrix} \xlongrightarrow{D_{\Gamma_{00}}^0} \begin{pmatrix}
	R(a)\\
	R(c)\\
	R(b)\\
	R(abc)\\
	R(d)\\
	R(acd)\\
	R(abd)\\
	R(bcd)
	\end{pmatrix} \xlongrightarrow{D_{\Gamma_{00}}^1} \begin{pmatrix}
	R\\
	R(ac)\\
	R(ab)\\
	R(bc)\\
	R(ad)\\
	R(cd)\\
	R(bd)\\
	R(abcd)
	\end{pmatrix}
\end{eqnarray}
where the operators are
\begin{equation}
	D_{\Gamma_{00}}^0 = \begin{pmatrix}
		\pi_{13} & x_2-x4 & x_3-x_1 & 0 & x_4 - x_2 & 0 & 0 & 0\\
		\pi_{24} & -(x_1 - x_3) & 0 & x_3-x_1 & 0 & x_4-x_2 & 0& 0\\
		\pi_{31} & 0 & -(x_1 - x_3) & -(x_2-x_4) & 0 & 0 & x_4-x_2 & 0\\
		0 & \pi_{31} & -\pi_{24} & \pi_{13} &0 & 0&0 & x_4-x_2 \\
		\pi_{42} &0& 0& 0&-(x_1 - x_3) & -(x_2-x_4)&-(x_3-x_1)&0\\
		0&\pi_{42} & 0 & 0 & -\pi_{24} & \pi_{13}& 0 & -(x_3-x_1) \\
		0&0&\pi_{42}&0&-\pi_{31}&0&\pi_{13}&x_2-x_4\\
		0&0&0&\pi_{42}&0&-\pi_{31}&\pi_{24}&-(x_1-x_3)
			\end{pmatrix},
\end{equation}
\begin{equation}
	D_{\Gamma_{00}}^1 = \begin{pmatrix}
		x_1 - x_3 & x_2-x_4&x_3-x_1&0 & x_4 - x_2 & 0 & 0 & 0\\
		\pi_{24} & -\pi_{13}& 0 & x_3-x_1& 0 & x_4-x_2 & 0& 0\\
		\pi_{31}&0&-\pi_{13}&-(x_2-x_4)& 0 & 0 & x_4-x_2 & 0\\
		0&\pi_{31}&-\pi_{24}&x_1-x_3 &0 & 0&0 & x_4-x_2\\
		\pi_{42} &0& 0& 0&-\pi_{13} & -(x_2-x_4) & -(x_3-x_1) & 0\\
		0&\pi_{42} & 0 & 0 &-\pi_{24} & x_1 - x_3 & 0 & -(x_3-x_1)\\
		0&0&\pi_{42}&0&-\pi_{31} & 0 & x_1 - x_3 & x_2-x_4\\
		0&0&0&\pi_{42}&0 & -\pi_{31} & \pi_{24} & -\pi_{13}
	\end{pmatrix}.
\end{equation}
After excluding the variables $x_3 = x_1$ and $x_4 = x_2$, we obtain the following simplified factorization:
\begin{equation}
	\Gamma_{00}' = \bigg[\begin{pmatrix}
		R'\\
		R'(ac)
	\end{pmatrix} \xrightarrow{\begin{pmatrix}
		x_1^N &0\\
		x_2^N &0\\
		\end{pmatrix}} \begin{pmatrix}
		R'(a)\\
		R'(c)
	\end{pmatrix}\xrightarrow{\begin{pmatrix}
		0&0\\
		x_2^N &-x_1^N
		\end{pmatrix}} \begin{pmatrix}
		R'\\
		R'(ac)
	\end{pmatrix}\bigg]
\end{equation}
where $R'=\mathbb{Q}[x_1,x_2,x_3,x_4]/\langle x_3-x_1, x_4-x_2\rangle=\mathbb{Q}[x_1,x_2]$. Its cohomology is given by:
\begin{equation}
	\mathcal{H}(\Gamma_{00}')=\mathcal{H}^0(\Gamma_{00}') = \begin{pmatrix}
		0\\
		\mathbb{Q}[x_1,x_2]/\langle x_1^N, x_2^N \rangle(ac)
	\end{pmatrix}.
\end{equation}
Here, for brevity, we write: $\mathbb{Q}[x_1,x_2]/\langle x_1^N, x_2^N \rangle = \mathbb{Q}[x_1,x_2,x_3,x_4]/\langle x_3-x_1, x_4-x_2, x_1^N, x_2^N\rangle$. After the corresponding quasi-isomorphism~\eqref{inv-map}, we obtain the cohomology of the factorization $\Gamma_{00}$:
\begin{equation}\label{homgamma00}
	\mathcal{H}(\Gamma_{00})=\mathcal{H}^0(\Gamma_{00}) = \begin{pmatrix}
		0\\
		\mathbb{Q}[x_1,x_2]/\langle x_1^N, x_2^N \rangle(ac)\\
		0(ab)\\
		-\mathbb{Q}[x_1,x_2]/\langle x_1^N, x_2^N \rangle(bc)\\
		\mathbb{Q}[x_1,x_2]/\langle x_1^N, x_2^N \rangle(ad)\\
		0(cd)\\
		\mathbb{Q}[x_1,x_2]/\langle x_1^N, x_2^N \rangle(bd)\\
		0(abcd)
	\end{pmatrix} = \begin{pmatrix}
	0\\
	\mathbb{Q}[x_1,x_2]/\langle x_1^N, x_2^N \rangle\{2-2N\}\\
	0\\
	-\mathbb{Q}[x_1,x_2]/\langle x_1^N, x_2^N \rangle\{2-2N\}\\
	\mathbb{Q}[x_1,x_2]/\langle x_1^N, x_2^N \rangle\{2-2N\}\\
	0\\
	\mathbb{Q}[x_1,x_2]/\langle x_1^N, x_2^N \rangle\{2-2N\}\\
	0
	\end{pmatrix}
\end{equation}
where in the last equality, we have restored gradings. The factorization $\Gamma_{10}$ is given by the following tensor product:
\begin{equation}\label{gamma10}
	\Gamma_{10} = [R\xrightarrow{u_1}R(a)\xrightarrow{x_1+x_2-x_3-x_4}R\,]\otimes[R\xrightarrow{u_2}R(c)\xrightarrow{x_1x_2-x_3x_4}R\,]\otimes[R\xrightarrow{\pi_{31}} R(b)\xrightarrow{x_3-x_1} R\,]\otimes[R\xrightarrow{\pi_{42}} R(d)\xrightarrow{x_4-x_2} R\,]
\end{equation}
where $u_1 = u_1(x_1,x_2,x_3,x_4)$ and $u_2 = u_2(x_1,x_2,x_3,x_4)$ are given by \eqref{u}. Introduce the notation $s_1 = x_1+x_2-x_3-x_4$  and $s_2 = x_1x_2-x_3x_4$. Let us expand the tensor product of factorizations:
\begin{eqnarray}
	\Gamma_{10}=\begin{pmatrix}
		R\\
		R(ac)\\
		R(ab)\\
		R(bc)\\
		R(ad)\\
		R(cd)\\
		R(bd)\\
		R(abcd)
	\end{pmatrix} \xlongrightarrow{D_{\Gamma_{10}}^0} \begin{pmatrix}
		R(a)\\
		R(c)\\
		R(b)\\
		R(abc)\\
		R(d)\\
		R(acd)\\
		R(abd)\\
		R(bcd)
	\end{pmatrix} \xlongrightarrow{D_{\Gamma_{10}}^1} \begin{pmatrix}
		R\\
		R(ac)\\
		R(ab)\\
		R(bc)\\
		R(ad)\\
		R(cd)\\
		R(bd)\\
		R(abcd)
	\end{pmatrix}
\end{eqnarray}
where the differentials are
\begin{equation}
	D_{\Gamma_{10}}^0 = \begin{pmatrix}
		u_1 & s_2 & x_3-x_1 & 0 & x_4 - x_2 & 0 & 0 & 0\\
		u_2 & -s_1 & 0 & x_3-x_1 & 0 & x_4-x_2 & 0 & 0\\
		\pi_{31} & 0 & -s_1 & -s_2 & 0 & 0 & x_4-x_2 & 0 \\
		0 & \pi_{31} & -u_2 & u_1 & 0 & 0 & 0 & x_4-x_2 \\
		\pi_{42}& 0 & 0 & 0 & -s_1 & -s_2 &-(x_3-x_1) & 0 \\
		0 & \pi_{42} & 0 & 0 & -u_2 & -u_1 & 0& -(x_3-x_1) \\
		0 & 0 & \pi_{42} & 0 & -\pi_{31} & 0 & u_1 & s_2 \\
		0 & 0 & 0 & \pi_{42} & 0 & -\pi_{31} & u_2 & -s_1
	\end{pmatrix},
\end{equation}
\begin{equation}
	D_{\Gamma_{10}}^1 = \begin{pmatrix}
		s_1 & s_2 &x_3-x_1 & 0 & x_4 - x_2 & 0 & 0 & 0\\
		u_2 & u_1 & 0& x_3-x_1 & 0 & x_4-x_2 & 0 & 0\\
		\pi_{31} & 0 & -u_1 & -s_2 & 0 & 0 & x_4-x_2 & 0 \\
		 0 & \pi_{31} & -u_2 & s_1  & 0 & 0 & 0 & x_4-x_2\\
		 \pi_{42}& 0 & 0 & 0&-u_1 & -s_2 & -(x_3-x_1) & 0\\
		 0 & \pi_{42} & 0 & 0&-u_2 & s_1 & 0 & -(x_3-x_1) \\
		  0 & 0 & \pi_{42} & 0&-\pi_{31} & 0 & s_1 & s_2  \\
		 	0 & 0 & 0 & \pi_{42}&0 & -\pi_{31} & u_2 & -u_1 \\
	\end{pmatrix}.
\end{equation}
After excluding the variables $x_3 = x_1$ and $x_4 = x_2$, we obtain the following simplified factorization:
\begin{equation}
	\Gamma_{10}'= \bigg[\begin{pmatrix}
		R'\\
		R'(ac)
	\end{pmatrix}\xrightarrow{\begin{pmatrix}
		u_1(x_1,x_2,x_1,x_2)&0\\
		u_2(x_1,x_2,x_1,x_2) & 0
		\end{pmatrix}} \begin{pmatrix}
		R'(a)\\
		R'(c)
	\end{pmatrix}\xrightarrow{\begin{pmatrix}
		0 & 0\\
		u_2(x_1,x_2,x_1,x_2) & 	u_1(x_1,x_2,x_1,x_2)
		\end{pmatrix}} \begin{pmatrix}
		R'\\
		R'(ac)
	\end{pmatrix}\bigg].
\end{equation}
Its cohomology is given by:
\begin{equation}
	\mathcal{H}(\Gamma_{10}') = \mathcal{H}^0(\Gamma_{10}') = \begin{pmatrix}
		0\\
		\mathbb{Q}[x_1,x_2]/\langle u_1, u_2 \rangle (ac)
	\end{pmatrix}
\end{equation}
where $\mathbb{Q}[x_1,x_2]/\langle u_1, u_2 \rangle = \mathbb{Q}[x_1,x_2,x_3,x_4]/\langle x_3-x_1, x_4-x_2, u_1, u_2 \rangle$ and the polynomials $u_1(x_1,x_2,x_1,x_2)$, $u_2(x_1,x_2,x_1,x_2)$ are homogeneous in $x_1$ and $x_2$ of degrees $N$ and $N-1$ respectively.

After the corresponding quasi-isomorphism~\eqref{inv-map}, we obtain the cohomology of the factorization $\Gamma_{10}$:
\begin{equation}\label{homgamma10}
	\mathcal{H}(\Gamma_{10})=\mathcal{H}^0(\Gamma_{10}) = \begin{pmatrix}
		0\\
		\mathbb{Q}[x_1,x_2]/\langle u_1, u_2 \rangle(ac)\\
		x_2 \mathbb{Q}[x_1,x_2]/\langle u_1, u_2 \rangle (ab)\\
		-\mathbb{Q}[x_1,x_2]/\langle u_1, u_2 \rangle (bc)\\
		x_3 \mathbb{Q}[x_1,x_2]/\langle u_1, u_2 \rangle (ad)\\
		-\mathbb{Q}[x_1,x_2]/\langle u_1, u_2 \rangle (cd)\\
		(x_3 - x_2) \mathbb{Q}[x_1,x_2]/\langle u_1, u_2 \rangle (bd)\\
		\frac{(x_2 (u_2(x_1,x_2,x_3,x_4)-u_2(x_1,x_2,x_3,x_2)) + (u_1(x_1,x_2,x_3,x_4)-u_2(x_1,x_2,x_3,x_2)))}{x_4-x_2}\mathbb{Q}[x_1,x_2]/\langle u_1, u_2 \rangle (abcd)
	\end{pmatrix}.
\end{equation}
Now, let us restore grading shifts:
\begin{equation}
	\mathcal{H}^0(\Gamma_{10}) = \begin{pmatrix}
		0\\
		\mathbb{Q}[x_1,x_2]/\langle u_1, u_2 \rangle\{3-2N\}\\
		x_2 \mathbb{Q}[x_1,x_2]/\langle u_1, u_2 \rangle \{1-2N\}\\
		-\mathbb{Q}[x_1,x_2]/\langle u_1, u_2 \rangle \{3-2N\}\\
		x_3 \mathbb{Q}[x_1,x_2]/\langle u_1, u_2 \rangle \{1-2N\}\\
		-\mathbb{Q}[x_1,x_2]/\langle u_1, u_2 \rangle \{3-2N\}\\
		(x_3 - x_2) \mathbb{Q}[x_1,x_2]/\langle u_1, u_2 \rangle \{1-2N\}\\
		\frac{(x_2 (u_2(x_1,x_2,x_3,x_4)-u_2(x_1,x_2,x_3,x_2)) + (u_1(x_1,x_2,x_3,x_4)-u_2(x_1,x_2,x_3,x_2)))}{x_4-x_2}\mathbb{Q}[x_1,x_2]/\langle u_1, u_2 \rangle \{5-4N\}
	\end{pmatrix}.
\end{equation}
The matrix factorization $\Gamma_{01}$ is given by the following tensor product:
\begin{equation}
	\Gamma_{01} = [R\xrightarrow{\pi_{13}}R(a)\xrightarrow{x_1-x_3}R\,]\otimes[R\xrightarrow{\pi_{24}}R(c)\xrightarrow{x_2-x_4}R\,]\otimes[R\xrightarrow{u_1} R(b)\xrightarrow{x_3+x_4-x_1-x_2} R\,]\otimes[R\xrightarrow{u_2} R(d)\xrightarrow{x_3 x_4 - x_1 x_2} R\,]\,.
\end{equation}
Its cohomology can be computed similarly:
\begin{equation}
	\mathcal{H}(\Gamma_{01})=\mathcal{H}^0(\Gamma_{01}) =\begin{pmatrix}
		0\\
		(x_1 - x_4) \mathbb{Q}[x_3,x_4]/\langle u_1, u_2 \rangle (ac)\\
		x_4 \mathbb{Q}[x_3,x_4]/\langle u_1, u_2 \rangle (ab)\\
		x_1 \mathbb{Q}[x_3,x_4]/\langle u_1, u_2 \rangle (bc)\\
		-\mathbb{Q}[x_3,x_4]/\langle u_1, u_2 \rangle (ad)\\
		-\mathbb{Q}[x_3,x_4]/\langle u_1, u_2 \rangle (cd)\\
		\mathbb{Q}[x_3,x_4]/\langle u_1, u_2 \rangle (bd)\\
		\frac{(x_4 (u_2(x_3,x_4,x_1,x_2)-u_2(x_3,x_4,x_1,x_4)) + (u_1(x_3,x_4,x_1,x_2)-u_2(x_3,x_4,x_1,x_4)))}{x_2-x_4}\mathbb{Q}[x_3,x_4]/\langle u_1, u_2 \rangle (abcd)
	\end{pmatrix}.
\end{equation}
Then, we restore gradings:
\begin{equation}
		\mathcal{H}^0(\Gamma_{01}) =\begin{pmatrix}
		0\\
		(x_1 - x_4) \mathbb{Q}[x_3,x_4]/\langle u_1, u_2 \rangle \{1-2N\}\\
		x_4 \mathbb{Q}[x_3,x_4]/\langle u_1, u_2 \rangle \{1-2N\}\\
		x_1 \mathbb{Q}[x_3,x_4]/\langle u_1, u_2 \rangle \{1-2N\}\\
		-\mathbb{Q}[x_3,x_4]/\langle u_1, u_2 \rangle \{3-2N\}\\
		-\mathbb{Q}[x_3,x_4]/\langle u_1, u_2 \rangle \{3-2N\}\\
		\mathbb{Q}[x_3,x_4]/\langle u_1, u_2 \rangle \{3-2N\}\\
		\frac{(x_4 (u_2(x_3,x_4,x_1,x_2)-u_2(x_3,x_4,x_1,x_4)) + (u_1(x_3,x_4,x_1,x_2)-u_2(x_3,x_4,x_1,x_4)))}{x_2-x_4}\mathbb{Q}[x_3,x_4]/\langle u_1, u_2 \rangle \{5-4N\}
	\end{pmatrix}
\end{equation}
where $\mathbb{Q}[x_3,x_4]/\langle u_1, u_2 \rangle = \mathbb{Q}[x_1,x_2,x_3,x_4]/\langle x_1-x_3, x_2-x_4,u_1, u_2 \rangle$.

The factorization $\Gamma_{11}$ is given by the following tensor product:
\begin{equation}
	\Gamma_{11}= [R\xrightarrow{u_1}R(a)\xrightarrow{s_1}R\,]\otimes[R\xrightarrow{u_2}R(c)\xrightarrow{s_2}R\,]\otimes[R\xrightarrow{\Bar{u}_1}R(b)\xrightarrow{-s_1}R\,]\otimes[R\xrightarrow{\Bar{u}_2}R(d)\xrightarrow{-s_2}R\,]
\end{equation}
where $u_1=u_1(x_1,x_2,x_3,x_4)$, $u_2=u_2(x_1,x_2,x_3,x_4)$, $\Bar{u}_1=u_1(x_3,x_4,x_1,x_2)$, $\Bar{u}_2=u_2(x_3,x_4,x_1,x_2)$ and $s_1 = x_1+x_2-x_3-x_4$, $s_2 = x_1 x_2 - x_3 x_4$.

 Let us expand the tensor product of factorizations:
\begin{equation}
	\Gamma_{11} = \begin{pmatrix}
		R\\
		R(ac)\\
		R(ad)\\
		R(cd)\\
		R(ab)\\
		R(bc)\\
		R(bd)\\
		R(abcd)
	\end{pmatrix}\xlongrightarrow{D_{\Gamma_{11}}^0} \begin{pmatrix}
	R(a)\\
	R(c)\\
	R(d)\\
	R(acd)\\
	R(b)\\
	R(abc)\\
	R(abd)\\
	R(bcd)
	\end{pmatrix} \xlongrightarrow{D_{\Gamma_{11}}^1} \begin{pmatrix}
	R\\
	R(ac)\\
	R(ad)\\
	R(cd)\\
	R(ab)\\
	R(bc)\\
	R(bd)\\
	R(abcd)
	\end{pmatrix}
\end{equation}
where the maps are
\begin{equation}
	D_{\Gamma_{11}}^0=\begin{pmatrix}
	u_1 & s_2 & -s_2 & 0 & -s_1 & 0 & 0 & 0\\
	u_2 & -s_1 & 0 & -s_2 & 0 & -s_1 & 0 & 0\\
	\Bar{u}_2 & 0 & -s_1 & -s_2 & 0 & 0 & -s_1 & 0\\
	0 & \Bar{u}_2 & -u_2 & u_1 & 0 & 0 & 0 & -s_1\\
	\Bar{u}_1 & 0 & 0 & 0 & -s_1 & -s_2 & s_1 & 0\\
	0 & \Bar{u}_1 & 0 & 0 & -u_2 & u_1 & 0 & s_1\\
	0 & 0 & \Bar{u}_1 & 0 & -\Bar{u}_2 & 0 & u_1 & s_2\\
	0 & 0 & 0 & \Bar{u}_1 & 0 & -\Bar{u}_2 & u_2 & -s_1
	\end{pmatrix},
\end{equation}
\begin{equation}
	D_{\Gamma_{11}}^1=\begin{pmatrix}
		s_1 & s_2 & -s_1 & 0 & -s_1 & 0 & 0 & 0\\
		-u_2 & s_1 & 0 & s_2 & 0 & -s_1 & 0 & 0\\
		\Bar{u}_2 & 0 & -u_1 & -s_2& 0 & 0 & -s_1 & 0\\
		 0 & \Bar{u}_2 & -u_2 & s_1 & 0 & 0 & 0 & -s_1\\
		\Bar{u}_1 & 0 & 0 & 0 & -u_1 & -s_2 & s_2 & 0\\
		0 & \Bar{u}_1 & 0 & 0 &-u_2 & s_1 & 0 & s_2\\
		0 & 0 & \Bar{u}_1 & 0 & -\Bar{u}_2 & 0 & s_1 & s_2\\
		0 & 0 & 0 & \Bar{u}_1 &-\Bar{u}_2 & 0 & s_1 & s_2
	\end{pmatrix}.
\end{equation}
In the factorization $R\xrightarrow{\Bar{u}_1}R\{b\}\xrightarrow{-s_1=x_3+x_4-x_1-x_2}R$, we eliminate the variable $x_3=x_1+x_2-x_4$ and obtain the following factorization:
\begin{equation}
	\Gamma_{11}' = \begin{pmatrix}
		R'\\
		R'(ac)\\
		R'(ad)\\
		R'(cd)
	\end{pmatrix} \xrightarrow{\begin{pmatrix}
			u_1' & s_2' & -s_2' & 0\\
			u_2' & 0 & 0 & -s_2'\\
			\Bar{u}_2' & 0 & 0 & -s_2'\\
			0 & \Bar{u}_2' & -u_2' & u_1'
		\end{pmatrix}} \begin{pmatrix}
		R'(a)\\
		R'(c)\\
		R'(d)\\
		R'(acd)
	\end{pmatrix} \xrightarrow{\begin{pmatrix}
			 0 & s_2' & -s_2' & 0\\
			u_2' & -u_1' & 0 & -s_2'\\
			\Bar{u_2}' & 0 &-u_1' & -s_2'\\
			0 & \Bar{u_2}' &-u_2' & 0
		\end{pmatrix}} \begin{pmatrix}
		R'\\
		R'(ac)\\
		R'(ad)\\
		R'(cd)
	\end{pmatrix}
\end{equation}
where the ring $R' = \mathbb{Q}[x_1,x_2,x_3,x_4]/\langle x_3+x_4-x_1-x_2 \rangle = \mathbb{Q}[x_1,x_2,x_4]$ and we get the functions: $u_1' = u_1(x_1,x_2,x_4-x_1-x_2,x_3)$, $u_2' = u_2(x_1,x_2,x_4-x_1-x_2,x_3)$, $\Bar{u}_1' = u_1(x_4-x_1-x_2,x_4,x_1,x_2)$, $\Bar{u}_2' = u_2(x_4-x_1-x_2,x_4,x_1,x_2)$ and $s_2' = x_1 x_2 - (x_4-x_1-x_2) x_4$.

Note that $\Bar{u}_2'= u_2'$, hence, the cohomology of $\Gamma_{11}'$ is:
\begin{equation}
	\mathcal{H}(\Gamma_{11}')=	\mathcal{H}^0(\Gamma_{11}') = \begin{pmatrix}
		0\\
		\mathbb{Q}[x_1,x_2,x_4]/\langle u_1', u_2', s_2 \rangle (ac)\\
		\mathbb{Q}[x_1,x_2,x_4]/\langle u_1', u_2', s_2 \rangle (ad)\\
		0 (cd)
	\end{pmatrix}.
\end{equation}
After applying the corresponding quasi-isomorphism, we obtain the cohomology of $\Gamma_{11}$:
\begin{equation}
	\mathcal{H}(\Gamma_{11})=	\mathcal{H}^0(\Gamma_{11}) = \begin{pmatrix}
		0\\
		\mathbb{Q}[x_1,x_2,x_4]/\langle u_1', u_2', s_2 \rangle (ac)\\
		\mathbb{Q}[x_1,x_2,x_4]/\langle u_1', u_2', s_2 \rangle (ad)\\
		0 (cd)\\
		0 (ab)\\
		-\mathbb{Q}[x_1,x_2,x_4]/\langle u_1', u_2', s_2 \rangle (bc)\\
		-\mathbb{Q}[x_1,x_2,x_4]/\langle u_1', u_2', s_2 \rangle(bd)\\
		-\frac{(\Bar{u}_2 - \Bar{u}_2') - (u_2 - u_2')}{x_3+x_4-x_1-x_2} \mathbb{Q}[x_1,x_2,x_4]/\langle u_1', u_2', s_2 \rangle (abcd)
	\end{pmatrix},
\end{equation}
and if the gradings are restored:
\begin{equation}
	\mathcal{H}^0(\Gamma_{11}) = \begin{pmatrix}
		0\\
		\mathbb{Q}[x_1,x_2,x_4]/\langle u_1', u_2', s_2 \rangle \{2-2N\}\\
		\mathbb{Q}[x_1,x_2,x_4]/\langle u_1', u_2', s_2 \rangle \{2-2N\}\\
		0 \\
		0 \\
		-\mathbb{Q}[x_1,x_2,x_4]/\langle u_1', u_2', s_2 \rangle \{2-2N\}\\
		-\mathbb{Q}[x_1,x_2,x_4]/\langle u_1', u_2', s_2 \rangle\{2-2N\}\\
		-\frac{(\Bar{u}_2 - \Bar{u}_2') - (u_2 - u_2')}{x_3+x_4-x_1-x_2} \mathbb{Q}[x_1,x_2,x_4]/\langle u_1', u_2', s_2 \rangle \{6-4N\}
	\end{pmatrix}.
\end{equation}
We now have a monocomplex of the cohomologies for the Hopf link:
\begin{equation}\label{Hom-Hopf-compl}
% https://tikzcd.yichuanshen.de/#N4Igdg9gJgpgziAXAbVABwnAlgFyxMJZAJgBoAGAXVJADcBDAGwFcYkQAjEAX1PU1z5CKMsWp0mrdgGMefEBmx4CRcqQCM4hizaIQ9OfyVCiAFg1bJukFB7iYUAObwioAGYAnCAFskZEDgQSOS87l6+iP6BSOqhIJ4+wTTRiADMcQkR6slBadyU3EA
\begin{tikzcd}
	&  & \mathcal{H}(\Gamma_{10})  \arrow{rrd}{d_{*1}} &  &   \\
	\mathcal{H}(\Gamma_{00}) \arrow{rru}{d_{0*}} \arrow{rrd}{d_{*0}} &  &               &  & \mathcal{H}(\Gamma_{11})  \\
	&  & \mathcal{H}(\Gamma_{01})  \arrow{rru}{d_{1*}} &  &  
\end{tikzcd}
\end{equation}
The differentials in this complex are composed of the block matrices $U^0$ and $U^1$ from \eqref{chi0}. As an example, we write the differential $d_{*0}$. The map $\chi_0$ acts between factorizations with indices $(a)$ and $(c)$ from the tensor products \eqref{gamma00} and \eqref{gamma10}, and acts as the identity on the remaining part. We obtain the following matrix for the differential in the basis \eqref{homgamma00},\eqref{homgamma10}:
\begin{equation}
	d_{*0} = \begin{pmatrix}
		U^0 & & & \\
		& U^1 & & \\
		& & U^1 & \\
		& & & U^0
	\end{pmatrix}:\; \mathcal{H}(\Gamma_{00}) \rightarrow \mathcal{H}(\Gamma_{10})\,.
\end{equation}
Here we fix $\mu = 0$ in~\eqref{chi0}. By writing out all the differentials explicitly, we can express the complex in the following form:
\begin{equation}
	% https://tikzcd.yichuanshen.de/#N4Igdg9gJgpgziAXAbVABwnAlgFyxMJZARgBpiBdUkANwEMAbAVxiRDpAF9T1Nd9CKAEykADFVqMWbAEZceIDNjwEiIoRPrNWiEAGN5vZQKIBmcpqk6QUQ4r4rByEZWpbpugGZ2l-VSlELNys2UR8HExQAFiDJbVCuCRgoAHN4IlBPACcIAFskQJAcCCRibkyc-MRC4qQhcpBsvNLqWsRTBqaqkSKS9uoGOhkYBgAFCP8QLKwUgAscOy6kAFZWvtFOyqRzXqQANk4KTiA
	\begin{tikzcd}
		&                         & \frac{\mathbb{Q}[x_1,x_2,x_3,x_4]}{\langle x_3-x_1, x_4-x_2, u_1, u_2 \rangle}\{4-4N\} \arrow{rd}{x_3-x_2} &             &   \\
		0 \arrow[r] &\frac{\mathbb{Q}[x_1,x_2,x_3,x_4]}{\langle x_3-x_1, x_4-x_2, x_1^N, x_2^N\rangle}\{4-4N\} \arrow{ru}{\mathds{1}} \arrow{rd}{\mathds{1}} & \bigoplus            & \frac{\mathbb{Q}[x_1,x_2,x_3,x_4]}{\langle x_3+x_4-x_1-x_2, u_1, u_2, x_1 x_2 - x_3 x_4 \rangle}\{2-4N\} \arrow[r] & 0 \\
		&                         & \frac{\mathbb{Q}[x_1,x_2,x_3,x_4]}{\langle x_1-x_3, x_2-x_4,u_1, u_2 \rangle}\{4-4N\} \arrow{ru}{-(x_1-x_4)} &             &  
	\end{tikzcd}
\end{equation}
where all maps should be understood as multiplications by the corresponding expressions. Let us now compute the kernels and images of the corresponding differentials:
\begin{equation}
	{\rm Ker}(d_0) = \langle x_1^k u_2(x_1,x_2,x_1,x_2)| k=0,...,N-1 \rangle\{4-4N\} \subset \frac{\mathbb{Q}[x_1,x_2,x_3,x_4]}{\langle x_3-x_1, x_4-x_2, x_1^N, x_2^N\rangle}\{4-4N\}\,,
\end{equation}
\begin{equation}
	{\rm Im}(d_0) = \frac{\mathbb{Q}[x_1,x_2,x_3,x_4]}{\langle x_3-x_1, x_4-x_2, u_1, u_2 \rangle}\{4-4N\} \oplus \frac{\mathbb{Q}[x_1,x_2,x_3,x_4]}{\langle x_3-x_1, x_4-x_2, u_1, u_2 \rangle}\{4-4N\}\,,
\end{equation}
\begin{equation}
	{\rm Ker}(d_1) = \frac{\mathbb{Q}[x_1,x_2,x_3,x_4]}{\langle x_3-x_1, x_4-x_2, u_1, u_2 \rangle}\{4-4N\} \oplus \frac{\mathbb{Q}[x_1,x_2,x_3,x_4]}{\langle x_3-x_1, x_4-x_2, u_1, u_2 \rangle}\{4-4N\}\,,
\end{equation}
\begin{equation}
	{\rm Im}(d_1) = (x_3-x_2) \frac{\mathbb{Q}[x_1,x_2,x_3,x_4]}{\langle x_3+x_4-x_1-x_2, u_1, u_2, x_1 x_2 - x_3 x_4 \rangle}\{2-4N\}\,,
\end{equation}
\begin{equation}
	{\rm Ker}(d_2) =  \frac{\mathbb{Q}[x_1,x_2,x_3,x_4]}{\langle x_3+x_4-x_1-x_2, u_1, u_2, x_1 x_2 - x_3 x_4 \rangle}\{2-4N\}\,.
\end{equation}
Thus, the cohomologies and their quantum dimensions are 
\begin{equation}
\begin{aligned}
	{\cal H}_0 &= {\rm Ker}(d_0) = \langle x_1^k u_2(x_1,x_2,x_1,x_2)| k=0,...,N-1 \rangle\{4-4N\} \; \Rightarrow \; \dim_q({\cal H}_0) = q^{4-4N} \cdot q^{2N-2} \cdot q^{N-1}[N] = q^{1-N}[N]\,, \\
	{\cal H}_1 &= {\rm Ker}(d_1)/{\rm Im}(d_0) = 0 \; \Rightarrow \; \dim_q({\cal H}_1) = 0 \,, \\
	{\cal H}_2 &= {\rm Ker}(d_2)/{\rm Im}(d_1) = \frac{\mathbb{Q}[x_1,x_2,x_3,x_4]}{\langle x_3-x_2,\, x_3+x_4-x_1-x_2,\, u_1,\, u_2,\, x_1 x_2 - x_3 x_4 \rangle}\{2-4N\} = \frac{\mathbb{Q}[x_1,x_2]}{\langle u_1,\, u_2  \rangle}\{2-4N\}\,, \\
	&\Rightarrow \; \dim_q({\cal H}_2) = q^{2-4N}\cdot q^{N-1} \cdot q^{N-2}[N][N-1]\,.
\end{aligned}
\end{equation}
The corresponding Poincare polynomial is:
\begin{equation}
	{\rm KhR}(t,q,q^N) = q^{1-N}[N] + t^2 q^{-2N-1}[N] [N-1]
\end{equation}
where recall that $[N]:=\frac{q^N - q^{-N}}{q-q^{-1}}$. Note that this answer coincides with the known expression from the literature~\cite{carqueville2014computing}.

\section{$N=2$ Khovanov--Rozansky cohomology}\label{sec:KhR-N=2}
Using the excluding variables theorem, it can be shown that for $N=2$, Khovanov--Rozansky cohomology reduces to the Khovanov cohomology from \cite{BN1}. Furthermore, we demonstrate how to construct a Koszul-like complex in Grassmann variables for $N=2$ according to the approach in \cite{Mor}. One can see a different proof in~\cite{hughes2014note}.

To do this, let us look at matrix factorization over $R=\mathbb{Q}[x_1,x_2,x_3,x_4]$ for MOY-vertex for $N=2$ (for simplicity we omit gradings in factorizations further):

% \begin{equation}
% \begin{picture}(300,50)(0,0)

%    \linethickness{0.26mm}

%     \put(-55,5){
%    \put(135,0){\vector(1,1){15}}
%    \put(150,0){\vector(-1,1){15}}
%    \put(142.5,7.5){\circle*{4.1}}
%    }

%     \put(0,0){\mbox{$\Gamma_\bullet=$}}   
    
% \end{picture}    
% \end{equation}

\begin{equation}
    \begin{split}
    \Gamma_\bullet \otimes M =
    \begin{array}{c}
         \begin{tikzpicture}[scale=1.0]
         	\draw[thick, postaction={decorate},decoration={markings, mark= at position 0.75 with {\arrow{stealth}}, mark= at position 0.25 with {\arrow{stealth}}}] (-0.5,-0.5) -- (0.5,0.5) node[left,pos=0.3] {$\scriptstyle x_4$} node[right,pos=0.7] {$\scriptstyle x_2$};
         	\draw[thick, postaction={decorate},decoration={markings, mark= at position 0.75 with {\arrow{stealth}}, mark= at position 0.25 with {\arrow{stealth}}}] (0.5,-0.5) -- (-0.5,0.5) node[right,pos=0.3] {$\scriptstyle x_3$} node[left,pos=0.7] {$\scriptstyle x_1$};
         	\draw[fill=\myred] (0,0) circle (0.08);
         	\begin{scope}[rotate=45]
         		\node[\myblue] at (1.1,0) {$\scriptstyle M$};
         		\draw[fill=\myblue, even odd rule] (0,0) circle (0.7)  (0,0) circle (0.9);
         	\end{scope}
         \end{tikzpicture}
    \end{array}
    = [R\xrightarrow[]{-3(x_3+x_4)} R\xrightarrow[]{x_1 x_2-x_3 x_4} R ]\otimes [R\xrightarrow[]{u_1(x_1,x_2,x_3,x_4)}R\xrightarrow[]{x_1+x_2-x_3-x_4}R]\otimes M\,, \\
        u_1(x_1,x_2,x_3,x_4)=x_1^2+x_2^2+(x_3+x_4)^2+x_1(x_3+x_4-x_2)+x_2(x_3+x_4)\,, \\
        u_2(x_1,x_2,x_3,x_4)= - 3(x_3 + x_4)
    \end{split}
\end{equation}
where we have written the explicit values of $u_1$ and $u_2$ for $N=2$. We have a closed diagram $\Gamma_\bullet\otimes M$, so that this factorization has the zero potential. Here we again denote factorizations and graphs by the same letters.

Using theorem from Section~\ref{sec:mat-fact}, exclude the variable $x_3$ by the relation $-3x_3-3x_4=0$. After this, we have: \begin{equation}
    \Gamma'_\bullet=[R\xrightarrow{u_1(x_1,x_2,-x_4,x_4)}R\xrightarrow{x_1+x_2}R]\langle 1 \rangle\,.
\end{equation}
Here the grading $\langle 1 \rangle$ appears because we factor by the first polynomial $a_1 = -3(x_3+x_4)$, see the note at the end of Section~\ref{sec:mat-fact}. After another exclusion by $x_1+x_2=0$, we obtain the following quasi-isomorphism:
\begin{equation}
    \Gamma_\bullet \otimes M \cong \left(M/\langle x_1 + x_2,\, x_3 + x_4 \rangle\right)\langle 1 \rangle\,.
\end{equation}
% \begin{equation}
%     M\otimes\Gamma_\bullet|_R\cong M |_{R'},\quad R'=R/(x_1=-x_2,x_3=-x_4)
% \end{equation}
We can draw this quasi-isomorphism in the picture:
\begin{equation}\label{n2}
\begin{picture}(300,30)(0,0)
    \linethickness{0.26mm}

    \put(85,30){\mbox{\scriptsize $x_1$}}
    \put(105,30){\mbox{\scriptsize $x_2$}}
    \put(85,-3){\mbox{\scriptsize $x_4$}}
    \put(105,-3){\mbox{\scriptsize $x_3$}}
    
{\setlength{\unitlength}{1.2pt}
    \put(-60,5){
    
   \put(135,0){\vector(1,1){15}}
   \put(150,0){\vector(-1,1){15}}
   \put(142.5,7.5){\color{\myred}\circle*{4.1}}
   }
}   

   \put(125,10){\mbox{$\cong$}}

{\setlength{\unitlength}{0.4pt}
\put(350,35){

\put(10,35){\mbox{\scriptsize $-x_2$}}
\put(70,35){\mbox{\scriptsize $x_2$}}
\put(10,-40){\mbox{\scriptsize $-x_4$}}
\put(70,-40){\mbox{\scriptsize $x_4$}}

\qbezier(30,20)(50,0)(70,20)
\qbezier(30,-20)(50,0)(70,-20)
}
}
   
\end{picture}
\end{equation}
% \begin{equation}
%     \begin{array}{c}
%          \begin{tikzpicture}[scale=0.6]
% 				\draw[thick,-stealth] (-0.5,-0.5) -- (0.5,0.5);
% 				\draw[thick,-stealth] (0.5,-0.5) -- (-0.5,0.5);
% 				\draw[fill=black] (0,0) circle (0.15);
%                 \node[above left] at (-0.5,0.5) {$\scriptstyle x_1$};
% 		\node[above right] at (0.5,0.5) {$\scriptstyle x_2$};
% 		\node[below right] at (0.5,-0.5) {$\scriptstyle x_3$};
% 		\node[below left] at (-0.5,-0.5) {$\scriptstyle x_4$};
% 		\end{tikzpicture} 
%     \end{array} \cong \begin{array}{c}
%          \begin{tikzpicture}[scale=0.5]
% 		\draw[thick] (-0.5,-0.5) to[out=45,in=135] (0.5,-0.5);
% 		\draw[thick] (-0.5,0.5) to[out=315,in=225] (0.5,0.5);
% 		\node[above left] at (-0.5,0.5) {$\scriptstyle -x_2$};
% 		\node[above right] at (0.5,0.5) {$\scriptstyle x_2$};
% 		\node[below right] at (0.5,-0.5) {$\scriptstyle x_4$};
% 		\node[below left] at (-0.5,-0.5) {$\scriptstyle -x_4$};
% 	\end{tikzpicture} 
%     \end{array}
% \end{equation}
From Section \ref{sec:excl-var-strand}, we know that we can exclude variables in diagrams $\Gamma_0 \xrightarrow{f} \Gamma_1$ until there is an equal number of them and $f$ transforms to a map $f$ with identified variables. Because of this and because of the locality of $\chi\,$, all maps of our hypercube can be defined from the following maps:
\begin{equation}
	m: \Gamma_0 \xrightarrow{\chi_0} \Gamma_1 \xrightarrow{f} \Gamma_1'
\end{equation}
\begin{equation}
	\Delta: \Gamma_1' \xrightarrow{g} \Gamma_1 \xrightarrow{\chi_1} \Gamma_0
\end{equation}
where
\begin{equation}\Gamma_0=
	\begin{array}{c}
		\begin{tikzpicture}[scale=0.75]
			\draw[thick,-stealth] (0.5,-0.5) to[out=90,in=270] (0.3,0) to[out=90,in=270] (0.5,0.5);
			\draw[thick,-stealth] (-0.5,-0.5) to[out=90,in=270] (-0.3,0) to[out=90,in=270] (-0.5,0.5);
			\draw[thick] (-0.5,0.5) to[out=90,in=90] (-0.8,0.5) -- (-0.8,-0.5) to[out=270,in=270] (-0.5,-0.5);
			\begin{scope}[xscale=-1]
				\draw[thick] (-0.5,0.5) to[out=90,in=90] (-0.8,0.5) -- (-0.8,-0.5) to[out=270,in=270] (-0.5,-0.5);
			\end{scope}
			\node[right] at (-0.63,-0.65) {$\scriptstyle x_1$};
			\node[right] at (-0.63,0.65) {$\scriptstyle x_1$};
			\node[left] at (0.65,-0.65) {$\scriptstyle x_2$};
			\node[left] at (0.65,0.65) {$\scriptstyle x_2$};
		\end{tikzpicture} 
	\end{array},\qquad \Gamma_1 = \begin{array}{c}
		\begin{tikzpicture}[scale=0.75]
			\draw[thick,-stealth] (0.5,-0.5) to[out=90,in=270] (-0.5,0.5);
			\draw[thick,-stealth] (-0.5,-0.5) to[out=90,in=270] (0.5,0.5);
			\draw[thick] (-0.5,0.5) to[out=90,in=90] (-0.8,0.5) -- (-0.8,-0.5) to[out=270,in=270] (-0.5,-0.5);
			\begin{scope}[xscale=-1]
				\draw[thick] (-0.5,0.5) to[out=90,in=90] (-0.8,0.5) -- (-0.8,-0.5) to[out=270,in=270] (-0.5,-0.5);
			\end{scope}
			\node[right] at (-0.63,-0.65) {$\scriptstyle x_1$};
			\node[right] at (-0.63,0.65) {$\scriptstyle x_1$};
			\node[left] at (0.65,-0.65) {$\scriptstyle x_2$};
			\node[left] at (0.65,0.65) {$\scriptstyle x_2$};
			\draw[fill=\myred] (0,0) circle (0.1);
		\end{tikzpicture}
	\end{array}
\end{equation}
and $f$, $g$ are quasi-isomorphisms from \eqref{n2}.

Let us introduce the ring $R=\mathbb{Q}[x_1,x_2]$ and write the factorizations $\Gamma_0$ and $\Gamma_1$ along with their cohomologies:
\begin{equation}
	\begin{split}
		\Gamma_0 = \begin{pmatrix}
			R\\
			R
		\end{pmatrix} \xrightarrow{\begin{pmatrix}
			x_1^2 & 0 \\
			x_2^2 & 0
			\end{pmatrix}} \begin{pmatrix}
			R\\
			R
		\end{pmatrix} \xrightarrow{\begin{pmatrix}
			0 & 0\\
			x_2^2 & -x_1^2
			\end{pmatrix}} \begin{pmatrix}
			R\\
			R
		\end{pmatrix},\\
		\mathcal{H}(\Gamma_0) = \mathcal{H}^0(\Gamma_0) = \begin{pmatrix}
			0\\
			\mathbb{Q}[x_1,x_2]/\langle x_1^2, x_2^2 \rangle 
		\end{pmatrix},
	\end{split}
\end{equation}
\begin{equation}
	\begin{split}
	\Gamma_1 = \begin{pmatrix}
		R\\
		R
	\end{pmatrix} \xrightarrow{\begin{pmatrix}
		3(x_1^2 + x_1 x_2 + x_2^2) & 0 \\
		-3(x_1+x_2) & 0
		\end{pmatrix}} \begin{pmatrix}
		R\\
		R
	\end{pmatrix} \xrightarrow{\begin{pmatrix}
		0 & 0 \\
		 -3(x_1+x_2) &  -3(x_1^2 + x_1 x_2 + x_2^2)
		\end{pmatrix}} \begin{pmatrix}
		R\\
		R
	\end{pmatrix}, \\
	\mathcal{H}(\Gamma_1) = \mathcal{H}^0(\Gamma_1) = \begin{pmatrix}
		0 \\
		\mathbb{Q}[x_1,x_2]/\langle 3(x_1^2 + x_1 x_2 + x_2^2), -3(x_1+x_2) \rangle 
	\end{pmatrix}.
	\end{split}
\end{equation}
After excluding the variable $x_2 = -x_1$, we obtain the following factorization $\Gamma_1'$ over the ring $R'=\mathbb{Q}[x_1,x_2]/\langle x_1+x_2 \rangle$ (we also account for the grading shift $\langle 1 \rangle$):
\begin{equation}
	\begin{split}
		\Gamma_1' = R' \xrightarrow{0} R' \xrightarrow{3x_1^2} R'\,, \\
		\mathcal{H}(\Gamma_1') = \mathcal{H}^0(\Gamma_1') = \mathbb{Q}[x_1,x_2]/\langle x_1+x_2, x_1^2 \rangle\,. 
	\end{split}
\end{equation}
Let us now describe the maps $m$ and $\Delta$ in detail:
\begin{equation}
	m: \begin{pmatrix}
		0\\
		\mathbb{Q}[x_1,x_2]/\langle x_1^2, x_2^2 \rangle
	\end{pmatrix} \xrightarrow{\mathds{1}} \begin{pmatrix}
	0 \\
	\mathbb{Q}[x_1,x_2]/\langle 3(x_1^2 + x_1 x_2 + x_2^2), -3(x_1+x_2) \rangle 
	\end{pmatrix} \; \mapsto \; \mathbb{Q}[x_1,x_2]/\langle x_1+x_2, x_1^2 \rangle\,, 
\end{equation}
\begin{equation}
	\Delta: \mathbb{Q}[x_1,x_2]/\langle x_1+x_2, x_1^2 \rangle \mapsto \begin{pmatrix}
		0 \\
		\mathbb{Q}[x_1,x_2]/\langle x_1+x_2, x_1^2 \rangle   
	\end{pmatrix} \xrightarrow{x_1-x_2}\begin{pmatrix}
	0\\
	\mathbb{Q}[x_1,x_2]/\langle x_1^2, x_2^2 \rangle 
	\end{pmatrix}. 
\end{equation}
 So, we can find the explicit form of the maps $m:\bigcirc \bigcirc \mapsto \bigcirc$ and $\Delta:\bigcirc \mapsto \bigcirc \bigcirc$, which act the same way as the maps in the Khovanov complex:
\begin{equation}
	m=\mathds{1}:\mathbb{Q}[x_1,x_2]/\langle x_1^2, x_2^2 \rangle \rightarrow \mathbb{Q}[x_1,x_2]/\langle x_1+x_2, x_1^2 \rangle\,,
\end{equation}
\begin{equation}
	\Delta=x_1 - x_2:\mathbb{Q}[x_1,x_2]/\langle x_1+x_2, x_1^2 \rangle \rightarrow \mathbb{Q}[x_1,x_2]/\langle x_1^2, x_2^2 \rangle\,.
\end{equation}
For convenience, we change the basis in the polynomial rings by setting $x_2 \to -x_2$ and restore the quantum grading. Then the differentials can be rewritten as:
\begin{equation}
	m=\mathds{1}:\mathbb{Q}[x_1,x_2]/\langle x_1^2, x_2^2 \rangle \{-2\}\rightarrow \mathbb{Q}[x_1,x_2]/\langle x_1-x_2, x_1^2 \rangle\{-1\}\,,
\end{equation}
\begin{equation}
	\Delta=x_1+ x_2:\mathbb{Q}[x_1,x_2]/\langle x_1-x_2, x_1^2 \rangle\{-1\} \rightarrow \mathbb{Q}[x_1,x_2]/\langle x_1^2, x_2^2 \rangle\{-2\}\,.
\end{equation}
Note that $\mathbb{Q}[x_1,x_2]/\langle x_1^2, x_2^2 \rangle \cong \mathbb{Q}[x_1]/x_1^2 \otimes \mathbb{Q}[x_2]/x_2^2$ and $\mathbb{Q}[x_1,x_2]/\langle x_1-x_2, x_1^2 \rangle \cong \mathbb{Q}[x_1]/x_1^2$. After renaming variables we can write:
\begin{equation}
	m=\mathds{1}:\mathbb{Q}[x_1]/\langle x_1^N\rangle \{-1\} \otimes \mathbb{Q}[x_1]/\langle x_1^N\rangle \{-1\} \rightarrow \mathbb{Q}[x_3]/\langle x_3^2 \rangle\{-1\}\,,
\end{equation}
\begin{equation}
	\Delta=x_1+ x_2:\mathbb{Q}[x_3]/\langle x_3^2 \rangle\{-1\} \rightarrow \mathbb{Q}[x_1]/\langle x_1^N\rangle \{-1\} \otimes \mathbb{Q}[x_1]/\langle x_1^N\rangle \{-1\}\,.
\end{equation}
The explicit action is:
\begin{equation}
	\begin{split}
	m=\mathds{1}: \; 1 \otimes 1 \mapsto 1,\quad x_1\otimes 1 \mapsto x_3, \quad 1 \otimes x_2 \mapsto x_3, \quad x_1\otimes x_2 \mapsto 0\,,
	\end{split}
\end{equation}
\begin{equation}
	\Delta=x_1-x_2: \; 1 \mapsto x_1\otimes 1 + 1 \otimes x_3, \quad x_3 \mapsto x_1\otimes x_2\,.
\end{equation}
That is, a closed cycle corresponds to a two-dimensional space $\mathbb{Q}[x]/\langle x^2 \rangle \{-1\} = \langle 1, x \rangle \{-1\}$, and note that $\dim_q(\mathbb{Q}[x]/\langle x^2 \rangle) = q [2]$. Let us introduce graded Grassmann variables $\vartheta_i$, $i=1,2$ ($\{\vartheta_i,\vartheta_j\}= \vartheta_i \vartheta_j + \vartheta_j \vartheta_i = 0$) and the space $V_2 = \langle \vartheta_1, \vartheta_2\rangle$ with $\deg (\vartheta_j) = q^{3-2j}$, so $\dim_q(V_2)=[2]$.

Now, we associate to closed cycles vector spaces generated by Grassmann variables $\bigcirc \rightarrow V_2$. The maps $m$ and $\Delta$ in these variables are rewritten as:
\begin{equation}
	m=\vartheta^{(3)}_1 \frac{\partial^2 }{\partial \vartheta^{(1)}_1 \partial \vartheta^{(2)}_1} + \vartheta^{(3)}_2 \left(\frac{\partial^2 }{\partial \vartheta^{(1)}_1 \partial \vartheta^{(2)}_2} + \frac{\partial^2 }{\partial \vartheta^{(1)}_2 \partial \vartheta^{(2)}_1}\right):\; V_2^{(1)} \otimes V_2^{(2)} \rightarrow V_2^{(3)}\,,
\end{equation}
\begin{equation}
	\Delta = \left(\vartheta^{(1)}_1 \vartheta^{(2)}_2 + \vartheta^{(1)}_2 \vartheta^{(2)}_1 \right) \frac{\partial}{\partial \vartheta^{(3)}_1} + \vartheta^{(1)}_2 \vartheta^{(2)}_2 \frac{\partial}{\partial \vartheta^{(3)}_2}:\; V_2^{(3)}\rightarrow V_2^{(1)} \otimes V_2^{(2)}\,. 
\end{equation}
So, we can construct a Koszul-like complex for Khovanov cohomology from \cite{Mor}. 

\section{Bipartite calculus for Khovanov--Rozansky cohomology}\label{sec:bip-red}

%\subsection{Bipartite reduction}

In this section, we show that for the case of bipartite links, the computation of Khovanov-Rozansky homologies simplifies significantly. To achieve this, we consider the bipartite vertex reduction introduced by Krasner \cite{K} and demonstrate that the bicomplex reduces to an ordinary complex of vector spaces. We then find the operators acting on these spaces.

Consider a reduction of a bipartite vertex from the Krasner work \cite{K}: 
\begin{equation}\label{bip-vert-compl}
\setlength{\unitlength}{0.65pt}
\begin{picture}(300,55)(200,-27.5)

\linethickness{0.2mm}

\put(40,28){\mbox{\footnotesize $x_1$}}
\put(110,28){\mbox{\footnotesize $x_3$}}
\put(40,-33){\mbox{\footnotesize $x_2$}}
\put(110,-33){\mbox{\footnotesize $x_4$}}

\qbezier(50,20)(55,9)(58,4) \qbezier(63,-4)(85,-40)(110,20)
\put(56,8){\vector(1,-2){2}} \put(90,-13){\vector(1,1){2}} \put(109,18){\vector(1,2){2}}
\qbezier(50,-20)(75,40)(97,4)  \qbezier(102,-4)(105,-9)(110,-20)
\put(104,-8){\vector(-1,2){2}} \put(70,13){\vector(-1,-1){2}} \put(51,-18){\vector(-1,-2){2}}

\put(125,-5){\mbox{\Large $:$}}

\put(5,0){

\put(120,0){

\put(22,28){\mbox{\footnotesize $x_1$}}
\put(70,28){\mbox{\footnotesize $x_3$}}
\put(22,-33){\mbox{\footnotesize $x_2$}}
\put(70,-33){\mbox{\footnotesize $x_4$}}

\put(50,10){\vector(1,0){4}}
\qbezier(30,20)(50,0)(70,20)
\qbezier(30,-20)(50,0)(70,-20)
\put(50,-10){\vector(-1,0){4}}
}

\put(205,-5){\mbox{$\{-2N\} \ \overset{S}{\longrightarrow}$}}

\put(50,0){

\put(135,0){

\put(100,0){
\put(18,28){\mbox{\footnotesize $x_1$}}
\put(60,28){\mbox{\footnotesize $x_3$}}
\put(18,-33){\mbox{\footnotesize $x_2$}}
\put(60,-33){\mbox{\footnotesize $x_4$}}
}

\put(135,0){\vector(0,-1){5}}
\put(150,0){\vector(0,1){5}}
\qbezier(125,20)(145,0)(125,-20)
\qbezier(160,20)(140,0)(160,-20)
}

\put(310,-5){\mbox{$\{-N-1\} \ \xrightarrow[]{x_1-x_4}$}}

\put(325,0){

\put(100,0){
\put(18,28){\mbox{\footnotesize $x_1$}}
\put(60,28){\mbox{\footnotesize $x_3$}}
\put(18,-33){\mbox{\footnotesize $x_2$}}
\put(60,-33){\mbox{\footnotesize $x_4$}}
}

\put(135,0){\vector(0,-1){5}}
\put(150,0){\vector(0,1){5}}
\qbezier(125,20)(145,0)(125,-20)
\qbezier(160,20)(140,0)(160,-20)
}

\put(500,-5){\mbox{$\{-N+1\}$}}
}
}

\end{picture}
\end{equation}

where the saddle map $S: \upd{} \xrightarrow[]{} \ler{}\langle1\rangle$ (as the map between factorizations) is defined by the two matrices:
\begin{equation}
    \begin{split}
    \label{saddle}
    S_0=\begin{pmatrix}
        e_{123} + e_{124} +(x_4 - x_3)r & 1 \\
        -e_{134} -e_{234} +(x_1 -x_2)r & 1
    \end{pmatrix},\quad S_1=\begin{pmatrix}
        -1 & 1 \\
        -e_{123} -e_{234} +(x_1 -x_4)r &-e_{134}-e_{124} + (x_3 - x_2)r 
    \end{pmatrix},\\
    e_{ijk} = \sum_{a+b+c = N-1} x_{i}^{a}x_{j}^{b}x_{k}^{c},\quad r \in \mathbb{Q}^{N-2}[x_1,x_2,x_3,x_4]
    \end{split}
\end{equation}
where $\mathbb{Q}^{N-2}[x_1,x_2,x_3,x_4]$ are polynomials of degree $N-2$ in $x_1,x_2,x_3,x_4$ variables. Formula~\eqref{bip-vert-compl} is the result of reductions of both the vertical and the horizontal morphisms. First, the matrix factorization of 2-dotted vertex can be reduced, see~\eqref{MOYloc_V}. Second, one can change bases in cohomologies in order to break the resulting monocomplex into direct sum of subcomplexes. Some subcomplexes are then eliminated because they are exact, i.e. do not contribute in the final cohomology. As the result, only three vertices remain. 

By applying this reduction to a link, we obtain a 3-hypercube (i.e. each edge has 3 vertices). Its vertices correspond to configurations of closed cycles, and its edges correspond to the maps $\pm S$ and $ \pm (x_i-x_j)$, which act on the respective cycles.

As in $N=2$ Khovanov--Rozansky cohomology, because of the locality of $\chi\,$, all maps of our hypercube can be defined from the following maps:
\begin{equation}
\begin{picture}(300,50)(160,-30)
    \put(125,10){\circle{20}}
    \put(112,10){\line(1,0){6}}
    \put(132,10){\line(1,0){6}}
    \put(105,10){\mbox{\small $x$}}
    \put(143,10){\mbox{\small $y$}}

    \put(160,7){\mbox{$\longrightarrow$}}

    \put(205,10){\circle{20}}
    \put(192,10){\line(1,0){6}}
    \put(185,10){\mbox{\small $x$}}

    \put(35,0){
    \put(205,10){\circle{20}}
    \put(192,10){\line(1,0){6}}
    \put(185,10){\mbox{\small $y$}}
    }

    \put(140,0){
    
    \put(205,10){\circle{20}}
    \put(192,10){\line(1,0){6}}
    \put(185,10){\mbox{\small $x$}}

    \put(35,0){
    \put(205,10){\circle{20}}
    \put(192,10){\line(1,0){6}}
    \put(185,10){\mbox{\small $y$}}

    \put(225,7){\mbox{$\longrightarrow$}}

    \put(150,0){
    \put(125,10){\circle{20}}
    \put(112,10){\line(1,0){6}}
    \put(132,10){\line(1,0){6}}
    \put(105,10){\mbox{\small $x$}}
    \put(143,10){\mbox{\small $y$}}
    }
    }
    
    }

    \put(0,-30){
    
    \put(125,10){\circle{20}}
    \put(112,10){\line(1,0){6}}
    \put(105,10){\mbox{\small $x$}}

    \put(145,7){\mbox{$\longrightarrow$}}

    \put(65,0){
    \put(125,10){\circle{20}}
    \put(112,10){\line(1,0){6}}
    \put(105,10){\mbox{\small $x$}}
    }

    \put(140,0){
    
    \put(205,10){\circle{20}}
    \put(192,10){\line(1,0){6}}
    \put(185,10){\mbox{\small $x$}}

    \put(35,0){
    \put(205,10){\circle{20}}
    \put(192,10){\line(1,0){6}}
    \put(185,10){\mbox{\small $y$}}

    \put(225,7){\mbox{$\longrightarrow$}}

    \put(70,0){
    \put(205,10){\circle{20}}
    \put(192,10){\line(1,0){6}}
    \put(185,10){\mbox{\small $x$}}
    
    \put(35,0){
    \put(205,10){\circle{20}}
    \put(192,10){\line(1,0){6}}
    \put(185,10){\mbox{\small $y$}}
    }
    }
    }
    
    }
    
    }
    
\end{picture}
\end{equation}
% \begin{equation}
%     \begin{split}
%         \bigcirc (two marks) \xrightarrow{} \bigcirc \bigcirc,\quad \bigcirc \bigcirc \xrightarrow{} \bigcirc (two marks)\\
%         \bigcirc \xrightarrow{} \bigcirc,\quad \bigcirc \bigcirc \xrightarrow[]{} \bigcirc\bigcirc
%     \end{split}
% \end{equation}
Let us calculate these maps. For convenience, we calculate several factorizations now:
\begin{equation}
\begin{split}
        \bigcirc\bigcirc=\begin{pmatrix}
            R\\
            R\{2-2N\}
        \end{pmatrix} \xrightarrow{\begin{pmatrix}
            x^N&0\\
            y^N&0
        \end{pmatrix}} \begin{pmatrix}
            R\{1-N\}\\
            R\{1-N\}
        \end{pmatrix} \xrightarrow{\begin{pmatrix}
            0&0\\
            y^N&-x^N
        \end{pmatrix}} \begin{pmatrix}
            R\\
            R\{2-2N\}
        \end{pmatrix}, \\
        \mathcal{H}(\bigcirc\bigcirc)=\mathcal{H}^0(\bigcirc\bigcirc)=\begin{pmatrix}
            0\\
            \mathbb{Q}[x,y]/\langle x^N,y^N\rangle\{2-2N\}
        \end{pmatrix},
\end{split}
\end{equation}
\begin{equation}
\begin{split}
        \text{--}\ \text{--}\hspace{-0.4cm}\bigcirc \ =\begin{pmatrix}
            R\\
            R\{2-2N\}
        \end{pmatrix} \xrightarrow{\begin{pmatrix}
            \pi_{xy}&x-y\\
            \pi_{yx}&x-y
        \end{pmatrix}} \begin{pmatrix}
            R\{1-N\}\\
            R\{1-N\}
        \end{pmatrix} \xrightarrow{\begin{pmatrix}
            y-x&x-y\\
             \pi_{yx}&- \pi_{xy}
        \end{pmatrix}} \begin{pmatrix}
            R\\
            R\{2-2N\}
        \end{pmatrix}, \\
        \mathcal{H}(\bigcirc)=\mathcal{H}^1(\bigcirc)=\begin{pmatrix}
             \mathbb{Q}[x]/\langle x^N\rangle\{1-N\}\\
            \mathbb{Q}[x]/\langle x^N\rangle\{1-N\}
        \end{pmatrix}.
\end{split}
\end{equation}
Let us find all operators that act between cycles:

\begin{enumerate}
    \item $m:\bigcirc\bigcirc\rightarrow \bigcirc\langle1\rangle \{N-1\}$
    
    %\medskip
    
    After applying the saddle map, we have the morphism:
     \begin{equation}
        \begin{pmatrix}
            0\\
            \mathbb{Q}[x,y]/\langle x^N,y^N\rangle\{2-2N\}
        \end{pmatrix} \xrightarrow[]{\Pi(S_0)} \begin{pmatrix}
             \mathbb{Q}[x]/\langle x^N\rangle\{1-N\}\\
            \mathbb{Q}[x]/\langle x^N\rangle\{1-N\}
        \end{pmatrix}.
    \end{equation}
    
    We can rewrite this map:
    \begin{equation}
        m=1: \mathbb{Q}[x,y]/\langle x^N,y^N\rangle\{2-2N\}\rightarrow\mathbb{Q}[x]/\langle x^N\rangle\{1-N\},\quad x^k\cdot y^{l} \rightarrow x^{k+l}\,.
    \end{equation}
    \item $\Delta: \bigcirc\xrightarrow{}\bigcirc\bigcirc\langle1\rangle\{N-1\}$
    
    After applying the saddle map, we have the morphism:
     \begin{equation}
        \begin{pmatrix}
             \mathbb{Q}[x]/\langle x^N\rangle\{1-N\}\\
            \mathbb{Q}[x]/\langle x^N\rangle\{1-N\}
        \end{pmatrix}
         \xrightarrow[]{\Pi(S_1)} 
        \begin{pmatrix}
            0\\
            \mathbb{Q}[x,y]/\langle x^N,y^N\rangle\{2-2N\}
        \end{pmatrix}.
    \end{equation}
    We rewrite this map as:
    \begin{equation}
        \Delta = \sum_{m+n = N-1}x^m y^n: \mathbb{Q}[x]/\langle x^N\rangle\{1-N\}\rightarrow \mathbb{Q}[x,y]/\langle x^N,y^N\rangle\{2-2N\},\quad x^{k} \rightarrow \sum_{m+n = N-1} x^{k+m} y^n\,.
    \end{equation}
    \item ${\rm Sh}:\bigcirc\xrightarrow[]{}\bigcirc\{2\}:$
    \begin{equation}
        {\rm Sh} =0:\mathbb{Q}[x]/\langle x^N\rangle\{1-N\}\xrightarrow[]{}\mathbb{Q}[x]/\langle x^N\rangle\{1-N\},\quad x^k\xrightarrow[]{}0\,.
    \end{equation}
    \item ${\rm Sh} : \bigcirc\bigcirc\rightarrow\bigcirc\bigcirc\{2\}:$
    \begin{equation}
        {\rm Sh} =x-y: \mathbb{Q}[x,y]/\langle x^N,y^N\rangle\{2-2N\}\rightarrow\mathbb{Q}[x,y]/\langle x^N,y^N\rangle\{2-2N\},\quad x^k\cdot y^l\rightarrow x^{k+1}y^l-x^{k}y^{l+1}\,.
    \end{equation}
\end{enumerate}
Note that $\mathbb{Q}[x,y]/\langle x^N,y^N\rangle=\mathbb{Q}[x]/\langle x^N\rangle\otimes \mathbb{Q}[y]/\langle y^N\rangle$, and after renaming variables, we can write this morphisms in a more simple way:
\begin{equation}
    \begin{split}
         m: \; \mathbb{Q}[x_1]/\langle x_1^N\rangle\{1-N\}\otimes \mathbb{Q}[x_2]/\langle x_2^N\rangle\{1-N\}\rightarrow\mathbb{Q}[x_3]/\langle x_3^N\rangle\{1-N\},\quad x_1^k\otimes x_2^{l} \mapsto x_3^{k+l}\,, 
    \end{split}
\end{equation}
\begin{equation}
     \Delta : \; \mathbb{Q}[x_3]/\langle x_3^N\rangle\{1-N\}\rightarrow \mathbb{Q}[x_1]/\langle x_1^N\rangle\{1-N\}\otimes \mathbb{Q}[x_2]/\langle x_2^N\rangle\{1-N\},\quad x_3^{k} \rightarrow \sum_{m+n = N-1} x_1^{k+m}\otimes x_2^n\,,
\end{equation}
\begin{equation}
    {\rm Sh}=0: \; \mathbb{Q}[x]/\langle x^N\rangle\{1-N\}\xrightarrow[]{}\mathbb{Q}[x]/\langle x^N\rangle\{1-N\},\quad x^k\xrightarrow[]{}0\,,
\end{equation}
\begin{equation}
\begin{split}
        {\rm Sh}: \; \mathbb{Q}[x_1]/\langle x_1^N\rangle\{1-N\}\otimes \mathbb{Q}[x_2]/\langle x_2^N\rangle\{1-N\}\rightarrow\mathbb{Q}[x_3]/\langle x_3^N\rangle\{1-N\}\otimes \mathbb{Q}[x_4]/\langle x_4^N\rangle\{1-N\}\,,\\
        x_1^k\otimes x_2^l\rightarrow x_3^{k+1}\otimes x_4^l-x_3^{k}\otimes x_4^{l+1}\,.
\end{split}
\end{equation}
It is important to note that when writing these maps, we work in the ring $\mathbb{Q}[x]/\langle x^N \rangle$, where $x^k = 0$ for $k \geq N$. This means all polynomial expressions in $x$ should be considered modulo $x^N$. The explicit action of these maps is:
\begin{equation}
	\begin{split}
		m: x_1^k \otimes x_2^l \mapsto x_3^{k+l}\quad k+l \leq N-1, \quad k,l\leq N-1\,, \\
	\end{split}
\end{equation}
\begin{equation}
	\begin{split}
		\Delta:  x_3^{k} \rightarrow \sum_{k+m\leq N-1} \sum_{m+n = N-1} x_1^{k+m}\otimes x_2^n =  \sum_{k+m\leq N-1} x_1^{k+m}\otimes x_2^{N-1-m} = \sum_{m=0}^{N-k} x_1^{k+m}\otimes x_2^{N-1-m}\,,
	\end{split}
\end{equation}
\begin{equation}
	\begin{split}
	{\rm Sh}: \; x_1^i\otimes x_2^j\rightarrow x_3^{i+1}\otimes x_4^j-x_3^{i}\otimes x_4^{j+1}\quad i,j \leq N-2\,, \\
	x_1^i\otimes x_2^j\rightarrow x_3^{i+1}\otimes x_4^{N-1}\quad i\leq N-2, j=N-1\,, \\
	x_1^i\otimes x_2^j\rightarrow x_3^{N-1}\otimes x_4^{j+1}\quad j\leq N-2, i=N-1\,.
	\end{split}
\end{equation}
Thus, we can now state that a closed cycle corresponds to the $N$-dimensional space $\mathbb{Q}[x]/\langle x^N\rangle\{1-N\}$ (and we do not need to compute the homology of the corresponding matrix factorizations), while a set of $n$ circles corresponds to their tensor product $\bigotimes\limits_{i=1}^n \mathbb{Q}[x_i]/\langle x_i^N\rangle\{1-N\}$. The maps $m$, $\Delta$, ${\rm Sh}$ act between these circles.

We now introduce a planar technique for computing the Khovanov-Rozansky cohomology of bipartite links according to the approach of \cite{2506.08721}. Instead of working with polynomial rings, it is more convenient to use vector spaces whose elements are odd (Grassmann) variables $\{\vartheta_i,\vartheta_j\}=\vartheta_i\vartheta_j+\vartheta_j\vartheta_i=0$:
\begin{equation}
	\begin{split}
		\mathbb{Q}[x]/\langle x^N\rangle\{1-N\}\rightarrow V_N=\langle\vartheta_1,...,\vartheta_N\rangle\,, \\
		\deg_q(x^k)=q^{2k}\quad \deg_q(\vartheta_j)=q^{N+1-2j}\,, \\
		\dim_q\mathbb{Q}[x]/\langle x^N\rangle\{1-N\}=[N]\quad \dim_qV_N=[N]\,, \\
		x^{k-1}\mapsto \vartheta_k\,.
	\end{split}
\end{equation}
We can now express the maps $m$, $\Delta$, Sh in these new variables using differential operators in the odd variables:
\begin{equation}
	m = \sum_{1\leq i ,j\leq N;\quad i+j \leq N+1} \vartheta_{i+j-1}\frac{\partial^2}{\partial\vartheta^{(1)}_{i}\partial\vartheta^{(2)}_j}: \; V_{N}^{(1)} \otimes V_{N}^{(2)}\rightarrow V_N\,,
\end{equation}
\begin{equation}
	\Delta = \sum_{i=1}^{N}(\sum_{j=0}^{N-i} \vartheta_{N-j}^{(1)}\vartheta_{i+j}^{(2)})\frac{\partial}{\partial \vartheta_{i}}: \; V_N \rightarrow V_{N}^{(1)} \otimes V_{N}^{(2)}\,,
\end{equation}
\begin{equation}
	{\rm Sh} = \sum_{i,j=1}^{N-1}({\vartheta}_{i+1}^{(3)}{\vartheta}_{j}^{(4)} - {\vartheta}_{i}^{(3)}\Bar{\vartheta}_{j+1}^{(4)})\frac{\partial^2}{\partial \vartheta_{i}^{(1)}\partial\vartheta_{j}^{(2)}} + \sum_{i=0}^{N-1}\vartheta_{i+1}^{(3)} \vartheta_{N}^{(4)}\frac{\partial^2}{\vartheta_{i}^{(1)} \vartheta_{N}^{(2)}} + \sum_{j=0}^{N-1}\vartheta_{N}^{(3)} \vartheta_{j+1}^{(4)} \frac{\partial^2}{\vartheta_{N}^{(1)} \vartheta_{j}^{(2)}} 
	: \; V_{N}^{(1)} \otimes V_{N}^{(2)} \rightarrow {V}_{N}^{(3)} \otimes {V}_{N}^{(4)}\,,
\end{equation}

\begin{equation}
	{\rm Sh} =0: \; V_N\rightarrow V_N\,.
\end{equation}
If we set $\vartheta_{N+1} \equiv 0$, then the map Sh can be rewritten in a more convenient form:
\begin{equation}
	{\rm Sh} = \sum_{i,j=1}^{N}(\Bar{\vartheta}_{i+1}^{(1)}\Bar{\vartheta}_{j}^{(2)} - \Bar{\vartheta}_{i}^{(1)}\Bar{\vartheta}_{j+1}^{(2)})\frac{\partial^2}{\partial \vartheta_{i}^{(1)}\partial\vartheta_{j}^{(2)}}: \; V_{N}^{(1)} \otimes V_{N}^{(2)} \rightarrow \Bar{V}_{N}^{(1)} \otimes \Bar{V}_{N}^{(2)}\,.
\end{equation}
Using these operators, we can construct the Koszul-like complex for Khovanov-Rozansky cohomology. Examples of calculation and the detailed algorithm of this approach see in~\cite{2506.08721}. 

It is important to note that in the case of the reduction for $N=2$ and under the bipartite reduction, we break the locality in the Khovanov-Rozansky complex. Now the differentials act not between tangles with marked variables $x_k$, but between closed circles where these variables are identified.

\section{Conclusion}\label{sec:conclusion}

In this paper, we have demonstrated that Krasner bipartite reduction~\cite{K} implies Khovanov--Rozansky cycle calculus for bipartite links from~\cite{2506.08721}. In addition, we have presented another explanation of equivalence between $N=2$ Khovanov--Rozansky matrix factorization technique~\cite{KhR} and the Khovanov cycle calculus~\cite{Kh}. 

We have also proved that one can exclude variables in MOY diagrams~\cite{KhR} what allows to simplify calculations of the Khovanov--Rozansky polynomials.  

The next steps of research could be as follows. 

\begin{enumerate}
	\item It is tempting to find out which reductions are applicable to other tangles and to built a Khovanov--Rozansky tangle calculus as it was done for the HOMFLY polynomials in~\cite{mironov2018tangle}.
	
	\item It is interesting to explore evolution properties of Khovanov--Rozansky polynomials and their jumps in evolution parameters, see for example~\cite{anokhina2019nimble,anokhina2024towards} for such phenomena in the Khovanov ($N=2$) case.
	
	\item In~\cite{anokhina2021khovanov,morozov2018knot,anokhina2024towards-2}, the efficient method to calculate colored Khovanov polynomials was proposed. It can be extended to the Khovanov--Rozansky case. 
\end{enumerate}

\section*{Acknowledgments}

We are grateful for enlightening discussions to D. Galakhov and A. Morozov. 

Funding for this publication was generously provided by the Priority 2030 Academic Leadership Initiative, contributing to the educational work of "Universities for a New Generation of Leaders", a project within the framework of the federal Youth and Children program.

\printbibliography

% \begin{thebibliography}{12}
% \bibitem{BN1} D. Bar-Natan, On Khovanov’s categorification of the Jones polynomial, Algebraic and
% Geometric Topology, 2 (2002) 337-370, arXiv:math.QA/0201043

% \bibitem{BN2} D. Bar-Natan, Khovanov’s cohomology for Tangles and Cobordisms, Geometry and Topology 9-33 (2005) 1443–1499,
% arXiv:math.GT/0410495.

% \bibitem{BN3} D. Bar-Natan, Fast Khovanov cohomology computations, arXiv:math.GT/0606318.

% \bibitem{KhR} M. Khovanov and L. Rozansky, ”Matrix factorizations and link cohomology”,
% arXiv:math/0401268.

% \bibitem{DM} V.Dolotin and A.Morozov Introduction to Khovanov Homologies
% I. Unreduced Jones superpolynomial, arXiv:1208.4994

% \bibitem{K} Daniel Krasner A computation in Khovanov-Rozansky, arXiv:0801.4018

% \bibitem{MOY} H. Murakami, T. Ohtsuki, and S. Yamada. Homfly polynomial via an invariant of colored plane graphs. Enseign.
% Math. (2), 44(3-4):325–360, 1998.

% \bibitem{2506.08721} A. Anokhina, E. Lanina, A. Morozov. Khovanov-Rozansky cycle calculus for bipartite links. arXiv:2506.08721

% \end{thebibliography}

\end{document}